\documentclass[11pt,a4paper]{article}
\pdfoutput=1
\usepackage{jheppub}

\usepackage{color}
\usepackage{amsmath}
\usepackage{pifont}   
\usepackage{verbatim}       
\usepackage{acronym} 
    
\usepackage{amsfonts}
\usepackage{amssymb}
\usepackage{mathrsfs} 
\usepackage{graphicx}
\usepackage{multirow} 
\usepackage{slashed} 
\usepackage{array}

\usepackage{graphicx}
\usepackage{epsfig}
\usepackage{lscape}
\usepackage{rotating}
\usepackage{epsf}
\usepackage{bm}
\usepackage{booktabs}
\usepackage{array}
\usepackage{caption}
\usepackage{subcaption}
\usepackage[utf8]{inputenc}
\usepackage{float}

\usepackage{multirow}
\usepackage{array,booktabs,colortbl,multirow}
\usepackage{colortbl,xcolor,colordvi,color}
\usepackage{rotating}
\allowdisplaybreaks

\title{Cosmology of the Twin Higgs without explicit $\mathbb{Z}_2$ breaking}

\author[a, b]{Hugues Beauchesne}
\author[a]{and Yevgeny Kats}

\affiliation[a]{Department of Physics, Ben-Gurion University, \\Beer-Sheva 8410501, Israel}
\affiliation[b]{Physics Division, National Center for Theoretical Sciences,\\ Taipei 10617, Taiwan}

\emailAdd{beauchesneh@phys.ncts.ntu.edu.tw, katsye@bgu.ac.il}

\abstract{The cosmology of the Twin Higgs requires the breaking of the $\mathbb{Z}_2$ symmetry, but it is still an open question whether this breaking needs to be explicit. In this paper, we study how the Mirror Twin Higgs could be modified to be compatible with current cosmological constraints without explicit $\mathbb{Z}_2$ breaking. We first present a simple toy model that can realize baryogenesis without explicit $\mathbb{Z}_2$ breaking or reaching temperatures that would lead to domain walls. The model can also either solve the $N_{\text{eff}}$ problem and bring the abundance of mirror atoms to an allowed level or provide the correct dark matter abundance. We then present another simple model that leads to mirror neutron dark matter and thus acceptable dark matter self-interactions. We also include in appendix a series of results on energy exchange between different sectors that might prove useful for other cosmological problems.}

\begin{document}

\maketitle

\section{Introduction}\label{Sec:Introduction}
The Twin Higgs \cite{Chacko:2005pe, Barbieri:2005ri} attempts to solve the little hierarchy problem by introducing partners that are neutral under the Standard Model (SM) gauge groups and is the prime example of Neutral Naturalness. Its simplest version is the Mirror Twin Higgs. In this model, a copy of every SM field is introduced. The principal difference is that these mirror partners are instead charged under new gauge groups that reflect those of the Standard Model. The Higgs doublet and its mirror partner can then be combined to write a potential that respects an approximate global $SU(4)$ symmetry. Its spontaneous breaking to $SU(3)$ results in seven (pseudo)-Goldstone bosons. Three are eaten by the massive SM gauge bosons and three by their partners. The remaining one corresponds to the experimentally observed Higgs boson. Its mass is protected at one loop by a $\mathbb{Z}_2$ interchange symmetry which ensures that the leading correction to the potential respects $SU(4)$ and hence does not contribute to the Higgs mass directly. The latter is then effectively protected by neutral partners. This symmetry imposes an equality between the Yukawa and gauge couplings of the two sectors. Without unacceptable tuning, the mirror partners are typically a factor of a few heavier than their SM equivalents.

It is a well established fact that the $\mathbb{Z}_2$ symmetry must be broken for the Twin Higgs to be compatible with the Higgs signal strengths measurements. During the early days of the model and often still to this day, this was done by introducing explicit soft $\mathbb{Z}_2$ breaking. The possibility of doing without this explicit $\mathbb{Z}_2$ breaking and having the symmetry only broken spontaneously is both aesthetically appealing and likely to facilitate UV completions. Refs.~\cite{Beauchesne:2015lva, Harnik:2016koz, Yu:2016bku, Yu:2016swa, Jung:2019fsp} demonstrated that, at least as far as collider constraints are concerned, it is possible to do so and in addition the amount of tuning required decreases.

At the same time, the Twin Higgs model can be consistent with cosmology, which is already the subject of a considerable literature \cite{Farina:2015uea, Chacko:2016hvu, Craig:2016lyx, Farina:2016ndq, Barbieri:2017opf, Csaki:2017spo, Chacko:2018vss, Badziak:2019zys, Harigaya:2019shz, Koren:2019iuv, Curtin:2019lhm, Curtin:2019ngc, Terning:2019hgj, Feng:2020urb, Beauchesne:2020mih, Curtin:2021alk, Curtin:2021spx, Chacko:2021vin}.\footnote{See also Refs.~\cite{Hodges:1993yb, Berezhiani:2000gw, Foot:2004pa, An:2009vq, Roux:2020wkp, khlopov1, khlopov2, khlopov3, khlopov4, Foot:2014uba, Foot:2016wvj} for examples of cosmology in Mirror World.} In a similar fashion to the Higgs signal strengths, the cosmology of the Twin Higgs also requires $\mathbb{Z}_2$ breaking. It was amply demonstrated that viable cosmology can be obtained via explicit $\mathbb{Z}_2$ breaking. However, one question that is still unanswered in the literature is whether cosmology requires this breaking to be explicit. Indeed, all previous works either included some sort of explicit $\mathbb{Z}_2$ breaking, be it some gauge or Yukawa couplings being different, some particles being absent in one sector or even more esoteric possibilities, or never presented a full model that could address all known issues while providing an adequate dark matter candidate. 

The fact however is that the cosmology of the Twin Higgs with explicit $\mathbb{Z}_2$ breaking already presents major challenges. With only spontaneous breaking, these challenges are exacerbated, as respecting this symmetry imposes additional constraints and reduces our set of tools to address them.

The first such challenge is baryogenesis. The $\mathbb{Z}_2$ symmetry being only broken spontaneously almost unavoidably leads to domain walls, which may overclose the Universe. As long as inflation lasts long enough, their density can thankfully be brought to acceptable levels. In addition, domain walls will not be reintroduced during reheating as long as the reheating temperature does not reach the $\mathbb{Z}_2$ restoration scale. However, most standard baryogenesis mechanisms take place at temperatures higher or not too far from the expected $\mathbb{Z}_2$ restoration temperature. One obvious way to solve this apparent conflict is for baryogenesis to take place below the $\mathbb{Z}_2$ restoration scale. However, this could be challenging in general for mechanisms like electroweak baryogenesis \cite{Kuzmin:1985mm, Shaposhnikov:1986jp, Shaposhnikov:1987tw} or leptogenesis \cite{Fukugita:1986hr}. This is even more difficult if one wishes for dark hadrons to represent dark matter via some realization of Asymmetric Dark Matter \cite{Petraki:2013wwa, Kaplan:2009ag, Zurek:2013wia}.

The second challenge is the contribution of the mirror photon and mirror neutrinos to the effective number of relativistic degrees of freedom $N_{\text{eff}}$, which is severely constrained by both the Cosmic Microwave Background (CMB) and Big Bang Nucleosynthesis (BBN) \cite{Fields:2019pfx, Aghanim:2018eyx}. With explicit $\mathbb{Z}_2$ breaking, such particles can be removed or made heavier, but this is non-trivial when $\mathbb{Z}_2$ is not broken explicitly. Note however that some existing solutions to the $N_{\text{eff}}$ problem only need some expectation values to differ between the two sectors and could in principle be accommodated without explicit $\mathbb{Z}_2$ breaking (see e.g. Refs.~\cite{Farina:2015uea, Chacko:2016hvu, Craig:2016lyx, Barbieri:2017opf}).

The third challenge is that, even if it could explain the observed dark matter abundance via Asymmetric Dark Matter, the standard Mirror Twin Higgs would lead to dark matter in the form of mirror atoms. The problem with this scenario is that these would display self-interactions similar to normal atoms. If dark atoms were to represent the entirety of dark matter, their self-interactions would be ruled out by orders of magnitude or would require tuning at an unacceptable level \cite{Kaplan:2009de, CyrRacine:2012fz, Cline:2013pca}. It is then crucial to be able to modify the model such that the dark matter takes a more acceptable form, such as mirror neutrons.

With this context in mind, the goal of the present paper is to study the feasibility of constructing cosmologically viable Twin Higgs Models without explicit $\mathbb{Z}_2$ breaking. The construction of a full model is rather ambitious and we will instead limit ourselves to studying whether it is possible to individually solve the three challenges mentioned above. More specifically, we will study whether it is possible to realize baryogenesis without reintroducing domain walls, whether the same process can also generate the correct dark matter abundance and/or solve the $N_{\text{eff}}$ problem and whether dark matter can be converted into an acceptable form.

The end result will be that it is indeed possible to overcome these challenges. This will be demonstrated by presenting two different models. The first one includes two Majorana fermions. The heaviest one is assumed to dominate the energy content of the Universe at early times. It then decays and produces a net amount of both baryons and mirror baryons, as well as some amount of the lighter Majorana fermion. As the Universe expands, the lighter Majorana fermion comes to dominate the energy abundance. Because of kinematical reasons, it then decays mainly to the Standard Model sector, thus reheating that sector. The model can provide the correct matter abundance while maintaining temperatures that are low enough not to reintroduce domain walls. It can also either solve the $N_{\text{eff}}$ problem and reduce the abundance of dark atoms to an acceptable level or provide the correct dark matter abundance.

The second model solves the remaining problem of dark matter self-interactions. A set of vector quarks is introduced in each sector. These mix with their respective up quarks via Yukawa interactions with the Higgs. This mixing has the effect of adding to the mass of the up quark of a given sector a contribution proportional to the vev of the Higgs of that sector cubed. This results in the mass of the mirror up quark increasing faster than the mass of the mirror down as the vev of the mirror Higgs increases. As such, the mirror proton can be made considerably heavier than the mirror neutron. Dark matter then consists of mirror neutrons and the abundance of mirror atoms can be brought to negligible levels. The smallness of the mass of the up quark ensures that the required amount of mixing is small enough to be comfortably below any current experimental constraints.

The article is organized as follows. The first model is introduced, its mechanism explained, its constraints discussed and its parameter space studied. The second model is then introduced, its constraints discussed, its parameter space studied and alternative models presented. An appendix presents some useful results on energy exchange between different sectors for cosmological evolution. Additional appendices discuss the decay asymmetry, scattering asymmetries, the evolution equations, the Higgs signal strengths and the computation of the dark atom abundance.

\section{Addressing baryogenesis, dark matter abundance and $N_{\text{eff}}$}\label{Sec:BNeffR}
We begin this paper by introducing a toy model which can potentially address baryogenesis, dark matter abundance and $N_{\text{eff}}$ while maintaining temperatures below the $\mathbb{Z}_2$ restoration scale. The model serves as a proof of principle and it goes without saying that variations are possible. This section contains a description of the model, an explanation of the mechanisms involved, a discussion of the different constraints and some summary scans of parameter space. To avoid obscuring the discussion with technicalities, all mathematical details concerning the cosmological evolution are relegated to Appendices~\ref{App:TA}, \ref{Sec:DecayAsymmetry}, \ref{Sec:ScatteringAsymmetry} and \ref{Sec:Evolution}.

\subsection{Model summary}\label{sSec:ModelSummary}
The field content of the model is as follows. First, a complete copy of the Standard Model is introduced. Fields from the SM sector are labelled with an $A$ and those of the mirror sector with a $B$.\footnote{When referring to an unspecified sector, we will use the index $M$.} The Higgs doublets are labelled as $H^M$ and obtain expectation values $\langle H^{M0} \rangle= v^M/\sqrt{2}$, with $v^B$ larger than $v^A$ by a factor of a few to satisfy Higgs signal strength requirements. How these vevs are acquired is irrelevant to the present discussion, but can be done via spontaneous breaking \cite{Beauchesne:2015lva, Harnik:2016koz, Yu:2016bku, Yu:2016swa, Jung:2019fsp}. In addition, several fields without SM equivalents are introduced. Every field labelled by $\chi$ is a left-handed Weyl spinor and those labelled by $\phi$ are complex scalars. The fields are
\begin{equation}\label{eq:FCModelC}
  \begin{aligned}
    \chi_{N_1} & : \left(\mathbf{1}, \mathbf{1}, 0,             \mathbf{1}, \mathbf{1}, 0\right), &
    \chi_{N_2} & : \left(\mathbf{1}, \mathbf{1}, 0,             \mathbf{1}, \mathbf{1}, 0\right),\\
    \phi^A     & : \left(\mathbf{3}, \mathbf{1}, -\frac{1}{3},  \mathbf{1}, \mathbf{1}, 0\right), & 
    \phi^B     & : \left(\mathbf{1}, \mathbf{1}, 0, \mathbf{3}, \mathbf{1}, -\frac{1}{3}\right),
  \end{aligned}
\end{equation}
where we used the notation
\begin{equation}\label{eq:GroupStructure}
  (SU(3)_A, SU(2)_A, U(1)_A, SU(3)_B, SU(2)_B, U(1)_B).
\end{equation}

The Lagrangian containing the interactions relevant to us can be separated into two parts. The first one involves the fermions and can be written as
\begin{equation}\label{eq:LagrangianModelC}
  \begin{aligned}
    \mathcal{L}_1 = & - \frac{1}{2} m_{N_1} \chi_{N_1} \cdot \chi_{N_1} - \frac{1}{2} m_{N_2} \chi_{N_2} \cdot \chi_{N_2} + \text{h.c.}\\
                    & + \lambda_{3ij}\left[(\phi^A)^\dagger \chi_{d_{iR}^A} \cdot \chi_{u_{jR}^A} + (\phi^B)^\dagger \chi_{d_{iR}^B} \cdot \chi_{u_{jR}^B} \right] + \text{h.c.}\\
                    & + \lambda_{4ij}\left[\phi^A \chi_{N_j} \cdot \chi_{d_{iR}^A} + \phi^B \chi_{N_j} \cdot \chi_{d_{iR}^B} \right] + \text{h.c.}
  \end{aligned}
\end{equation}
This can be rewritten in terms of Majorana spinors $N_i$ as
\begin{equation}\label{eq:LagrangianModelC4Fermions}
  \begin{aligned}
    \mathcal{L}_1 = & - \frac{1}{2} m_{N_1} \bar{N}_1 N_{1} - \frac{1}{2} m_{N_2} \bar{N}_2 N_{2}\\
                    & + \lambda_{3ij}\left[(\phi^A)^\dagger \bar{d}_i^A P_L (u^A_j)^c + (\phi^B)^\dagger \bar{d}_i^B P_L (u^B_j)^c \right] + \text{h.c.}\\
                    & + \lambda^*_{4ij}\left[(\phi^A)^\dagger \bar{N}_j P_R d_i^A + (\phi^B)^\dagger \bar{N}_j P_R d_i^B\right] + \text{h.c.}
  \end{aligned}
\end{equation}
It is easy to verify that this Lagrangian allows for both baryon number and CP violation.

The second part of the Lagrangian is responsible for providing different masses to $\phi^A$ and $\phi^B$ without explicit $\mathbb{Z}_2$ breaking. This can be done in several ways. First, the Lagrangian could contain the term
\begin{equation}\label{eq:mTMdiffex1}
  -\lambda_0\left[|H^A|^2 |\phi^A|^2 +  |H^B|^2 |\phi^B|^2\right].
\end{equation}
Replacing the Higgs doublets by their expectation values will affect the masses of $\phi^A$ and $\phi^B$ differently. Alternatively, new scalar fields could be introduced and play a similar role to $H^M$. This can be done for example by introducing another Higgs doublet and its partner or by introducing a real scalar and its partner. In these cases, the different vevs can be obtained again by spontaneous breaking of the $\mathbb{Z}_2$ symmetry. Since there are so many possibilities and since this is sufficient, we will simply work with the effective Lagrangian
\begin{equation}\label{eq:LagrangianModelC2}
  \begin{aligned}
    \mathcal{L}_2 & = - m_\phi^2\left[|\phi^A|^2 + |\phi^B|^2 \right] - \Delta m_{\phi^A}^2|\phi^A|^2 - \Delta m_{\phi^B}^2|\phi^B|^2\\
                  & = - m_{\phi^A}^2|\phi^A|^2 - m_{\phi^B}^2|\phi^B|^2.
  \end{aligned}
\end{equation}

\subsection{Description of the mechanism}\label{sSec:DescriptionMechanism}
\begin{figure}[t!]
  \centering
   \captionsetup{justification=centering}
    \begin{subfigure}{0.495\textwidth}
    \centering
    \caption{$\rho_i/\rho_{\text{tot}}$}
    \includegraphics[width=\textwidth]{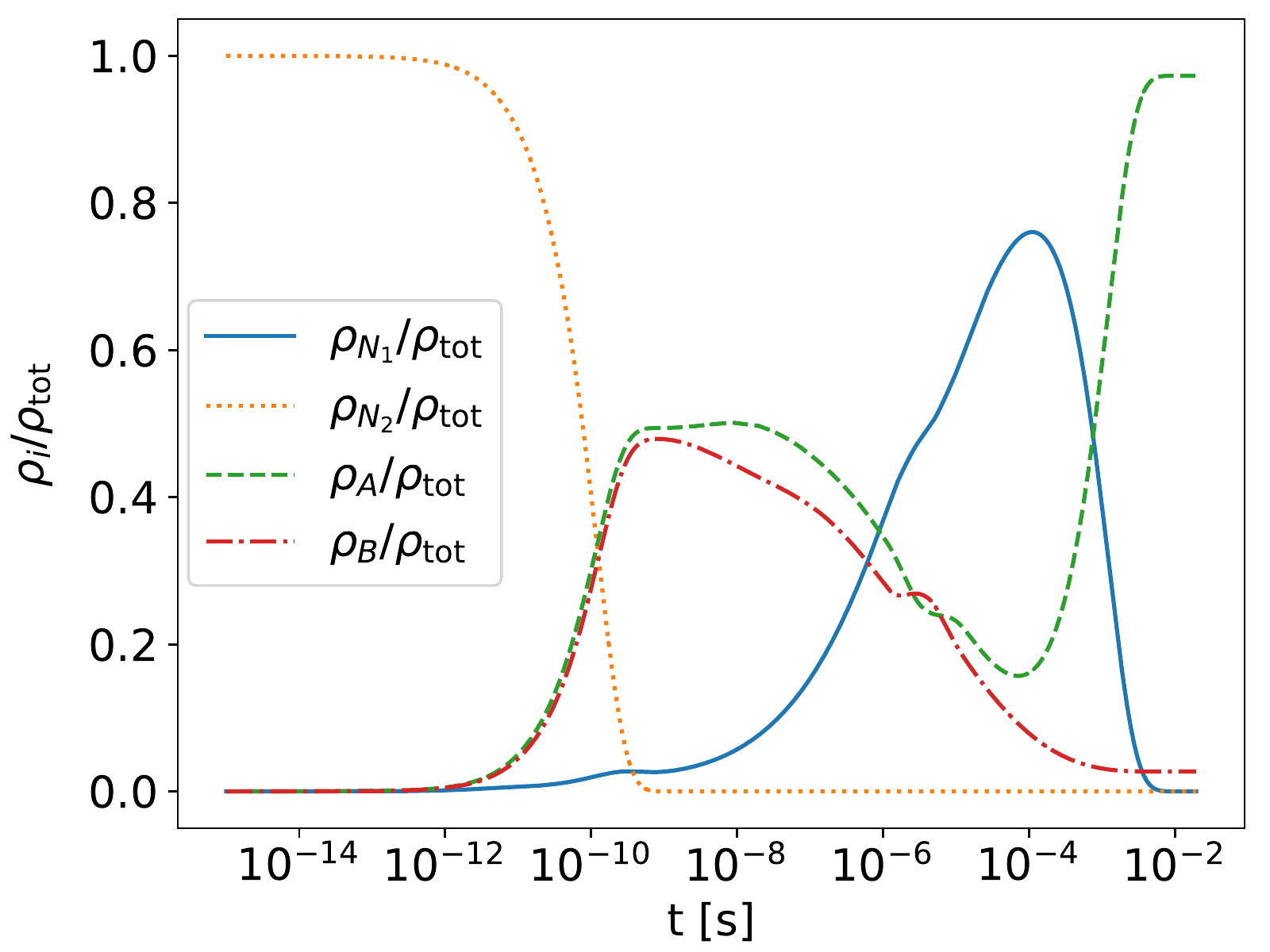}
    \label{fig:BAbundance}
  \end{subfigure}
  \begin{subfigure}{0.495\textwidth}
    \centering
    \caption{$|\Delta Y_{B_M}|$}
    \includegraphics[width=\textwidth]{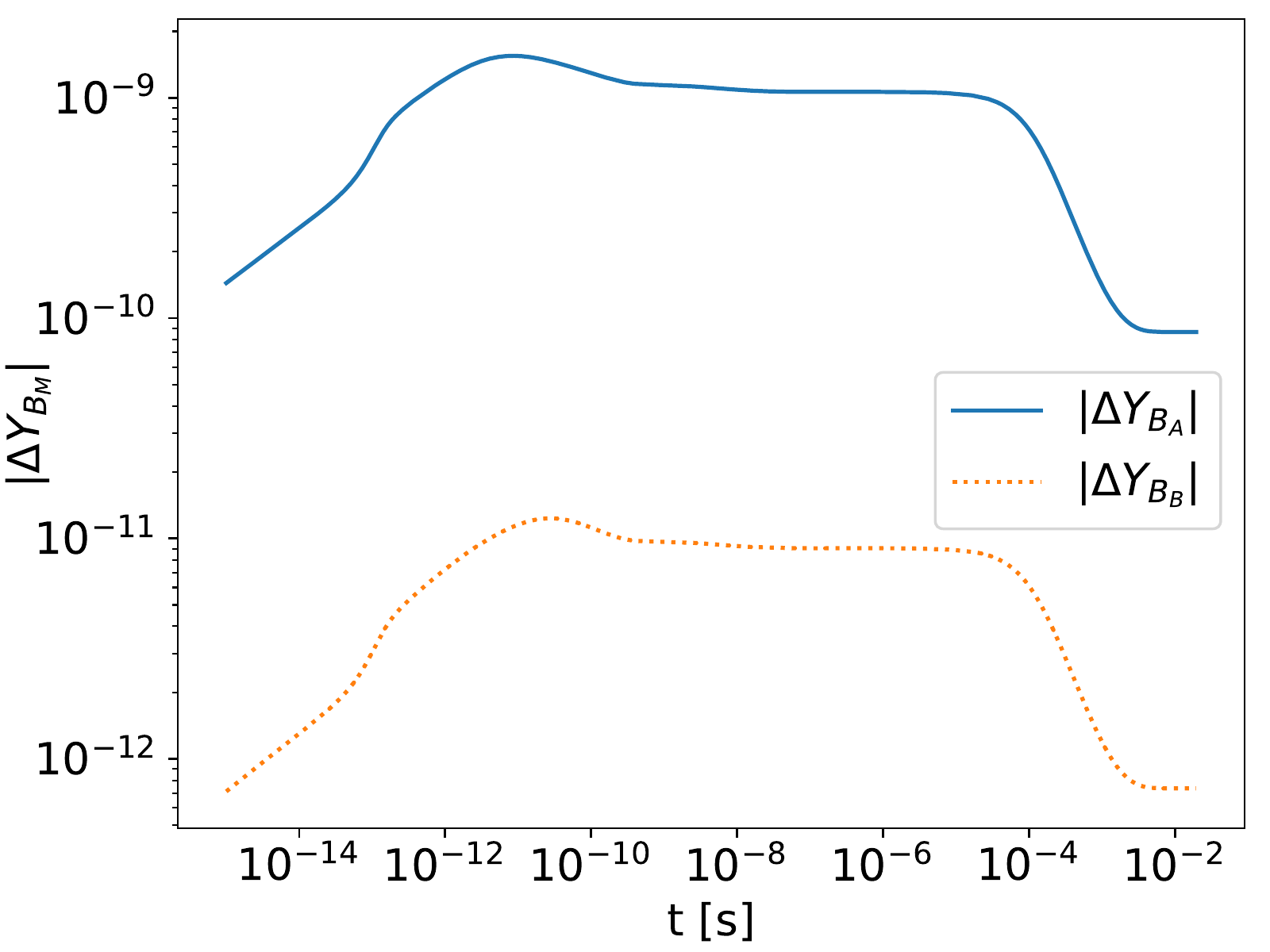}
    \label{fig:BAsymmetry}
  \end{subfigure}
  \vspace{-0.2cm}
  \begin{subfigure}{0.495\textwidth}
    \centering
    \caption{$T_i$}
    \includegraphics[width=\textwidth]{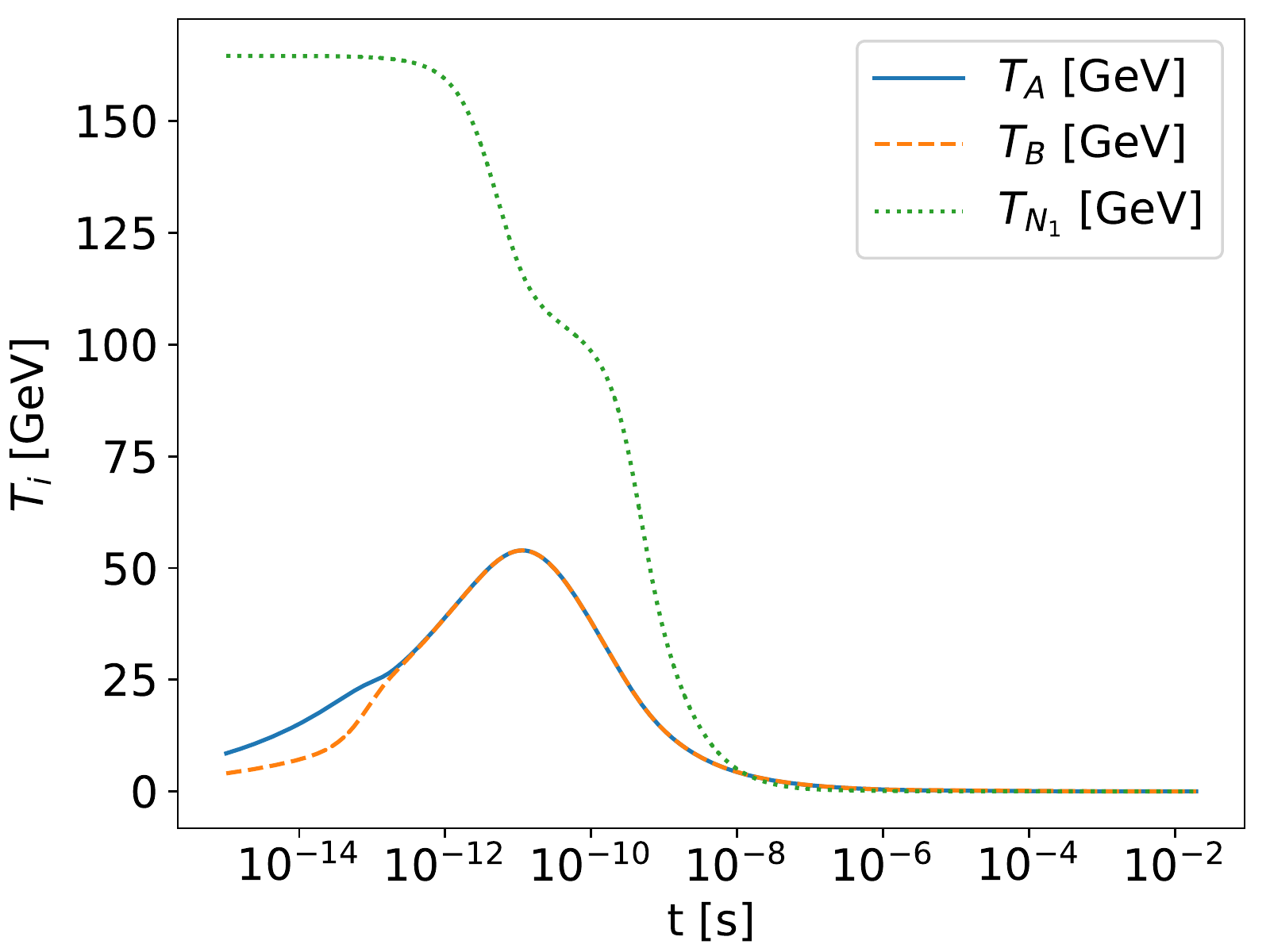}
    \label{fig:BTemperatures}
  \end{subfigure}
  \begin{subfigure}{0.495\textwidth}
    \centering
    \caption{$T_B/T_A$}
    \includegraphics[width=\textwidth]{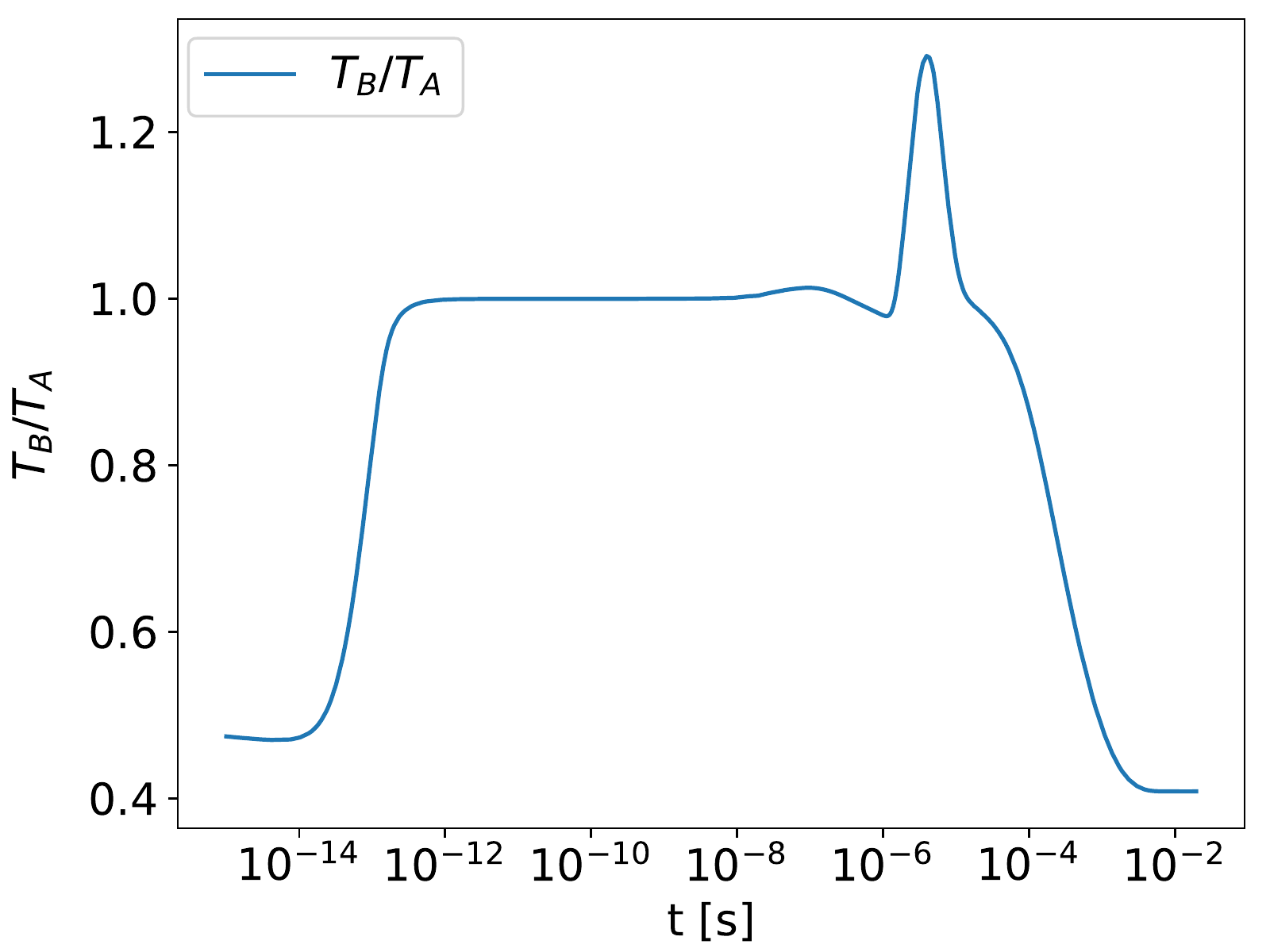}
    \label{fig:BTemperatureRatio}
  \end{subfigure}
  \captionsetup{justification=justified}
\caption{Evolution of the (a) energy fractions, (b) asymmetries, (c) temperatures and (d) temperature ratio for the benchmark of Sec.~\ref{sSec:DescriptionMechanism}.}\label{fig:BenchmarkM1}
\end{figure}

We now proceed to describe how this model solves the issues it is designed to address. We refer to Fig.~\ref{fig:BenchmarkM1} for illustration of the evolution of different relevant quantities in a given benchmark. The parameters are taken as
\begin{equation}\label{eq:ValuesPlotIllustration}
  \begin{tabular}{llll}
    $m_{N_1} = 150 \text{ GeV}$,   & $m_{N_2} = 1500  \text{ GeV}$, & $m_{\phi^A} = 15267 \text{ GeV}$, & $m_{\phi^B} = 25000 \text{ GeV}$ \\
    $v^B/v^A = 5$,                 & $\lambda_{323} = 0.05$,        & $\lambda_{431} = 0.05$,           & $\lambda_{432} = 0.0005\,e^{i \pi/4}$.
  \end{tabular}
\end{equation}
All unspecified $\lambda_{3ij}$ and $\lambda_{4ij}$ are set to zero. The initial density of $N_2$ is set to $10^7$ $\text{GeV}^3$ and its temperature to zero. All other initial densities are set to zero. The mass of $\phi^A$ is chosen to reproduce the correct baryon abundance. This benchmark also leads to an $r_T = T_B/T_A$ of 0.408, which satisfies the bounds on $N_{\text{eff}}$ as will be discussed in Sec.~\ref{sSec:Constraints}. The abundance of dark baryons is $\Omega_{B}= 6.02 \times 10^{-4}$. Even assuming all dark baryons are mirror atoms, this is still considerably below experimental bounds, as will be discussed in Sec.~\ref{sSec:Constraints}. The value of $v^B/v^A$ satisfies the Higgs signal strengths, which are discussed in Appendix~\ref{Sec:HiggsSS}. The temperature of sector $M$ is labelled as $T_M$, its energy density as $\rho_M$, its entropy density as $s_M$, its net baryon density as $\Delta B_M$ and
\begin{equation}\label{eq:PlotQuantities}
  \Delta Y_{B_M} = \frac{\Delta B_M}{s_A}.
\end{equation}
This benchmark is not special and a summary exploration of the parameter space will be performed in Sec.~\ref{sSec:ParameterSpace}.\footnote{Since the $\phi^M$ scalars are always treated as heavy, the results can easily be rescaled by making the transformations $m_{\phi^M}\to \alpha m_{\phi^M}$, $\lambda_{3ij} \to \alpha\lambda_{3ij}$ and $\lambda_{4ij} \to \alpha\lambda_{4ij}$, where $\alpha$ is a real constant.} The whole process can be separated into three qualitative phases.

Initially, the energy content of the Universe is dominated by the heavier Majorana fermion $N_2$. This situation could easily be realized through the decay of the inflaton if its coupling to $N_2$ is considerably stronger than its other couplings. Early on, $N_2$ starts to decay. It can decay to the $A$ sector mainly via three channels: to three quarks, to three antiquarks or to $N_1$, a quark and an antiquark. Similar decays to the mirror sector are also present. Because of the presence of a third decay channel in each sector and in conjunction with the Nanopoulos-Weinberg theorem  \cite{Nanopoulos:1979gx}, $N_2$ can present an asymmetry in its decay to baryons and antibaryons and similarly for its decay to the mirror sector. In practice, this comes from the interference of the diagrams of Fig.~\ref{fig:FMasymmetry}. The asymmetry in the $B$ sector can be adjusted by changing the ratio $m_{\phi^B}/m_{\phi^A}$. Once the $N_2$ are mostly decayed, the Universe is populated with particles from the $A$ and $B$ sectors as well as some $N_1$. Baryon asymmetries are also present in both sectors.

As time passes, the expansion of the Universe dissolves the energy densities of the different particles. As the $A$ and $B$ sectors are radiation dominated, their energy densities scale as $a^{-4}$, where $a$ is the scale factor. Since $N_1$ is mostly non-relativistic, its energy density instead scales as $a^{-3}$ and soon comes to dominate the energy abundance.

Finally, the $N_1$ population starts to decay. In principle, $N_1$ could decay to particles of either sector. However, there exists a sizable region of parameter space, which includes the benchmark, where the decay to the $B$ sector is strongly suppressed because of kinematics. In the benchmark, $N_1$ can decay to the $A$ sector as an off-shell top, a bottom and a strange. It however cannot decay to a mirror top, a mirror bottom and a mirror strange or even two mirror bottoms, a mirror strange and a mirror $W$ as both the mirror top and mirror $W$ are too heavy to be produced on-shell. This results in the $N_1$ population transferring its energy almost exclusively to the $A$ sector and thus a relative reheating of that sector. This constitutes the main mechanism through which the $N_{\text{eff}}$ problem is solved. This is also why the $N_i$ were assumed to couple mainly to up-type quarks of the third generation, as having the decay of $N_1$ only being possible to one sector is easy to accomplish thanks to the large mass of the top quark. The decay of $N_1$ does not generate any sizable asymmetry and in fact partially dissolves the asymmetries by injecting entropy.

The end result of this mechanism is an $A$ sector with a net population of baryons and a $B$ sector with a much smaller net population of mirror baryons. This both explains baryogenesis and satisifies the bounds on dark matter self-interactions associated with the dark atoms. The fact that the mirror sector is much cooler also ensures that the $N_{\text{eff}}$ constraints are satisfied. The temperature of each sector is also maintained at all times considerably below the electroweak scale.

\subsection{Constraints}\label{sSec:Constraints}

\subsubsection*{$N_{\text{eff}}$}
The number of effective relativistic degrees of freedom $N_{\text{eff}}$ is measured by Planck to be $2.99\pm 0.17$ \cite{Aghanim:2018eyx}. During BBN and the creation of the cosmic microwave background, the numbers of relativistic degrees of freedom in both sectors are about the same, which puts a limit on
\begin{equation}\label{eq:rT}
  r_T = \frac{T_B}{T_A} = \left(\frac{\Delta N_{\text{eff}}}{7.4}\right)^{1/4}
\end{equation}
of $\sim$ 0.44 at 95\% CL, where we used the fact that the SM value of $N_{\text{eff}}$ is 3.046.

\subsubsection*{Fraction of dark atoms}
Without additional model building, dark matter would take the form of mirror atoms. However, a too large fraction of dark atoms $X_{\text{DA}}$ is excluded by limits on dark matter self-interactions. Ref.~\cite{Fan:2013yva} claims that this fraction can still be as high as  about 10\%, though the amount of uncertainty on this number is rather unclear. In addition, Ref.~\cite{Chacko:2018vss} claims that the limit on $X_{\text{DA}}$ might be brought to the few percent level in the not-so-distant future. When relevant, we will present contours of $X_{\text{DA}}$ and emphasize that the region above 10\% is disfavoured.

\subsubsection*{Big Bang Nucleosynthesis}
If $N_1$ is sufficiently long-lived, it will disturb BBN by injecting energetic hadrons and modify the observed abundances of primordial elements. Unfortunately, the cosmology of the model is rather exotic and no study of the decay of metastable particles during BBN perfectly mimics it. As such, we will simply ask that the lifetime of $N_1$ be below 0.1~s, which is the typical bound (see for example Refs.~\cite{Kawasaki:2004qu, Jedamzik:2006xz, Jedamzik:2009uy}). Considering that BBN limits are generally not strongly dependent on parameters such as the mass of the metastable particle and its branching ratio to hadronic channels, a more advanced treatment is not expected to change this constraint much.

\subsubsection*{Higgs signal strengths}
The limits on the Higgs signal strengths are applied using the results of Appendix~\ref{Sec:HiggsSS}.

\subsubsection*{Direct collider searches}
The only new coloured particle in the model is the colour-triplet scalar $\phi^A$. Its pair production at the $13$~TeV LHC in the mass range we consider ($\gtrsim 5$~TeV) is negligible ($\ll 1$ event). The fermions $N_1$ and $N_2$ are gauge singlets and do not need to have any significant couplings that involve pairs of light quarks, so their direct production is irrelevant too. In part of the parameter space, $N_1$ can be produced in top quark decays, but the branching fraction is highly suppressed by the mass of $\phi^A$, the small couplings and phase space. Part of the $B$ sector particles can be produced in Higgs decays (and escape the detectors invisibly), but the resulting effect on the visible branching fractions of the Higgs is too small to be seen in the current datasets (see Appendix~\ref{Sec:HiggsSS}).

\subsection{Parameter space and comments}\label{sSec:ParameterSpace}
We now provide some summary scans and comment on various properties of the model.

\begin{figure}[t!]
  \centering
   \captionsetup{justification=centering}
    \begin{subfigure}{0.495\textwidth}
    \centering
    \caption{$\Omega_A$}
    \includegraphics[width=\textwidth]{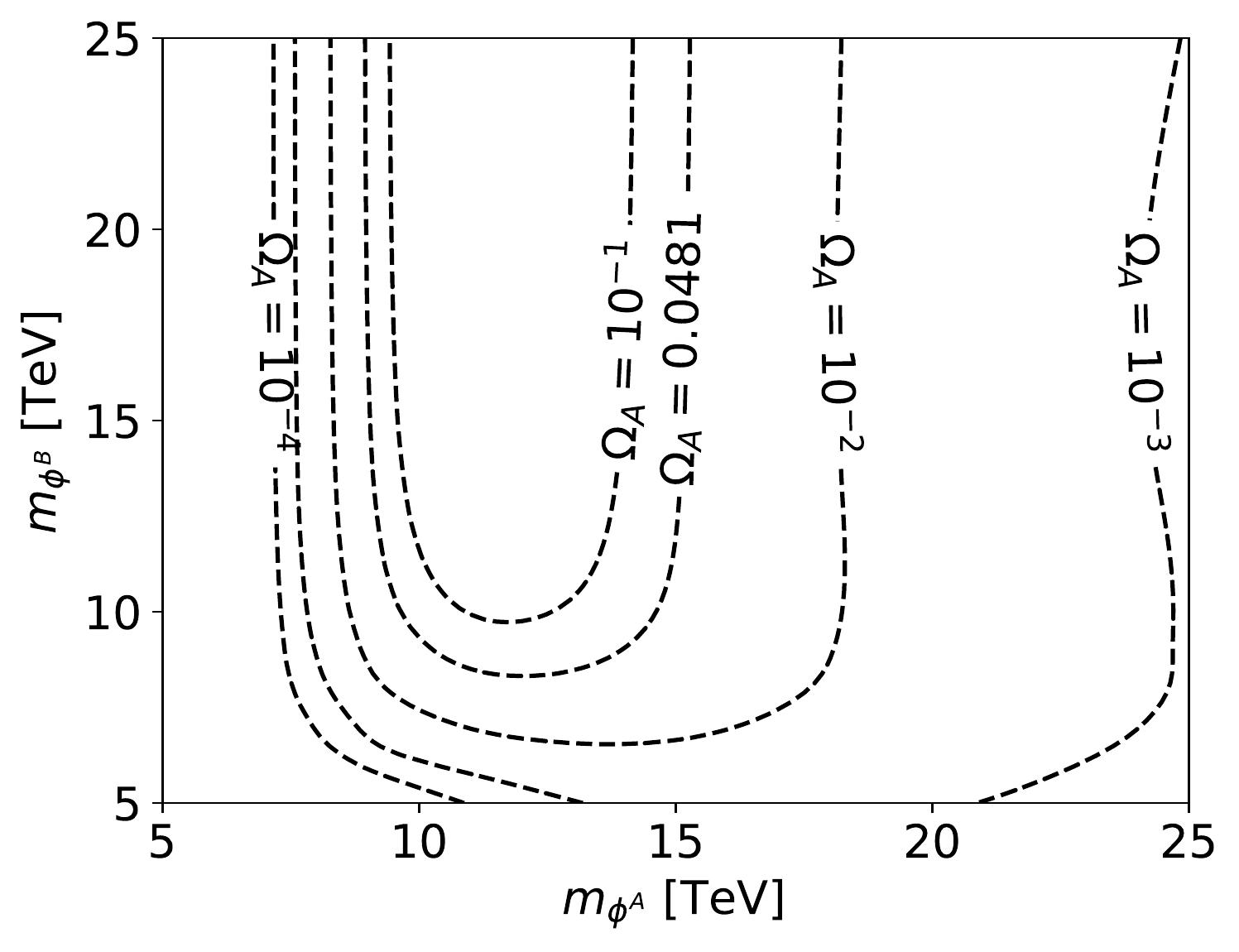}
    \label{fig:AbundanceA}
  \end{subfigure}
  \begin{subfigure}{0.495\textwidth}
    \centering
    \caption{$\Omega_B$}
    \includegraphics[width=\textwidth]{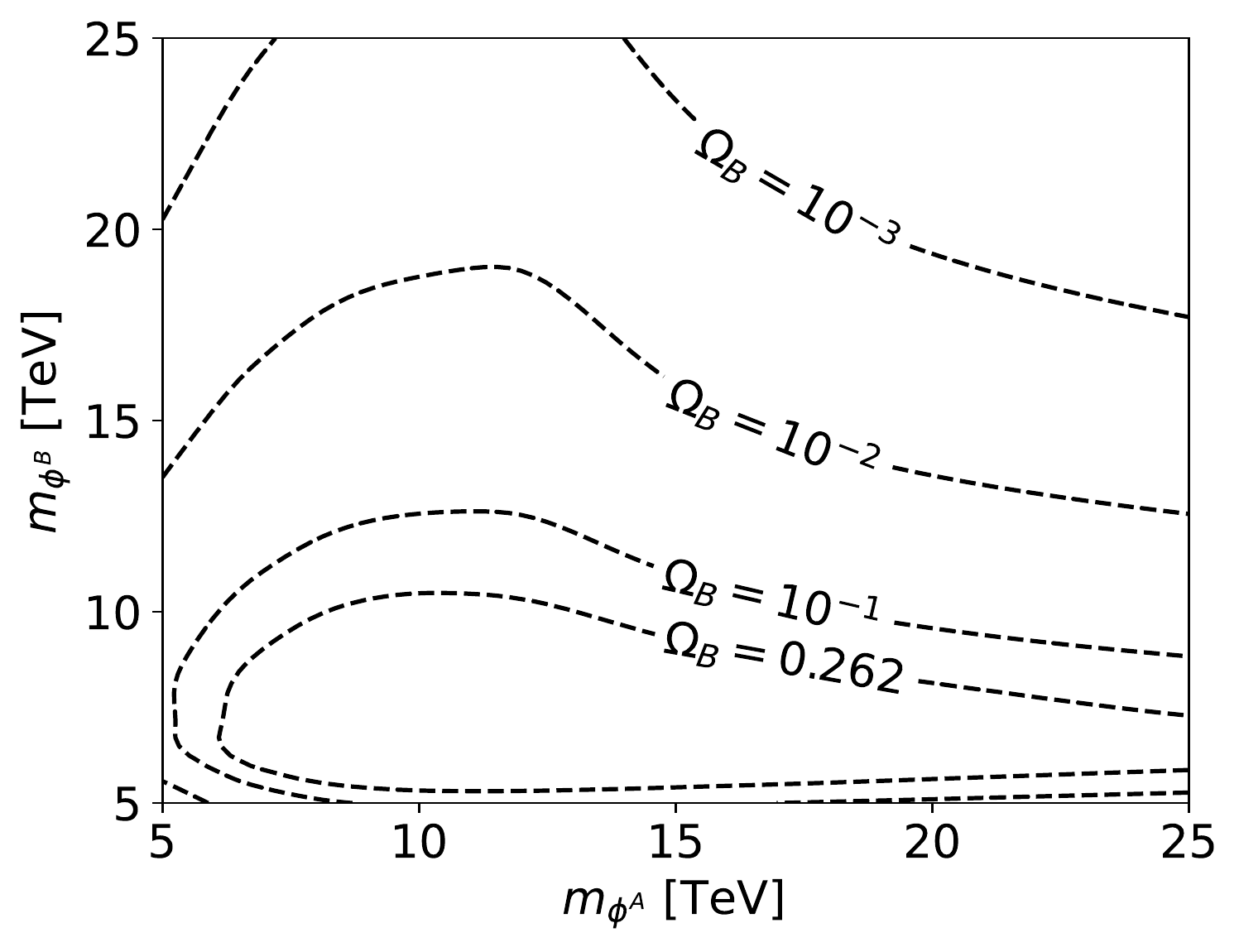}
    \label{fig:AbundanceB}
  \end{subfigure}
  \vspace{-0.2cm}
  \begin{subfigure}{0.495\textwidth}
    \centering
    \caption{$T_B/T_A$}
    \includegraphics[width=\textwidth]{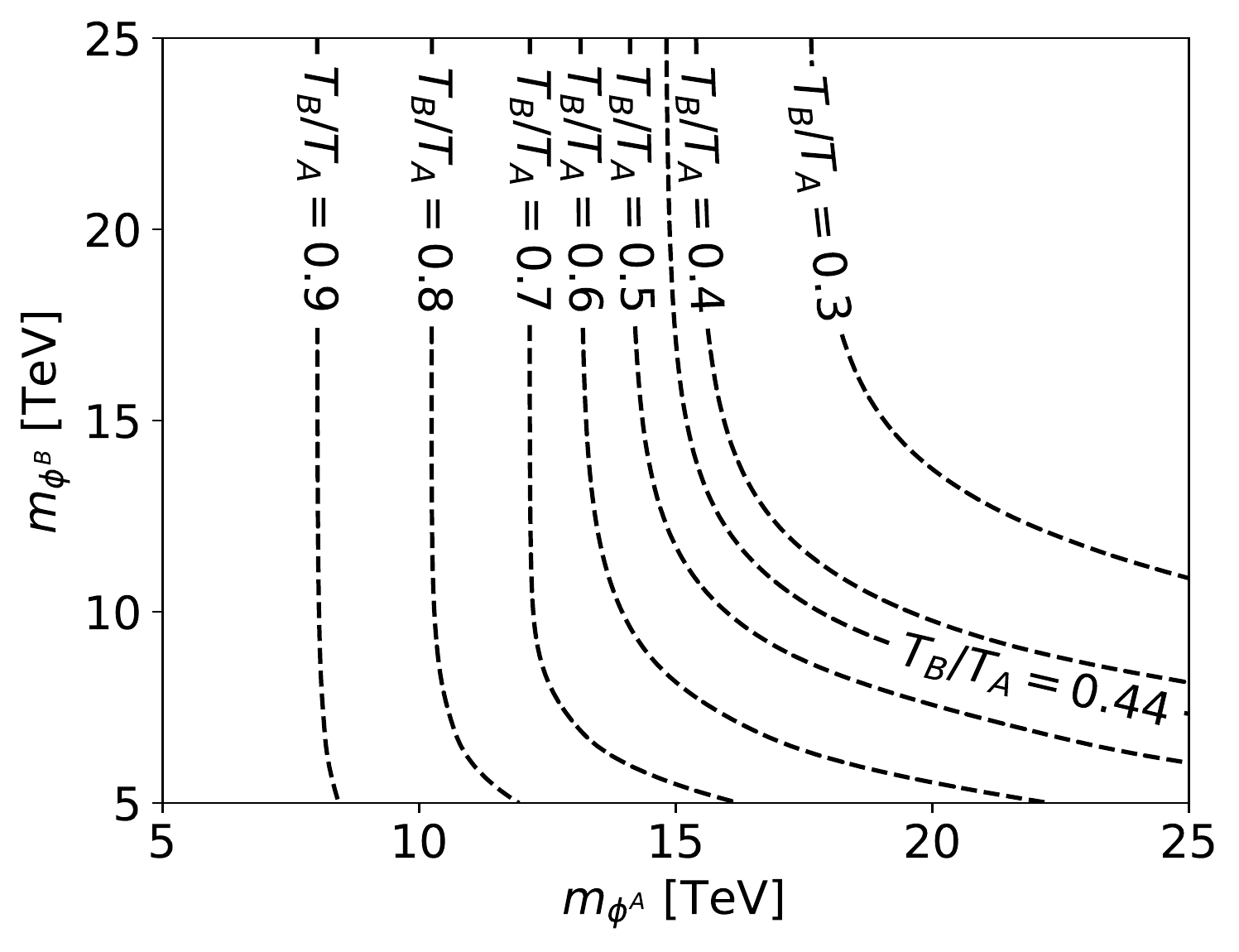}
    \label{fig:TBonTA}
  \end{subfigure}
  \begin{subfigure}{0.495\textwidth}
    \centering
    \caption{$T_A^{\text{max}}$}
    \includegraphics[width=\textwidth]{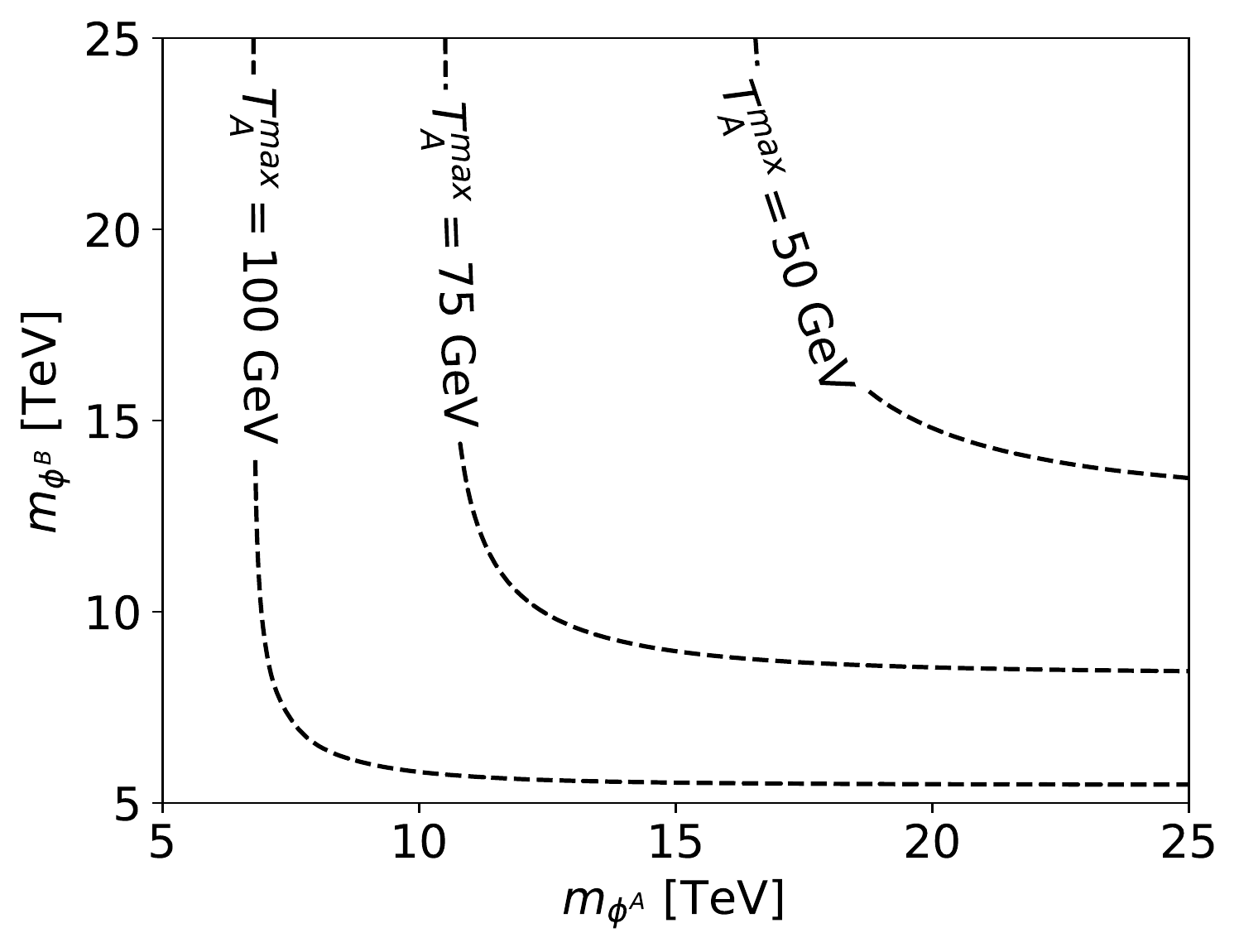}
    \label{fig:TAMax}
  \end{subfigure}
  \captionsetup{justification=justified}
\caption{Contours of constant (a) $\Omega_A$, (b) $\Omega_B$, (c) $T_B/T_A$ and (d) $T_A^{\text{max}}$ for the benchmark of Eq.~\eqref{eq:ValuesPlotIllustration2} with $v^B/v^A = 5$ and $m_{N_1} = 150$~GeV.}\label{fig:Scan1}
\end{figure}

\begin{figure}[t!]
  \centering
   \captionsetup{justification=centering}
    \begin{subfigure}{0.495\textwidth}
    \centering
    \caption{$v^B/v^A = 5$, $m_{N_1} = 140$~GeV}
    \includegraphics[width=\textwidth]{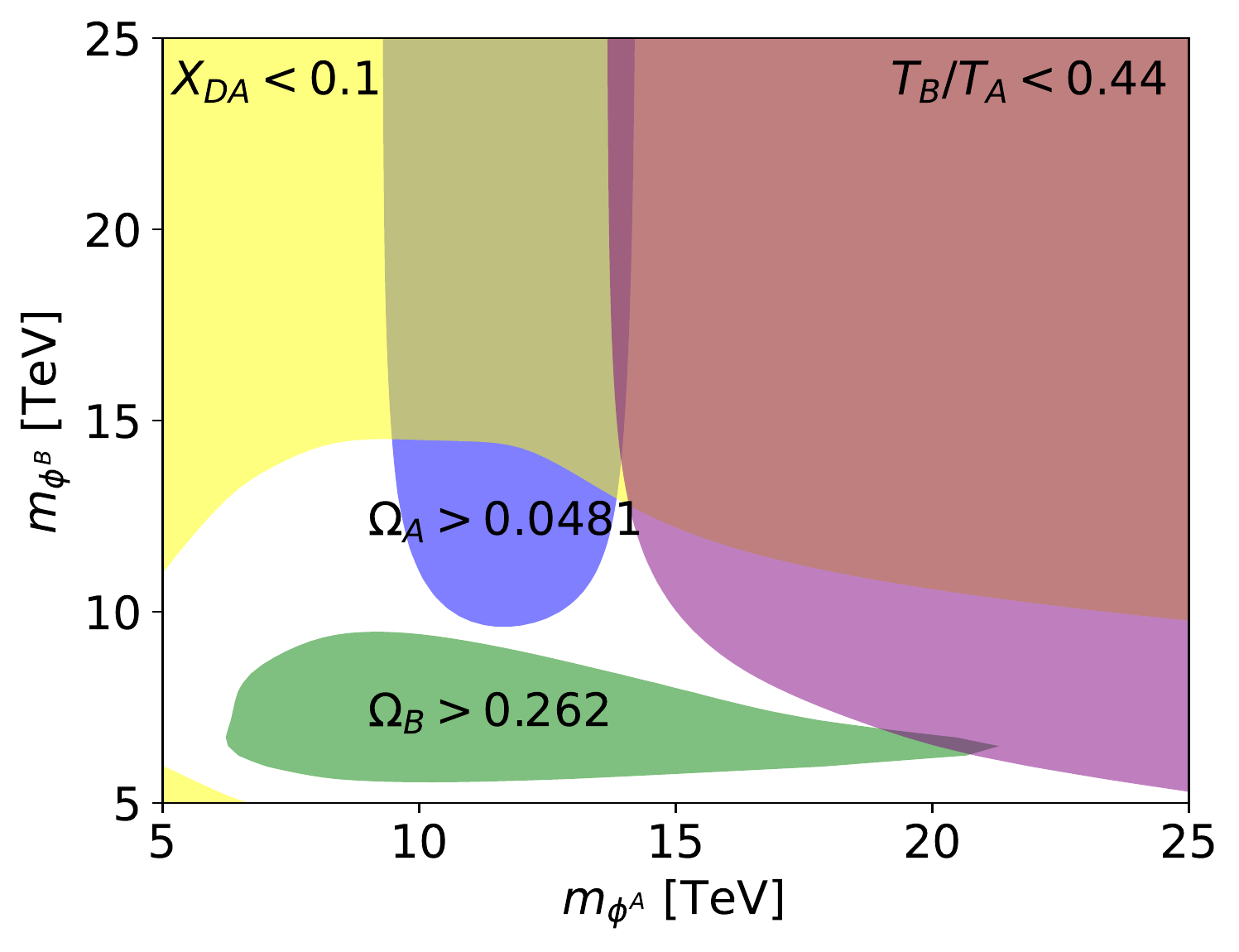}
    \label{fig:F5}
  \end{subfigure}
  \begin{subfigure}{0.495\textwidth}
    \centering
    \caption{$v^B/v^A = 6$, $m_{N_1} = 140$~GeV}
    \includegraphics[width=\textwidth]{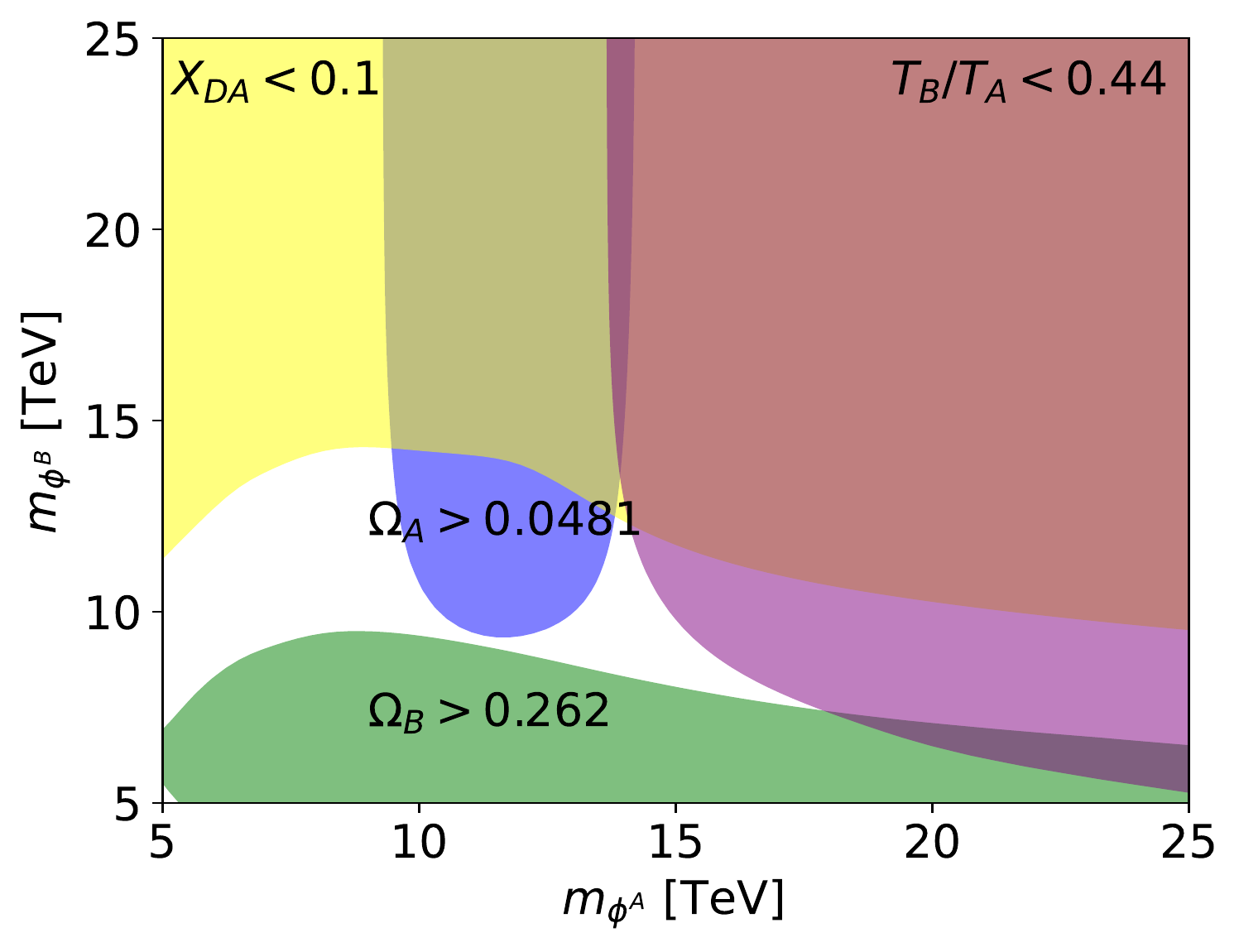}
    \label{fig:F16}
  \end{subfigure}
  \vspace{-0.5cm}
  \begin{subfigure}{0.495\textwidth}
    \centering
    \caption{$v^B/v^A = 5$, $m_{N_1} = 160$~GeV}
    \includegraphics[width=\textwidth]{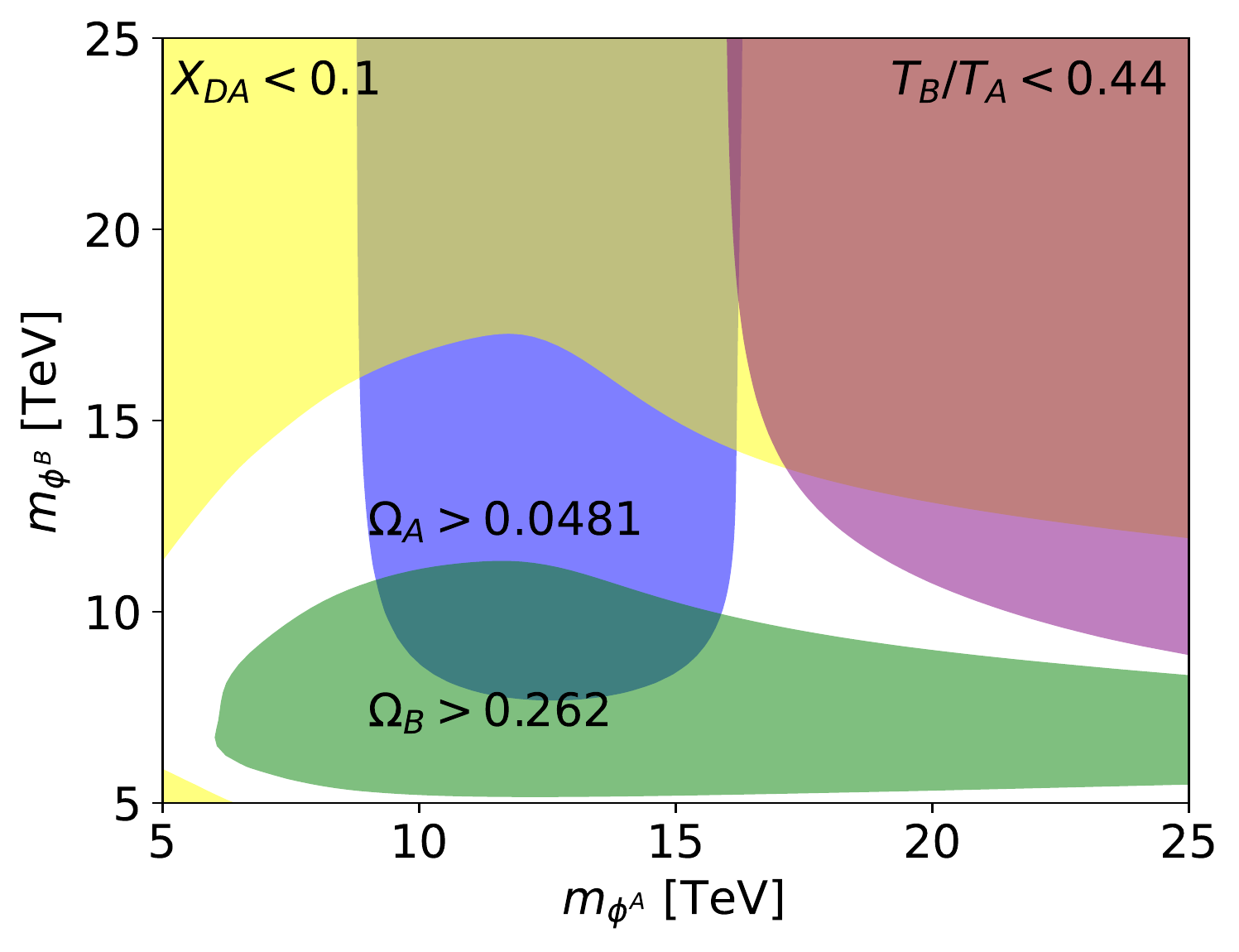}
    \label{fig:F7}
  \end{subfigure}
  \begin{subfigure}{0.495\textwidth}
    \centering
    \caption{$v^B/v^A = 6$, $m_{N_1} = 160$~GeV}
    \includegraphics[width=\textwidth]{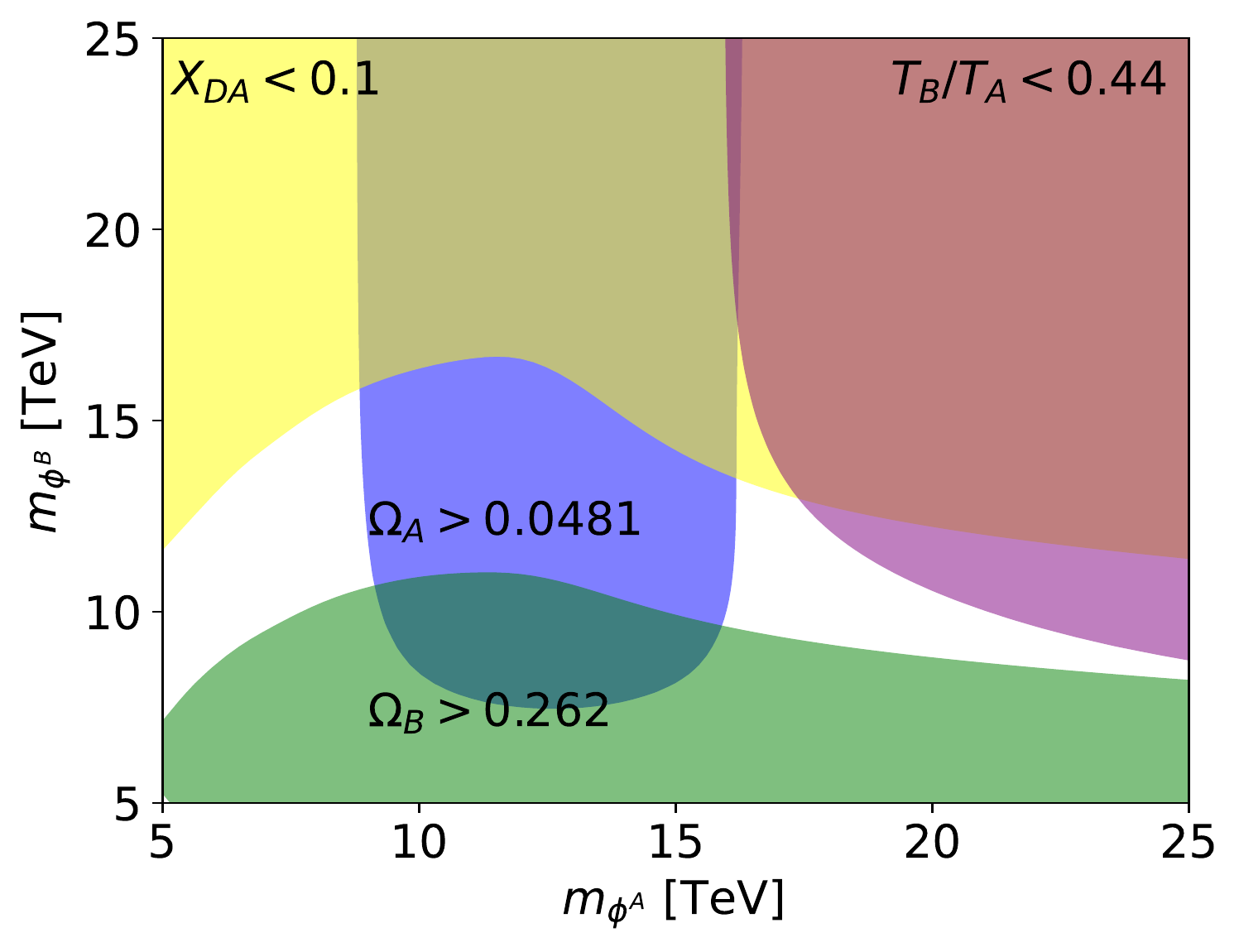}
    \label{fig:F18}
  \end{subfigure}
  \captionsetup{justification=justified}
  \vspace{-0.5cm}
  \begin{subfigure}{0.495\textwidth}
    \centering
    \caption{$v^B/v^A = 5$, $m_{N_1} = 180$~GeV}
    \includegraphics[width=\textwidth]{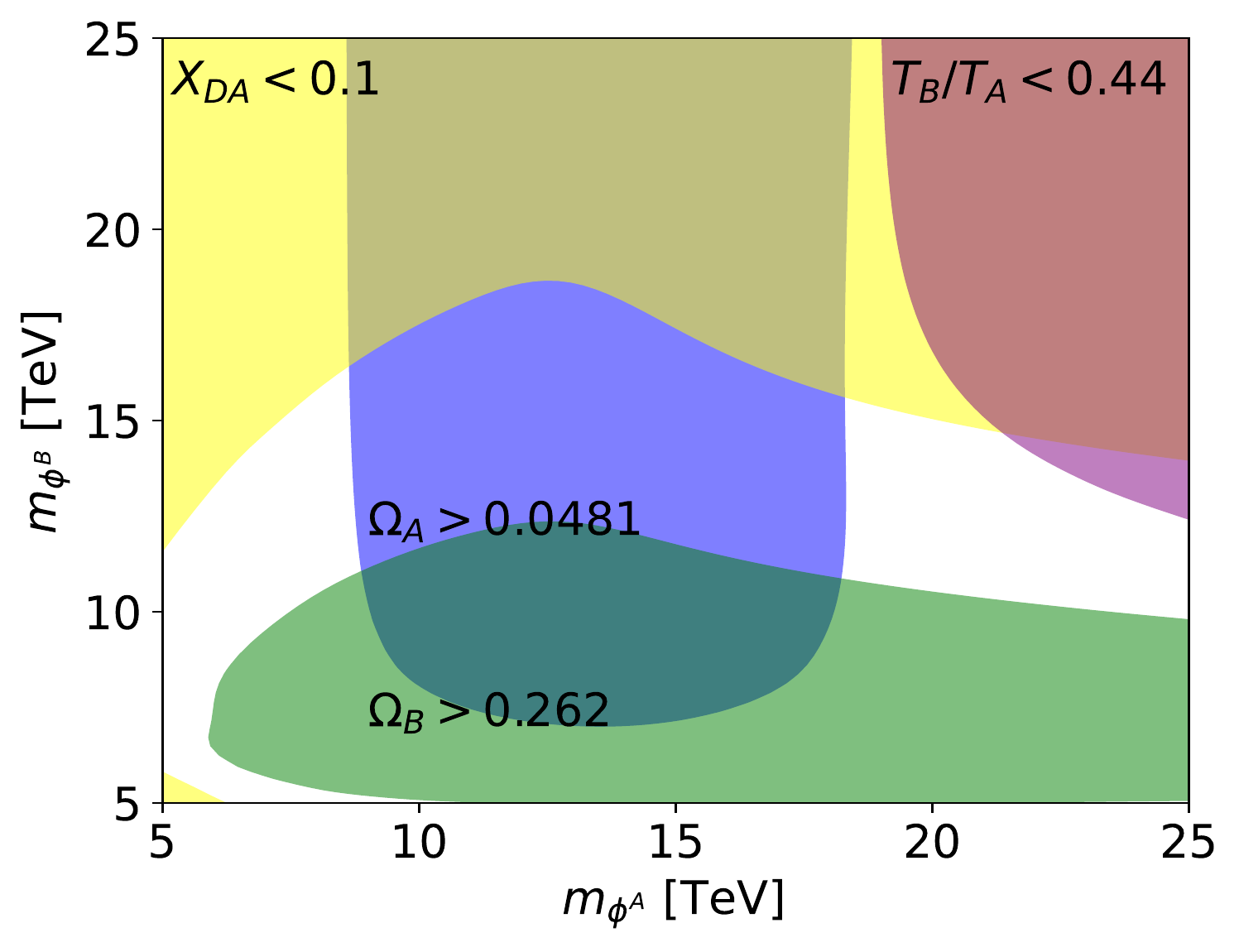}
    \label{fig:F9}
  \end{subfigure}
  \begin{subfigure}{0.495\textwidth}
    \centering
    \caption{$v^B/v^A = 6$, $m_{N_1} = 180$~GeV}
    \includegraphics[width=\textwidth]{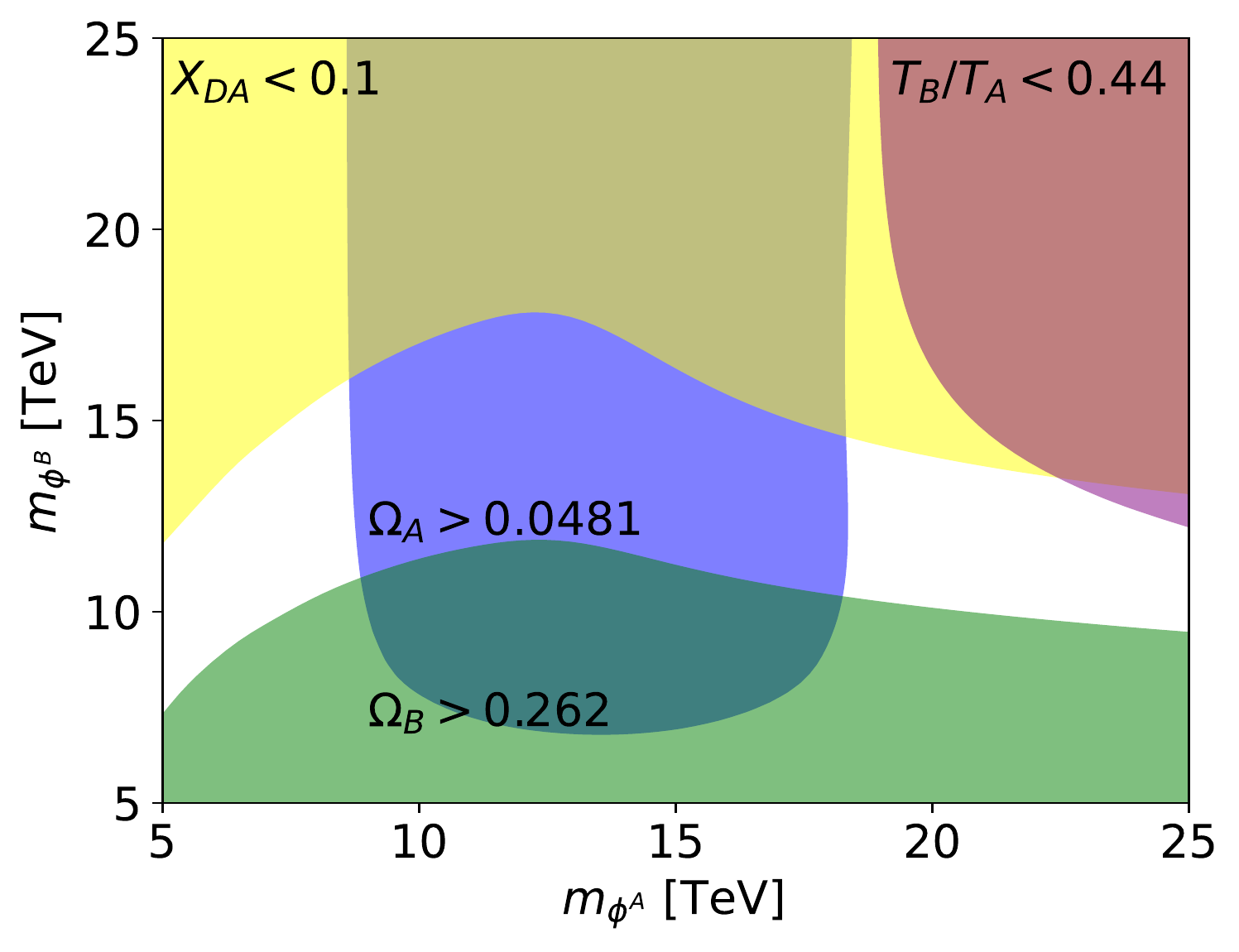}
    \label{fig:F20}
  \end{subfigure}
  \captionsetup{justification=justified}
\caption{Regions of sufficient matter abundance (blue), sufficient dark matter abundance (green), sufficiently low $\Delta N_{\text{eff}}$ (purple) and sufficiently low dark atom abundance (yellow).}\label{fig:RegionPlot1}
\end{figure}

\begin{figure}[t!]
  \centering
   \captionsetup{justification=centering}
    \begin{subfigure}{0.495\textwidth}
    \centering
    \caption{$v^B/v^A = 7$, $m_{N_1} = 140$~GeV}
    \includegraphics[width=\textwidth]{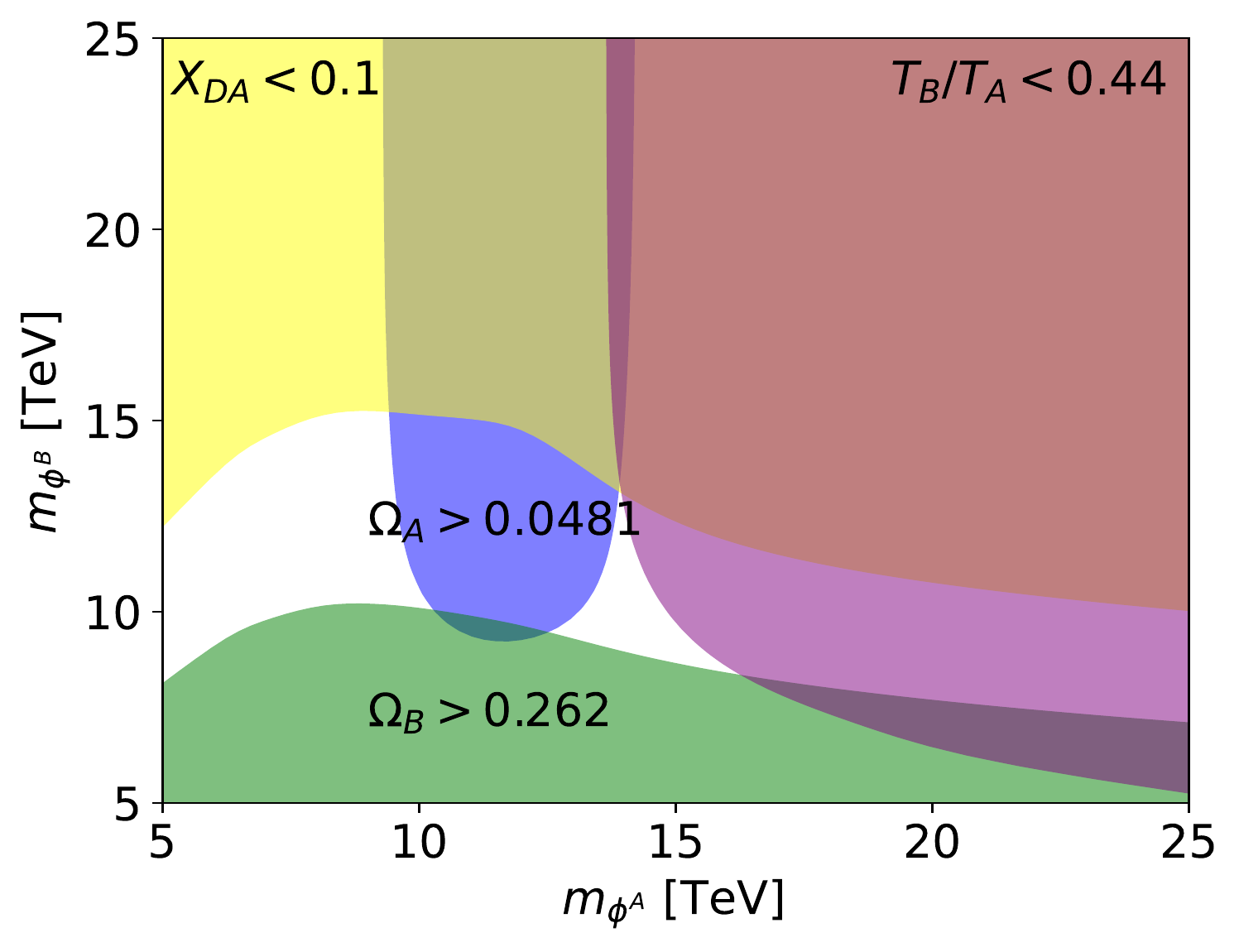}
    \label{fig:F27}
  \end{subfigure}
  \begin{subfigure}{0.495\textwidth}
    \centering
    \caption{$v^B/v^A = 8$, $m_{N_1} = 140$~GeV}
    \includegraphics[width=\textwidth]{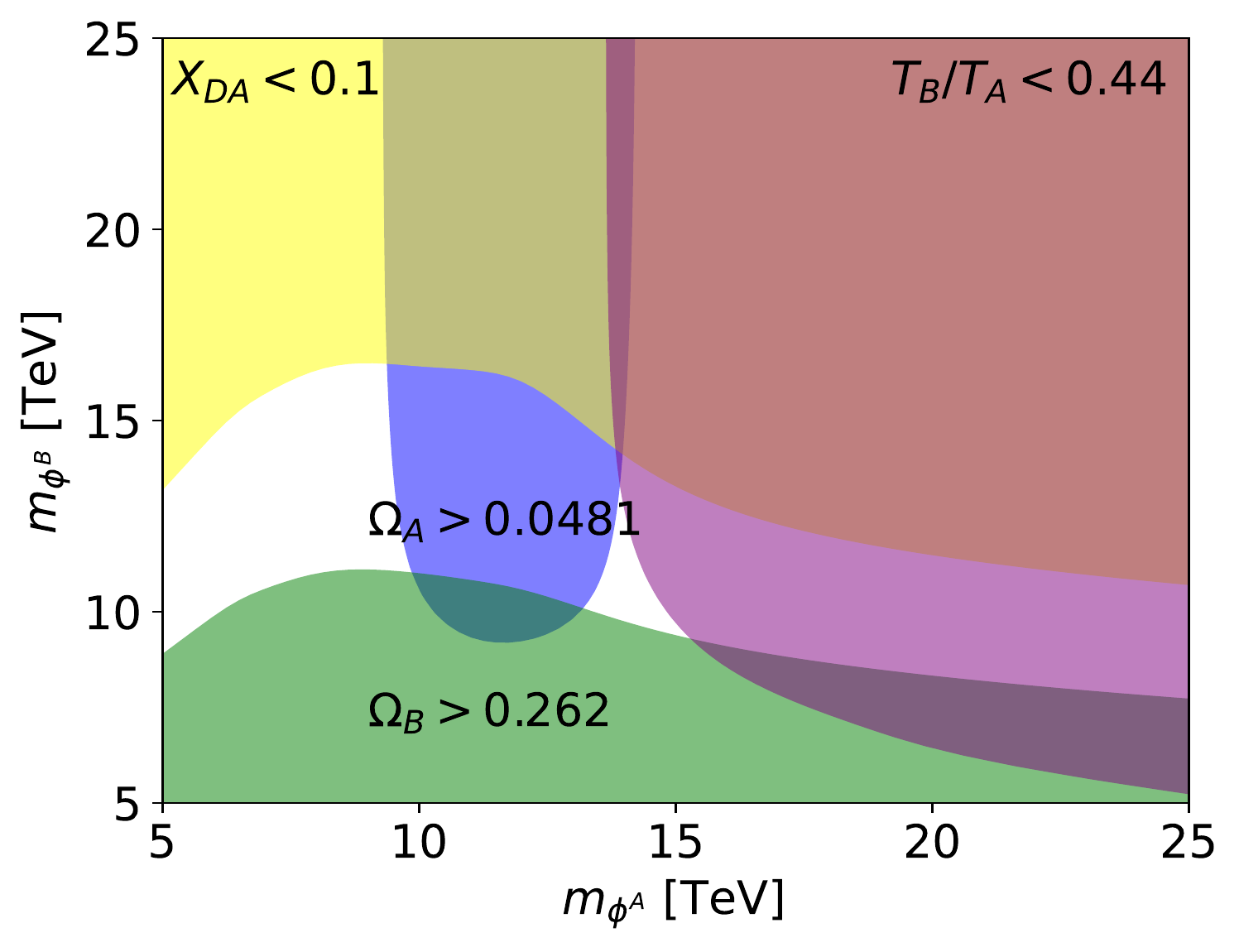}
    \label{fig:F38}
  \end{subfigure}
  \vspace{-0.5cm}
  \begin{subfigure}{0.495\textwidth}
    \centering
    \caption{$v^B/v^A = 7$, $m_{N_1} = 160$~GeV}
    \includegraphics[width=\textwidth]{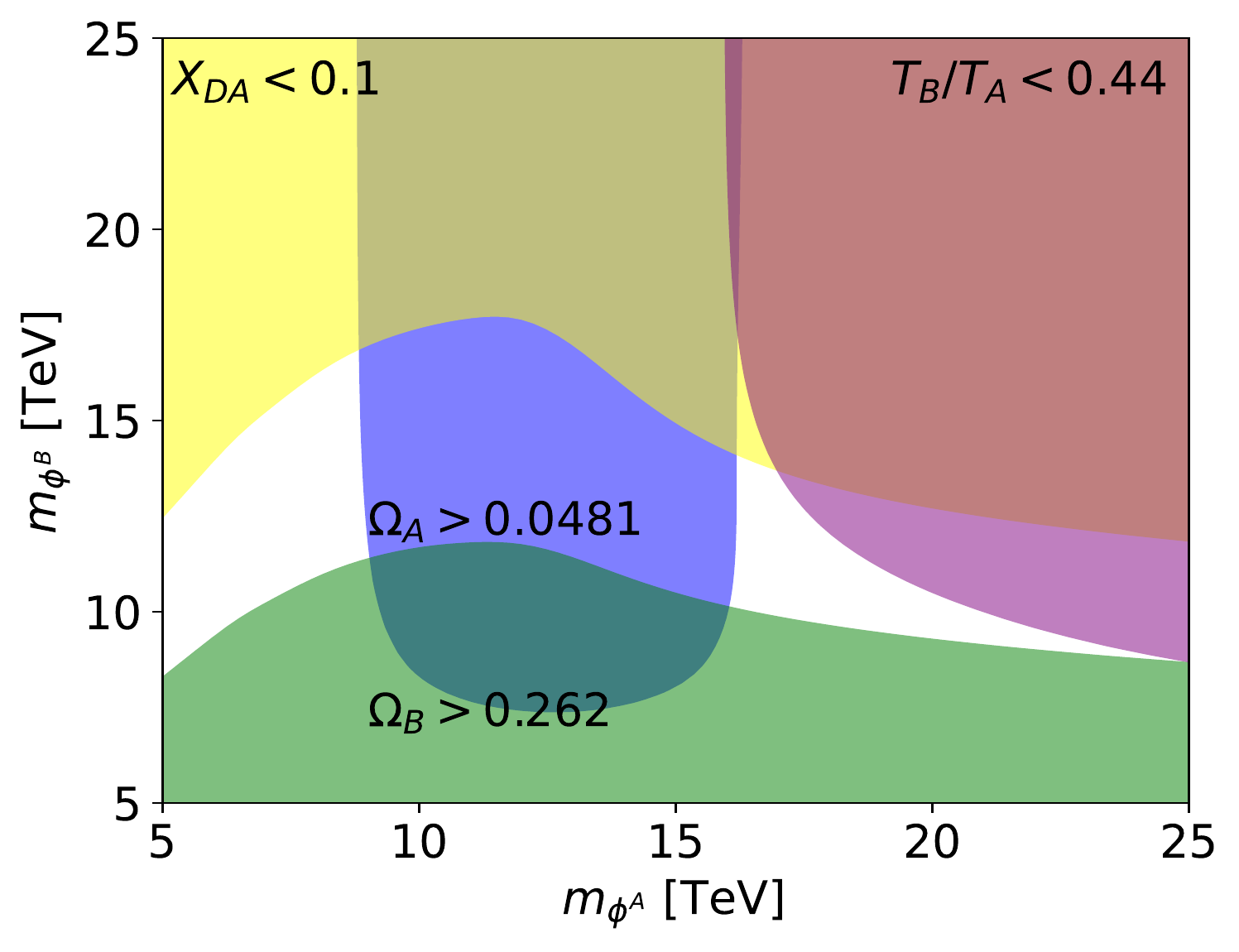}
    \label{fig:F29}
  \end{subfigure}
  \begin{subfigure}{0.495\textwidth}
    \centering
    \caption{$v^B/v^A = 8$, $m_{N_1} = 160$~GeV}
    \includegraphics[width=\textwidth]{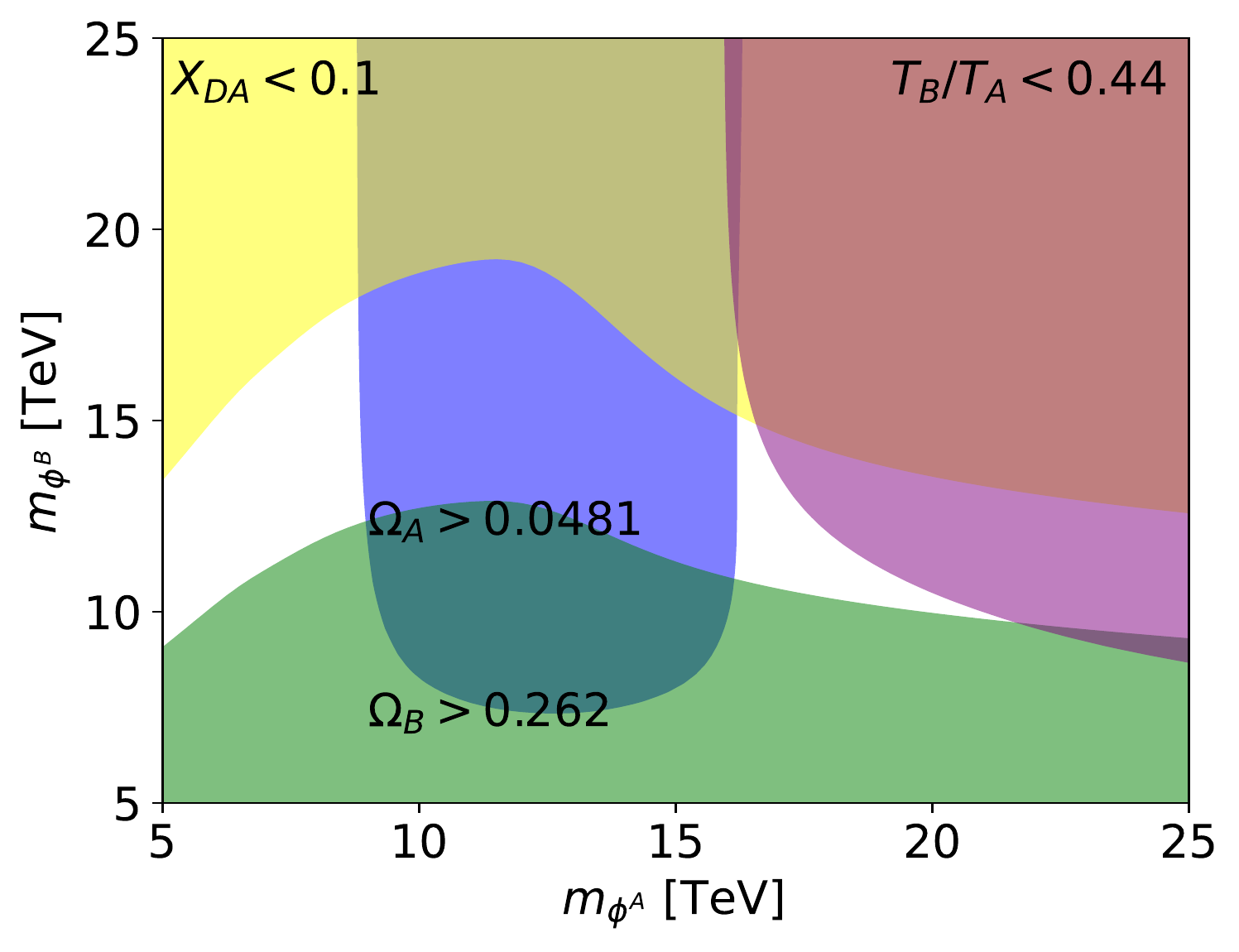}
    \label{fig:F40}
  \end{subfigure}
  \captionsetup{justification=justified}
  \vspace{-0.5cm}
  \begin{subfigure}{0.495\textwidth}
    \centering
    \caption{$v^B/v^A = 7$, $m_{N_1} = 180$~GeV}
    \includegraphics[width=\textwidth]{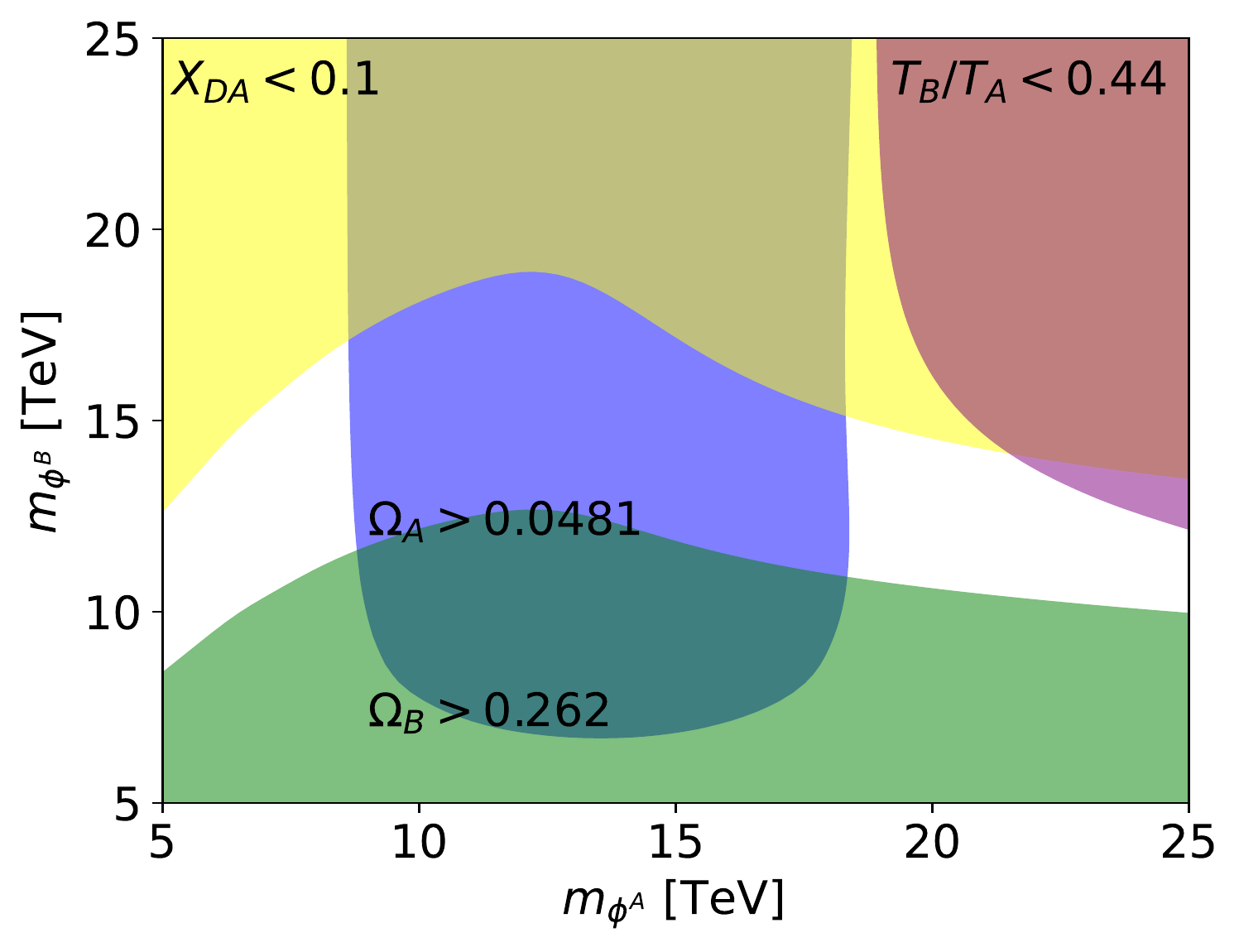}
    \label{fig:F31}
  \end{subfigure}
  \begin{subfigure}{0.495\textwidth}
    \centering
    \caption{$v^B/v^A = 8$, $m_{N_1} = 180$~GeV}
    \includegraphics[width=\textwidth]{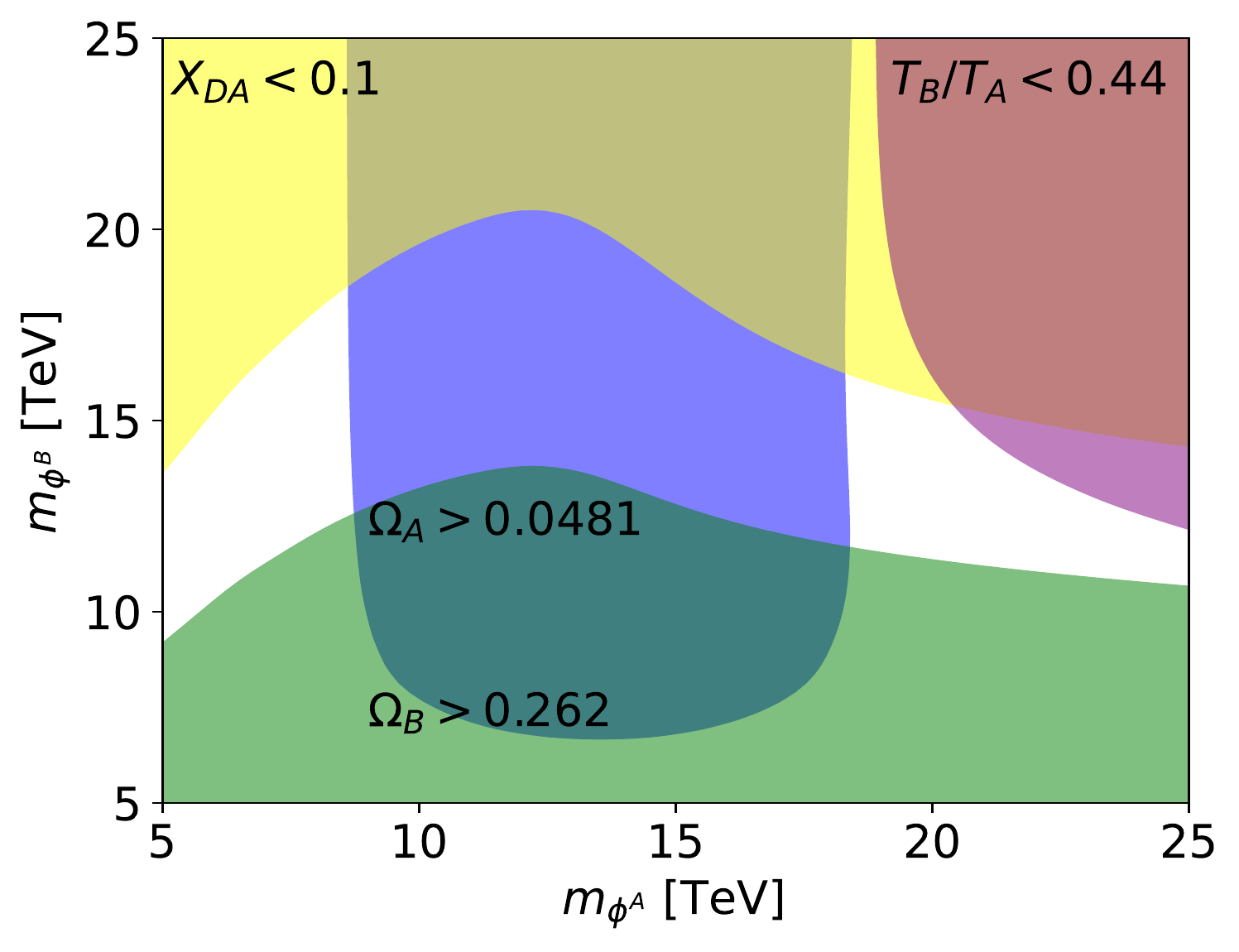}
    \label{fig:F42}
  \end{subfigure}
  \captionsetup{justification=justified}
\caption{Regions of sufficient matter abundance (blue), sufficient dark matter abundance (green), sufficiently low $\Delta N_{\text{eff}}$ (purple) and sufficiently low dark atom abundance (yellow).}\label{fig:RegionPlot2}
\end{figure}

Fig.~\ref{fig:Scan1} shows contours of different relevant quantities as a function of $m_{\phi^A}$ and $m_{\phi^B}$ for $v^B/v^A = 5$ and $m_{N_1} = 150$~GeV. The other parameters are set to
\begin{equation}\label{eq:ValuesPlotIllustration2}
  m_{N_2} = 1500  \text{ GeV}, \qquad \lambda_{323} = 0.05, \qquad \lambda_{431} = 0.05, \qquad \lambda_{432} = 0.0005\,e^{i \pi/4}.
\end{equation}
All other $\lambda$ couplings are set to zero. The initial density of $N_2$ is set to $10^7$ $\text{GeV}^3$ and its temperature to zero. All other initial densities are set to zero. Figs.~\ref{fig:RegionPlot1} and \ref{fig:RegionPlot2} show in different colours the regions of parameter space that provide a sufficient amount of matter, a sufficient amount of dark matter, a sufficiently low $\Delta N_{\text{eff}}$ or a sufficiently low dark atom abundance. These plots use the same parameters as Fig.~\ref{fig:Scan1}, except for different values of $v^B/v^A$ and $m_{N_1}$. The region excluded by BBN is outside the plot to the right.

As can be seen, there are regions of parameter space that can individually provide the correct matter abundance, the correct dark matter abundance or a sufficiently low $\Delta N_{\text{eff}}$. At low $m_{\phi^A}$, the amount of normal matter is small because wash-out effects erase most of the asymmetry. At high $m_{\phi^A}$, the amount of matter is also small because the decay asymmetry is suppressed. The amount of matter is therefore optimized for an intermediary value of $m_{\phi^A}$. The same discussion applies to the $B$ sector. The only difference is that the wash-out effects are less important because they involve the mirror top, which is very heavy and thus suppresses wash-out. The contribution to $N_{\text{eff}}$ simply decreases as $m_{\phi^A}$ and $m_{\phi^B}$ become larger, as this increases the lifetime of $N_1$ and decreases the efficiency of processes that destroy $N_1$ around the time $N_2$ decays.

In addition, there are regions that can meet several of these requirements at the same time. Two of them are especially interesting. First, the blue/green region provides a sufficient amount of matter and dark matter. It however does not provide a sufficiently low $\Delta N_{\text{eff}}$ and the dark matter self-interactions are too large. These two issues could however be addressed by additional model buildings. Second, the blue/purple/yellow region provides a sufficient amount of matter and a sufficiently low $\Delta N_{\text{eff}}$. It also has the benefit of leading to an amount of mirror atoms sufficiently low to pass the dark matter self-interactions bounds. This region is especially interesting, as it only requires an additional source of dark matter for it to provide a complete valid cosmology. Assuming the mechanism responsible for the production of this extra dark matter is over when $N_2$ starts to decay, our mechanism wouldn't be affected much by the small amount of dark matter that would lead to the current density. Otherwise, the consequences of the production of this extra dark matter are too model dependent to make a general statement. More extensive scans of the parameters of Eq.~\eqref{eq:ValuesPlotIllustration2} and initial conditions did not reveal any region that could at the same time provide the correct matter and dark matter abundance with a sufficiently low $\Delta N_{\text{eff}}$. If such a region exists, it is most likely unnatural or requires a large reheating temperature.

Different branching ratios of $N_2$ to the $A$ and $B$ sectors have relatively little impact on the final temperature ratio. This is because the $A$ and $B$ sectors acquire sufficiently high temperatures to reach thermal equilibrium via Higgs boson exchange. The initial temperature difference is simply erased. Also, this thermal equilibration between the $A$ and $B$ sectors makes it necessary for $N_1$ to decay after the two sectors have decoupled. However, the amount of time required for $N_1$ to dominate the energy abundance is typically much larger and generally controls the required lifetime of $N_1$.

The process $N_1 N_1 \to \bar{b}^M b^M$ is responsible for the destruction of a large fraction of the $N_1$. This is why relatively so few of them are present in Fig.~\ref{fig:BAbundance} immediately after the decay of $N_2$, despite its relatively large branching ratio to $N_1 b^M \bar{b}^M$.

The peak that can be seen in Fig.~\ref{fig:BTemperatureRatio} at $10^{-5}$~s corresponds to the $B$ sector QCD phase transition quickly followed by the $A$ sector QCD phase transition. Despite its rather striking nature, it affects relatively little the final results as the $A$ and $B$ sectors have decoupled by this point. If the decoupling of the two sectors had taken place between the two QCD phase transitions, this could have contributed to a partial solution of the $N_{\text{eff}}$ problem, which was explored in Ref.~\cite{Farina:2015uea}.

Arguably the most crucial question concerning baryogenesis is whether it can be done while maintaining temperatures low enough not to reintroduce domain walls. A rigorous answer to that question however requires the knowledge of the full potential and not simply $v^A$ and $v^B$ as we have only provided. As such, this question cannot be fully answered here. However, it can be seen in Fig.~\ref{fig:TAMax} that the reheating temperatures can be comfortably below the electroweak scale. As long as $N_1$ is sufficiently light, this reheating temperature is mostly independent of $m_{N_1}$ and $v^B/v^A$. Barring any esoteric model building, the $\mathbb{Z}_2$ symmetry restoration temperature of a given Twin Higgs model should be far higher than such temperatures and domain walls should not be a problem. We also mention that these temperatures are considerably above the lower bound for reheating from BBN \cite{Hannestad:2004px}.

\section{Mirror neutrons as dark matter}\label{Sec:NDM}
As explained before, dark matter cannot realistically take the form of dark atoms in the Mirror Twin Higgs. In this section, we discuss a model without explicit $\mathbb{Z}_2$ breaking in which dark matter consists of mirror neutrons. The model is summarized and the constraints discussed. Two alternative models are then presented.

\subsection{Model summary}\label{ssSec:VLQModel}
One easy albeit not necessarily obvious way to make the mirror proton heavier than the mirror neutron is via the inclusion of vector quarks. Introduce the vector fermions
\begin{equation}\label{eq:VLQfieldContent}
  \begin{aligned}
    U^A & : \left(\mathbf{3},       \mathbf{1}, \frac{2}{3}, \mathbf{1}, \mathbf{1}, 0          \right), &
    U^B & : \left(\mathbf{1},       \mathbf{1}, 0,           \mathbf{3}, \mathbf{1}, \frac{2}{3}\right), \\
    Q^A & : \left(\mathbf{3},       \mathbf{2}, \frac{1}{6}, \mathbf{1}, \mathbf{1}, 0          \right), &
    Q^B & : \left(\mathbf{1},       \mathbf{1}, 0,           \mathbf{3}, \mathbf{2}, \frac{1}{6}\right).
  \end{aligned}
\end{equation}
The part of the Lagrangian that controls the up-type quark masses is
\begin{equation}\label{eq:LagVLQ}
  \begin{aligned}
    \mathcal{L} = & - y_u \left[(\tilde{H}^A)^\dagger \bar{u}^A P_L q^A + (\tilde{H}^B)^\dagger \bar{u}^B P_L q^B\right]  +\text{h.c.}\\
                  & - Y_Q \left[(\tilde{H}^A)^\dagger \bar{u}^A P_L Q^A + (\tilde{H}^B)^\dagger \bar{u}^B P_L Q^B\right]  +\text{h.c.}\\
                  & - Y_U \left[(\tilde{H}^A)^\dagger \bar{U}^A P_L q^A + (\tilde{H}^B)^\dagger \bar{U}^B P_L q^B\right]  +\text{h.c.}\\
                  & - Y_V \left[(\tilde{H}^A)^\dagger \bar{U}^A P_R Q^A + (\tilde{H}^B)^\dagger \bar{U}^B P_R Q^B\right]  +\text{h.c.}\\
                  & - M_U\left(\bar{U}^A U^A + \bar{U}^B U^B \right) - M_Q\left(\bar{Q}^A Q^A + \bar{Q}^B Q^B \right),
  \end{aligned}
\end{equation}
where $\tilde{H}^M = i\sigma_2 (H^M)^*$ and where we considered only the first generation. The mass of the lightest eigenstate $\hat{u}_1^M$ of sector $M$ is then given approximately by
\begin{equation}\label{eq:mlightestVLQ}
  m_{\hat{u}^M_1} \approx \frac{y_u v^M}{\sqrt{2}} + \frac{Y_U Y_Q Y_V^* (v^M)^3}{2\sqrt{2} M_Q M_U}.
\end{equation}
A similar mixing could take place for the down quark, but we will assume it to be negligible.

The main point of this mechanism is the correction to $m_{\hat{u}^M_1}$ that goes as $(v^M)^3$. Assuming this term is negligible for the down quark, the presence of this correction ensures that the mass of the mirror up quark increases more rapidly than the mirror down quark as $v^B$ increases. Barring any experimental constraints, this is sufficient to make the mirror up quark heavier than the mirror down quark and results in the mirror proton being heavier than the mirror neutron, which decreases the abundance of mirror atoms.

As a more technical aside, the presence of the $(v^M)^3$ term is especially interesting. Ref.~\cite{Beauchesne:2020mih} studied the mirror neutron as a dark matter candidate in the Mirror Twin Two Higgs Doublet Model (MT2HDM). One of the major challenges was that increasing the mirror vevs could certainly increase the splitting between the mass of the mirror proton and mirror neutron, but it also reduces the mirror Fermi constant. This reduction has the effect of making processes that convert mirror protons to mirror neutrons freeze-out earlier. These two effects partially cancel each other, either requiring $\tan\beta^A$ to be small or forcing certain parameters to be closer to their experimental limits. This dependence on $(v^M)^3$ ensures that this cancellation is much weaker, avoiding the main issue of the MT2HDM.

\subsection{Constraints}\label{sSec:VLQConstraints}

\subsubsection*{$N_{\text{eff}}$, Higgs signal strengths and fraction of dark atoms}
The constraints on $N_{\text{eff}}$ and the Higgs signal strengths are applied as in Sec.~\ref{sSec:Constraints}. The fraction of dark atoms is computed using the procedure of Appendix~\ref{App:XDAComp}.

\subsubsection*{Precision measurements}
The mixing of chiral quarks with vector quarks also affects electroweak precision measurements. The masses of the up-like quarks coming from Eq.~\eqref{eq:LagVLQ} are 
\begin{equation}\label{eq:UpquarkMasses}
  \mathcal{L} \supset -\begin{pmatrix} \bar{u}^M & \bar{U}^M & \bar{U}^M_Q  \end{pmatrix}\begin{pmatrix} \frac{y_u v^M}{\sqrt{2}} & 0 & \frac{Y_Q v^M}{\sqrt{2}} \\ \frac{Y_U v^M}{\sqrt{2}} & M_U & 0 \\ 0 & \frac{Y_V^* v^M}{\sqrt{2}} & M_Q \end{pmatrix} \begin{pmatrix} P_L u^M \\ P_L U^M \\ P_L U^M_Q \end{pmatrix} + \text{h.c.},
\end{equation}
where $U^M_Q$ ($D^M_Q$) is the positively (negatively) charged part of $Q^M$. This can be diagonalized by performing the basis change
\begin{equation}\label{eq:BVLQ}
  \begin{pmatrix} P_L u^M \\ P_L U^M \\ P_L U^M_Q \end{pmatrix} = R^M_L \begin{pmatrix} P_L\hat{u}^M_1 \\ P_L\hat{u}^M_2 \\ P_L\hat{u}^M_3\end{pmatrix}, \qquad
  \begin{pmatrix} P_R u^M \\ P_R U^M \\ P_R U^M_Q \end{pmatrix} = R^M_R \begin{pmatrix} P_R\hat{u}^M_1 \\ P_R\hat{u}^M_2 \\ P_R\hat{u}^M_3\end{pmatrix},
\end{equation}
where $\hat{u}_i^M$ are the mass eigenstates of sector $M$ ordered from lightest to heaviest. We consider three types of precision measurements.

First, the $S$ and $T$ parameters can be computed using the results of Refs.~\cite{Anastasiou:2009rv, Lavoura:1992np, Chen:2003fm, Carena:2007ua} (see also Refs.~\cite{Chen:2017hak, Cheung:2020vqm} for their use in relation to vector fermions)
\begin{equation}\label{eq:ST}
  \begin{aligned}
    S = &\frac{N_c}{2\pi}\sum_{i,j}\Big\{\left(|A^L_{ij}|^2 +  |A^R_{ij}|^2\right)\psi_+(y_i, y_j) + 2\text{Re}\left(A^L_{ij}A^{R*}_{ij}\right)\psi_-(y_i, y_j)\\
        & -\frac{1}{2}\left[\left(|X^L_{ij}|^2 +  |X^R_{ij}|^2\right)\chi_+(y_i, y_j) + 2\text{Re}\left(X^L_{ij}X^{R*}_{ij}\right)\chi_-(y_i, y_j)\right]\Big\},\\
    T = &\frac{N_c}{16\pi s_W^2 c_W^2}\sum_{i,j}\Big\{\left(|A^L_{ij}|^2 +  |A^R_{ij}|^2\right)\theta_+(y_i, y_j) + 2\text{Re}\left(A^L_{ij}A^{R*}_{ij}\right)\theta_-(y_i, y_j)\\
        & -\frac{1}{2}\left[\left(|X^L_{ij}|^2 +  |X^R_{ij}|^2\right)\theta_+(y_i, y_j) + 2\text{Re}\left(X^L_{ij}X^{R*}_{ij}\right)\theta_-(y_i, y_j)\right]\Big\},
  \end{aligned}
\end{equation}
where $N_c$ is the number of colours, $s_W$ $(c_W)$ is the sin (cos) of the weak mixing angle, $y_i = m_i^2/m_Z^2$,
\begin{equation}\label{eq:FuncBasisST}
  \begin{aligned}
    \psi_+(y_1, y_2)   &= \frac{1}{3} - \frac{1}{9}\ln\frac{y_1}{y_2}\\
    \psi_-(y_1, y_2)   &= -\frac{y_1 + y_2}{6\sqrt{y_1 y_2}}\\
    \chi_+(y_1, y_2)   &= \frac{5(y_1^2 + y_2^2) - 22y_1 y_2}{9(y_1 - y_2)^2} + \frac{3y_1 y_2 (y_1 + y_2) - y_1^3 - y_2^3}{3(y_1 - y_2)^3}\ln\frac{y_1}{y_2}\\
    \chi_-(y_1, y_2)   &= -\sqrt{y_1 y_2} \left[\frac{y_1 + y_2}{6y_1 y_2} - \frac{y_1 + y_2}{(y_1 - y_2)^2}  + \frac{2y_1 y_2}{(y_1 - y_2)^3}\ln\frac{y_1}{y_2}\right]\\
    \theta_+(y_1, y_2) &= y_1 + y_2 - \frac{2y_1 y_2}{y_1 - y_2}\ln\frac{y_1}{y_2}\\
    \theta_-(y_1, y_2) &= 2\sqrt{y_1 y_2} \left[\frac{y_1 + y_2}{y_1 - y_2}\ln\frac{y_1}{y_2} - 2\right],
  \end{aligned}
\end{equation}
with
\begin{equation}\label{eq:XA}
  \begin{aligned}
    A^L_{\hat{u}_i d}        =& (R_L^A)_{1i}^*, \quad A^R_{\hat{u}_i d} = 0, \quad A^L_{\hat{u}_i D_Q} = (R_L^A)_{3i}^*, \quad A^R_{\hat{u}_i D_Q} = (R_R^A)_{3i}^*,\\
             X^L_{dd}        =& -1, \quad X_{dd}^R = 0, \quad X_{D_Q D_Q}^L = X_{D_Q D_Q}^R = -1,\\
   X^L_{\hat{u}_i \hat{u}_j} =& \left(1 - \frac{4}{3} s_W^2\right)(R_L^A)_{1i}^* (R_L^A)_{1j} + \left(1 - \frac{4}{3} s_W^2\right)(R_L^A)_{3i}^* (R_L^A)_{3j}\\
                              & - \frac{4}{3} s_W^2 (R_L^A)_{2i}^* (R_L^A)_{2j} + \frac{4}{3} s_W^2 \delta_{ij},\\
   X^R_{\hat{u}_i \hat{u}_j} =& - \frac{4}{3} s_W^2 (R_R^A)_{1i}^* (R_R^A)_{1j} + \left(1 - \frac{4}{3} s_W^2\right)(R_R^A)_{3i}^* (R_R^A)_{3j}\\
                              & - \frac{4}{3} s_W^2 (R_R^A)_{2i}^* (R_R^A)_{2j} + \frac{4}{3} s_W^2 \delta_{ij}.
  \end{aligned}
\end{equation}
The new physics contributions to the oblique parameters are then
\begin{equation}\label{eq:DeltaST}
  \Delta S = S - S_{\text{SM}}, \qquad \Delta T = T - T_{\text{SM}},
\end{equation}
with
\begin{equation}\label{eq:STSM}
  S_{\text{SM}} = \frac{N_c}{6\pi}\left[1 - \frac{1}{3}\ln \frac{m_{u^A}^2}{m_{d^A}^2} \right], \qquad
  T_{\text{SM}} = \frac{N_c}{16\pi s_W^2 c_W^2}\theta_+(y_{u^A}, y_{d^A}).
\end{equation}

Second, the weak nuclear charges of ${}^{133}\text{Cs}$ and ${}^{204}\text{Tl}$ from atomic parity violation are computed using the results of Ref.~\cite{Okada:2012gy}. This gives a contribution from new physics of
\begin{equation}\label{eq:APV}
  \begin{aligned}
    \delta Q_W =& (2Z + N)\Bigg[\left(1 - \frac{4}{3} s_W^2\right)|(R_L^A)_{11}|^2 + \left(1 - \frac{4}{3} s_W^2\right)|(R_L^A)_{31}|^2 -\frac{4}{3} s_W^2|(R_L^A)_{21}|^2\\
                & - \frac{4}{3} s_W^2|(R_R^A)_{11}|^2 + \left(1 - \frac{4}{3} s_W^2\right)|(R_R^A)_{31}|^2 -\frac{4}{3} s_W^2|(R_R^A)_{21}|^2 - 1 +\frac{8}{3}s_W^2\Bigg],
  \end{aligned}
\end{equation}
where $Z$ and $N$ are respectively the number of protons and neutrons in an element.

Third, mixing of the chiral up with vector quarks leads to violation of the unitarity of the CKM matrix. The first row is the most precisely measured and the sum of the absolute values of its elements squared becomes
\begin{equation}\label{eq:CKMunitarityViolaton1}
  |V_{ud}|^2 + |V_{us}|^2 + |V_{ub}|^2 = 1 - \delta V,
\end{equation}
where
\begin{equation}\label{eq:CKMunitarityViolaton2}
  \delta V = 1 - |(R_L^A)_{11}^2|.
\end{equation}

\subsubsection*{Direct collider searches}
Searches for vector partners of the light quarks have been performed by ATLAS and CMS in the 8~TeV, 20~fb$^{-1}$ dataset~\cite{ATLAS:2015lpr,CMS:2017asf} and excluded pair production of such quarks up to masses of $845$~GeV or lower, depending on branching fractions. Even though dedicated searches for such signatures have not yet been done on the full currently available dataset, it is reasonable to assume that vector quarks with masses $\sim 1.5$~TeV and higher are still consistent with the data, given that recent dedicated searches for vector partners of the heavy quarks, whose decays include $b$ jets (which is an easier signature) set limits only up to $\sim 1.6$~TeV~\cite{ATLAS-CONF-2021-024,CMS:2020ttz}. Additionally, scenarios with large mixing have significant cross sections for single and pair production of vector quarks via electroweak processes, which can be constrained by various LHC measurements~\cite{Buckley:2020wzk}.

\subsection{Parameter space and comments}\label{sSec:ParameterSpaceVLQ}
\begin{figure}[t!]
  \centering
   \captionsetup{justification=centering}
    \begin{subfigure}{0.495\textwidth}
    \centering
    \caption{$X_{\text{DA}}$}
    \includegraphics[width=\textwidth]{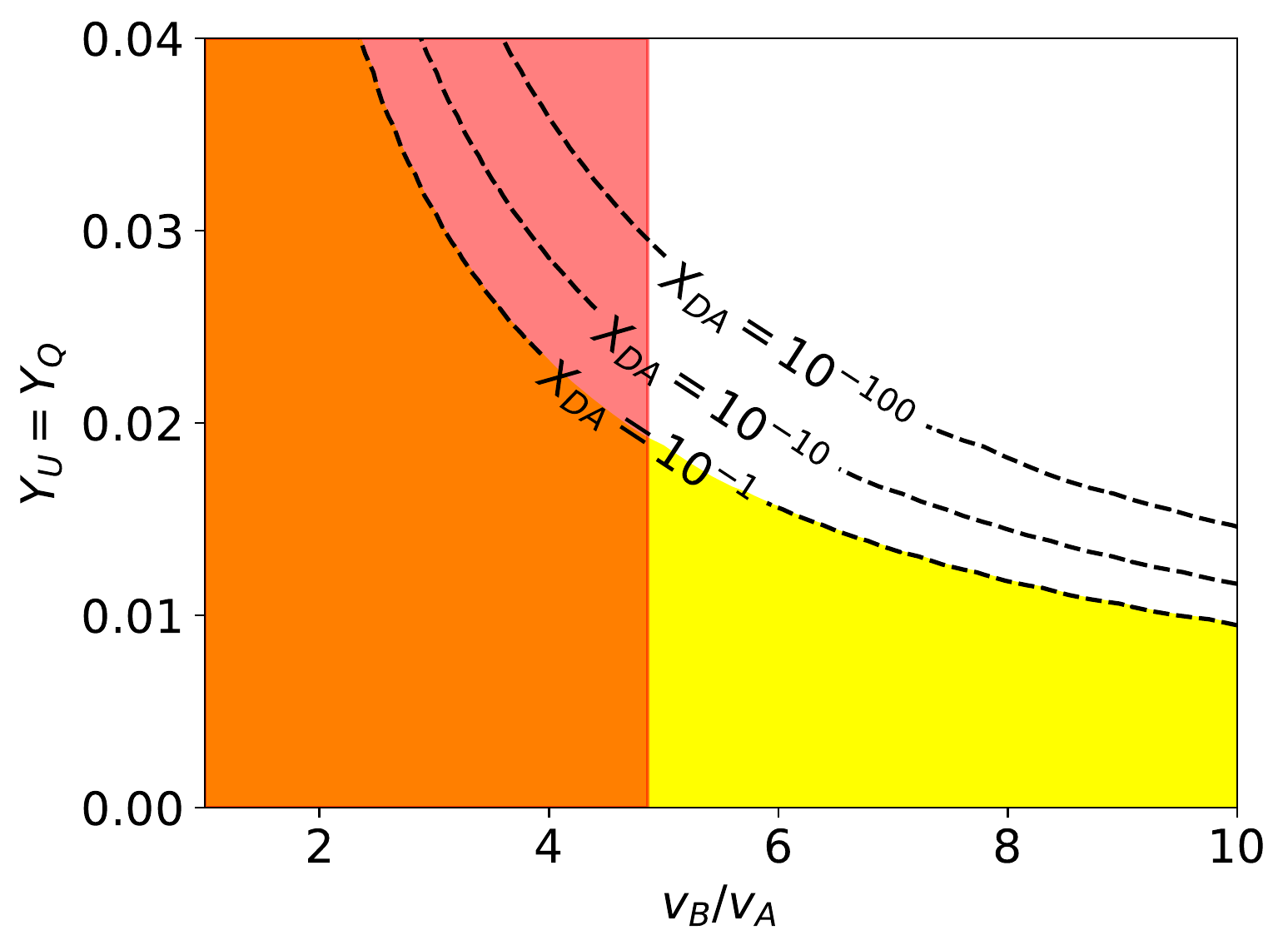}
    \label{fig:VLQ1XDA}
  \end{subfigure}
  \begin{subfigure}{0.495\textwidth}
    \centering
    \caption{$\delta Q_W ({}^{133}\text{Cs})$}
    \includegraphics[width=\textwidth]{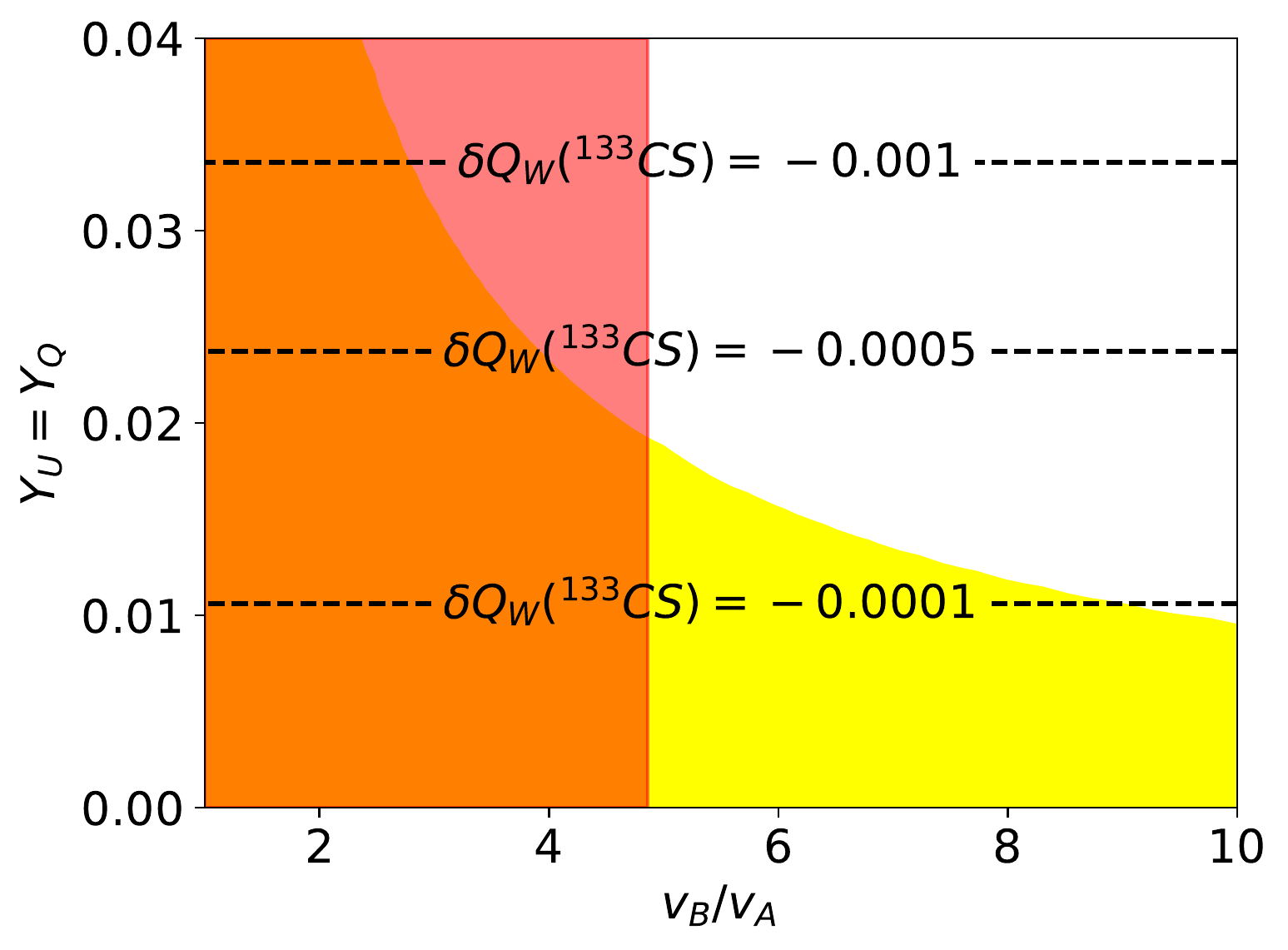}
    \label{fig:VLQ1CS}
  \end{subfigure}
  \vspace{-0.2cm}
  \begin{subfigure}{0.495\textwidth}
    \centering
    \caption{$\delta Q_W ({}^{204}\text{Tl})$}
    \includegraphics[width=\textwidth]{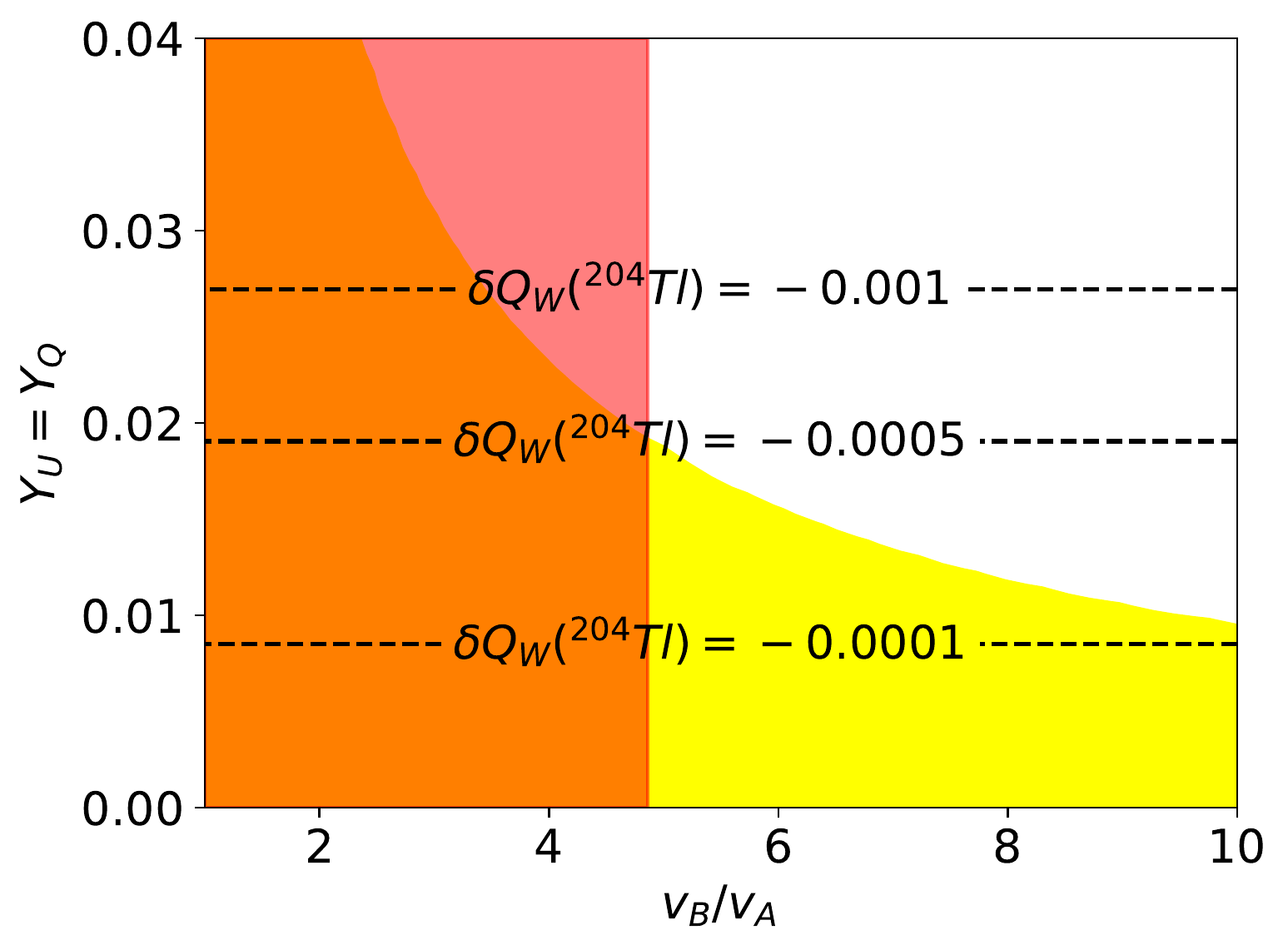}
    \label{fig:FVLQ1Tl}
  \end{subfigure}
  \begin{subfigure}{0.495\textwidth}
    \centering
    \caption{$\delta V$}
    \includegraphics[width=\textwidth]{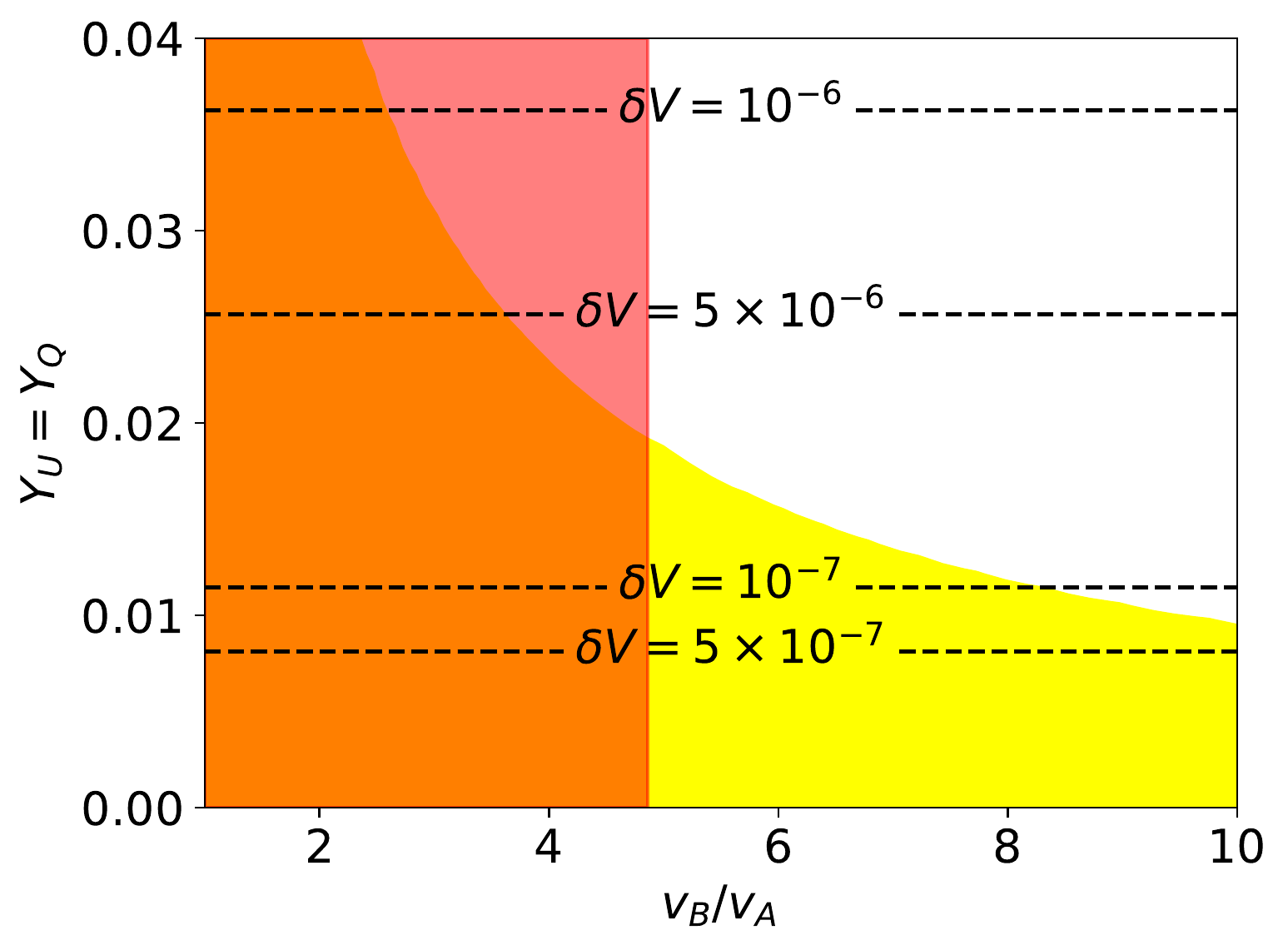}
    \label{fig:VLQ1deltaV}
  \end{subfigure}
  \vspace{-0.2cm}
  \begin{subfigure}{0.495\textwidth}
    \centering
    \caption{$t$}
    \includegraphics[width=\textwidth]{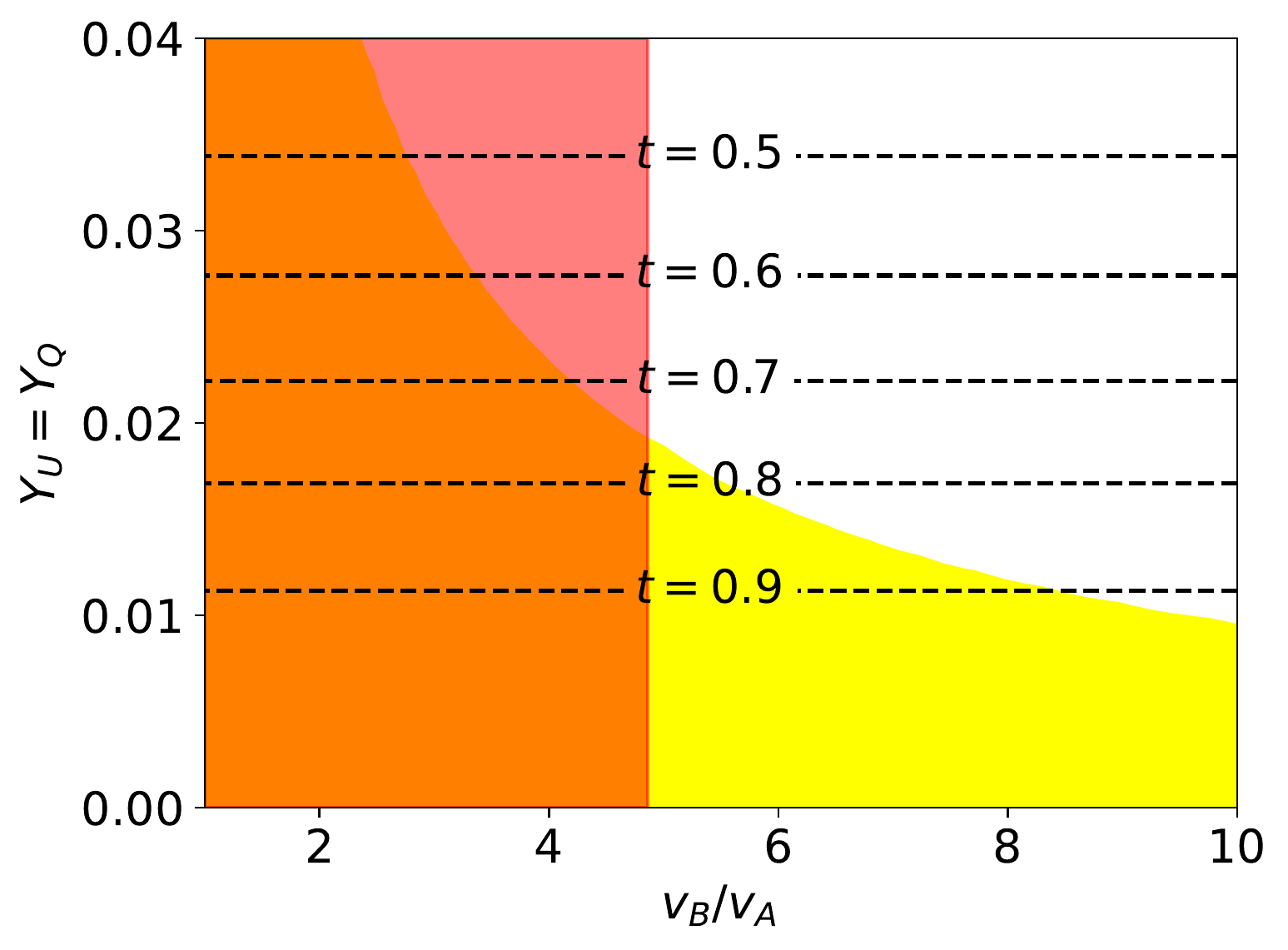}
    \label{fig:VLQ1t}
  \end{subfigure}
  \captionsetup{justification=justified}
\caption{Constraints on the vector quarks model with $M_U = 2$~TeV, $M_Q = 3$~TeV, $Y_V=1$ and $r_T = 0.4$. The red region is excluded at 95\% confidence level by the Higgs signal strengths and the yellow one is disfavoured by dark matter self-interactions bounds.}\label{fig:VLQ1}
\end{figure}

\begin{figure}[t!]
  \centering
   \captionsetup{justification=centering}
    \begin{subfigure}{0.495\textwidth}
    \centering
    \caption{$r_T = 0.2$}
    \includegraphics[width=\textwidth]{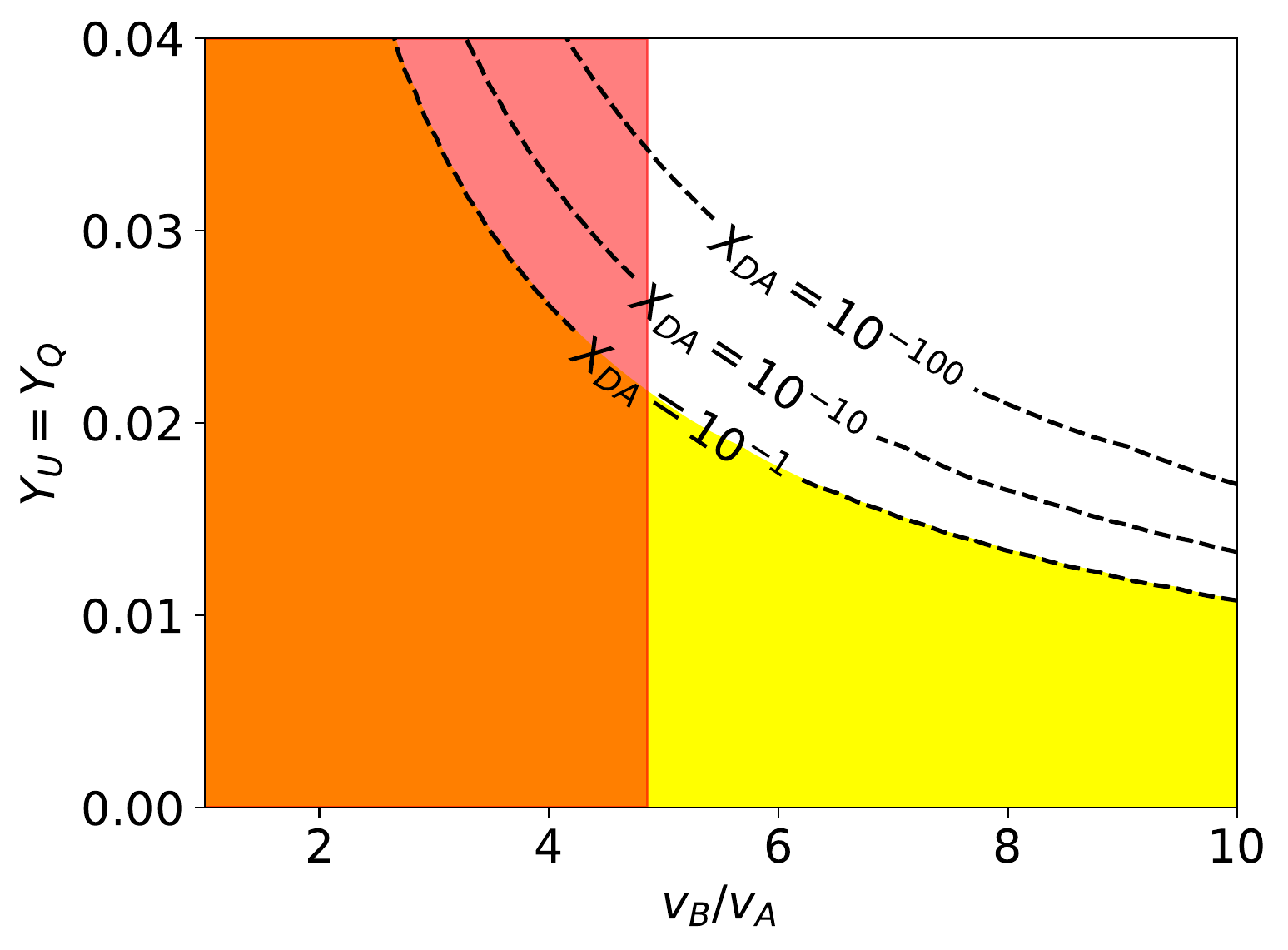}
    \label{fig:VLQ1XDA2}
  \end{subfigure}
  \begin{subfigure}{0.495\textwidth}
    \centering
    \caption{$r_T = 0.1$}
    \includegraphics[width=\textwidth]{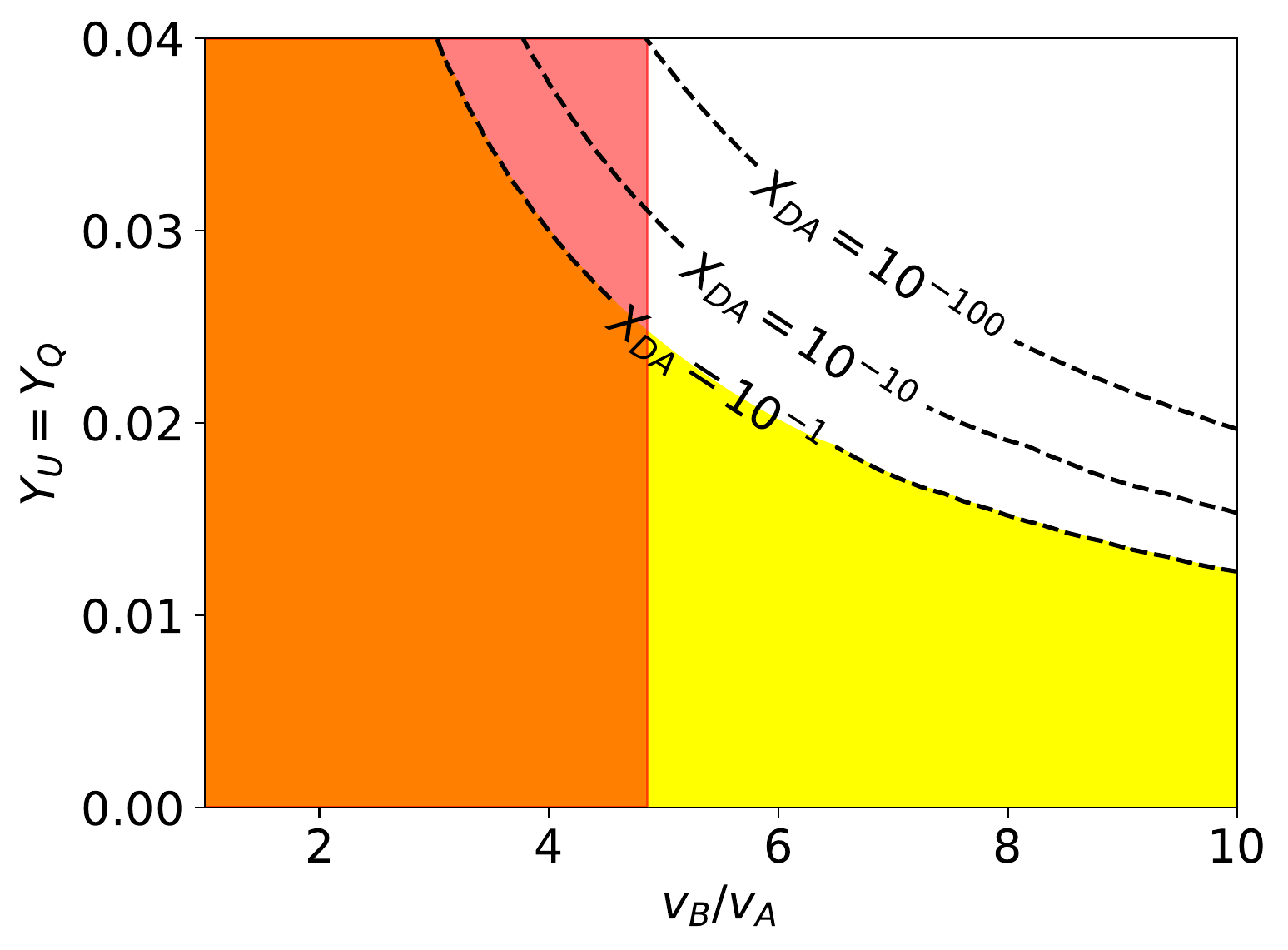}
    \label{fig:VLQ1XDA3}
  \end{subfigure}
  \captionsetup{justification=justified}
\caption{Similar to Fig.~\ref{fig:VLQ1} but for different values of $r_T$.}\label{fig:VLQ2}
\end{figure}

Fig.~\ref{fig:VLQ1} shows the allowed parameter space as a function of $v^B/v^A$ and $Y_U = Y_Q$ for $M_U = 2$~TeV, $M_Q = 3$~TeV, $Y_V=1$ and $r_T = 0.4$. The parameter $y_u$ is adjusted to reproduce the correct mass of the up quark. This choice of parameters ensures that no regions of the plots are excluded by the precision measurements, but is not in any way uncharacteristic. Contours corresponding to the dark atom abundance and the contributions to the different electroweak precision measurements of Sec.~\ref{sSec:VLQConstraints} are also shown. The oblique parameters are almost constant over the region shown and are given by $\Delta S \sim 6.6 \times 10^{-4}$ and $\Delta T \sim 2.6 \times 10^{-3}$. Fig.~\ref{fig:VLQ2} shows contours of $X_{\text{DA}}$ for other values of $r_T$.

As can be seen, the fraction of dark atoms can easily be brought to extremely low levels. This can be done while leading to contributions to experimental measurements well below any current limits. In addition, mixing of the chiral up quark with vector quarks could in principle contribute enough to its mass that it might require some amount of tuning for it to remain light. As such, we can define a measure of tuning as
\begin{equation}\label{eq:Delta}
  \Delta = \max_{p\in \mathcal{P}}\left\{\left|\frac{d\ln m_{\hat{u}^A_1}}{d\ln p}\right|\right\},
\end{equation}
where $\mathcal{P} = \{y_u, Y_Q, Y_U, Y_V, M_U, M_Q, v_A\}$. The tuning is then given by $t=1/\Delta$. As can be seen in Fig.~\ref{fig:VLQ1t}, all constraints can be satisfied without $t$ needing to be small.

Do note that a sufficiently large splitting between the masses of the mirror down and mirror up could eventually lead to the spin-3/2 baryon $d^B d^B d^B$ being lighter than the mirror neutron. This would reintroduce the dark atoms problem. A naive estimate in combination with the lattice results of Ref.~\cite{Zanotti:2003fx} reveals that this takes place at values of $Y_U = Y_Q$ and $v^B/v^A$ much larger than those required to obtain a sufficiently low $X_{\text{DA}}$.

In simple terms, the mechanism works so well because the mass of the up quark is so small that it can be considerably modified without introducing much mixing.

\subsection{Alternative models}\label{sSec:AlternativeModels}
In this section, we describe two alternative models to obtain mirror neutrons as dark matter candidates. We only present the models and leave detailed studies of their constraints for future work.

Both models are inspired by Ref.~\cite{Beauchesne:2020mih}, which showed that mirror atoms could be brought to acceptable abundances in the MT2HDM with explicit $\mathbb{Z}_2$ breaking. The idea of the paper was to introduce two Higgs doublets $H_1^A$ and $H_2^A$ and their partners $H_1^B$ and $H_2^B$. By assumption, $H_2^A$ provides mass to the up-type quarks and $H_1^A$ to the down-type quarks. The masses of the quarks of the $A$ and $B$ sectors obey the following relation
\begin{equation}\label{eq:mirrorQuarkMassRatio}
  \frac{m_{u^B}}{m_{d^B}} = \frac{\tan\beta^B}{\tan\beta^A}\frac{m_{u^A}}{m_{d^A}},
\end{equation}
where $\tan\beta^M = \langle H_2^M \rangle /  \langle H_1^M \rangle$. As such, taking a sufficiently large $\tan \beta^B /\tan\beta^A$ leads to a mirror proton heavier than the mirror neutron and should decrease the abundance of dark atoms. The main challenge however is that increasing the mirror vevs decreases the mirror Fermi constant and makes the processes that convert mirror protons to mirror neutrons freeze-out earlier. It is then necessary to go to relatively low $\tan\beta^A$ or be willing to accept a mass of the up quark closer to its experimental upper limit. In the end, the model is compatible with current bounds and does not require additional tuning besides the one necessary to pass the Higgs signal strengths requirements.

The correct structure of vevs was obtained in Ref.~\cite{Beauchesne:2020mih} by including soft masses that explicitly broke the $\mathbb{Z}_2$ symmetry. The idea of the models of this section is to obtain a similar vevs structure without any explicit $\mathbb{Z}_2$ breaking.

In the first model, a pair of new real scalars $S^A$ and $S^B$ are introduced. The following potential can then be introduced
\begin{equation}\label{eq:LagAlt1}
  \begin{aligned}
    V = & - \mu_S^2\left((S^A)^2 + (S^B)^2\right) + \lambda_S \left((S^A)^2 + (S^B)^2\right)^2 + \alpha (S^A)^2 (S^B)^2\\
        & + \lambda_1\left((S^A)^2 |H_1^A|^2 + (S^B)^2 |H_1^B|^2\right) + \lambda_2\left((S^A)^2 |H_2^A|^2 + (S^B)^2 |H_2^B|^2\right).
  \end{aligned}
\end{equation}
Assuming $\alpha > 0$, the $\mathbb{Z}_2$ symmetry will be broken spontaneously by the first line. At tree level, only one of $S^A$ or $S^B$ will get a vev and we can assume it to be $S^B$. The second line of Eq.~\eqref{eq:LagAlt1} then effectively acts as soft $\mathbb{Z}_2$ breaking masses that can be adjusted to reproduce the results of the MT2HDM with explicit $\mathbb{Z}_2$ breaking.

The second model is based on Ref.~\cite{Beauchesne:2015lva}. The following potential is introduced
\begin{equation}\label{eq:LagAlt2}
  \begin{aligned}
    V = & - \mu_1^2\left(|H_1^A|^2 + |H_1^B|^2\right) + \lambda_1\left(|H_1^A|^2 + |H_1^B|^2\right)^2 + \alpha_1|H_1^A|^2|H_1^B|^2\\
        & - \mu_2^2\left(|H_2^A|^2 + |H_2^B|^2\right) + \lambda_2\left(|H_2^A|^2 + |H_2^B|^2\right)^2 + \alpha_2|H_2^A|^2|H_2^B|^2\\
        & -B_\mu\left((H_1^A)^\dagger H_2^A + (H_1^B)^\dagger H_2^B \right) + \text{h.c}.
  \end{aligned}
\end{equation}
First, assume $B_\mu$ is zero. If $\alpha_i$ is positive and the other negative, the $i$ Higgs spontaneously breaks the $\mathbb{Z}_2$ symmetry by obtaining a vev in only one sector, which can be taken to be the $B$ sector. The other Higgs obtains a vev that maintains the $\mathbb{Z}_2$ symmetry. Once the $B_\mu$ term is turned on, the $\mathbb{Z}_2$ breaking is transmitted from the broken to the unbroken Higgs sector. It was shown that such a vev structure can pass the Higgs signal strengths requirements. There are then two standard behaviors:
\begin{equation}\label{eq:vevModel2}
  \begin{aligned}
    (1) \quad \alpha_1 > 0 \quad \text{and} \quad \alpha_2 < 0: \qquad \tan\beta^A > 1, \qquad \tan\beta^B < 1,\\
    (2) \quad \alpha_1 < 0 \quad \text{and} \quad \alpha_2 > 0: \qquad \tan\beta^A < 1, \qquad \tan\beta^B > 1.
  \end{aligned}
\end{equation}
The first possibility is the exact opposite of what is required. The second possibility however leads to a $\tan \beta^B /\tan\beta^A$ that can be considerably larger than one and at the same time a low $\tan\beta^A$. All the tools necessary to obtain a sufficiently low abundance of dark atoms are then present. The main drawback is that the model leads to a low $\tan\beta^A$, which can complicate UV completions.

\section{Conclusion}\label{Sec:Conclusion}
The Twin Higgs attempts to solve the little hierarchy problem by introducing a mirror copy of the Standard Model related by a $\mathbb{Z}_2$ symmetry. Because of the measurements of the Higgs signal strengths and cosmology, the $\mathbb{Z}_2$ symmetry must however be broken. The possibility of only breaking this symmetry spontaneously is certainly aesthetically appealing. It was already demonstrated that this can be done for the Higgs signal strengths, but it remained an open question as to whether this could be done for cosmology. As such, the goal of this paper was to determine whether it is possible to create Twin Higgs models in which the $\mathbb{Z}_2$ symmetry is only broken spontaneously that can successfully lead to baryogenesis, provide the correct dark matter abundance, solve the $N_{\text{eff}}$ problem and provide a viable dark matter candidate.

We found that it is indeed possible to create models that address the above issues. To demonstrate this, we built and studied two of them. In the first model, a pair of Majorana fermions is introduced. In the early Universe, the heaviest Majorana fermion dominates the energy abundance. It then decays, producing a net amount of baryons and mirror baryons as well as some amount of the lighter Majorana fermion. The latter eventually comes to dominate the energy abundance of the Universe. Because of the masses of the particles involved, the lighter Majorana fermion then decays almost exclusively to the Standard Model sector thus reheating it. This model can provide the correct matter abundance without reaching temperatures that would reintroduce domain walls. It can also either solve the $N_{\text{eff}}$ problem and generate an acceptably low abundance of dark atoms or provide the correct dark matter abundance.

The second model attempts to convert the dark matter to a form compatible with limits on dark matter self-interactions. This is done by introducing vector quarks that mix with the up quark of their respective sector via Yukawa interactions involving the Higgs. This contributes to the mass of the up quark of a given sector a term proportional to the vev of the Higgs of that sector cubed. This can easily make the mirror up heavier than the mirror down and thus result in a mirror proton heavier than the mirror neutron. The dark matter then takes the form of mirror neutrons and the amount of mirror atoms can be brought to negligible levels. All considered experimental constraints can easily be satisfied and the model can be combined with the first one without adverse side effects.

As a closing word, the models presented in this paper indeed show that the challenges associated to the cosmology of the Twin Higgs without explicit $\mathbb{Z}_2$ breaking can be solved individually and sometimes multiple at a time. However, whether there exists a simple model that can solve all of these problems at the same time is still an open question.

\acknowledgments
This research was supported in part by the Israel Science Foundation (grant no.\ 780/17) and the United States - Israel Binational Science Foundation (grant no.\  2018257). This work was supported by the Ministry of Science and Technology, National Center for Theoretical Sciences of Taiwan.

\appendix

\section{Thermal averages and energy exchange rates}\label{App:TA}
In this appendix, we present the computations for the thermally averaged cross sections and energy exchange rates. This is done by expanding the work of Ref.~\cite{Edsjo:1997bg}, from which we reuse the notation.

\subsection{General approach to energy exchange}\label{sApp:TAGeneral}
We focus on $2\to 2$ processes of the form
\begin{equation}\label{eq:2to2}
  i j \to m n,
\end{equation}
where $i$, $j$, $m$ and $n$ are a set of particles not necessarily of distinct species. We will refer to a generic particle from this set by a lower case Greek letter. The mass and number of internal degrees of freedom of particle $\alpha$ are labelled respectively as $m_\alpha$ and $g_\alpha$. In a fixed `laboratory' frame, the momentum of particle $\alpha$ is labelled as $p_\alpha$, its energy as $E_\alpha$ and its three-momentum as $\mathbf{p_\alpha}$. A convenient and complete basis for these energies is
\begin{equation}\label{eq:Energyvariables}
    E_+ = E_i + E_j, \quad E_- = E_i - E_j, \quad E_+' = E_m + E_n, \quad E_-' = E_m - E_n.
\end{equation}
All thermal averages we will be concerned with are of the form
\begin{equation}\label{eq:TAGenerric}
\langle F(s, E_+, E_-) v_{ij}\rangle_{i j \to m n}^{T_i, T_j} = \frac{\int \frac{d^3 \mathbf{p}_i}{(2\pi)^3} \frac{d^3 \mathbf{p}_j}{(2\pi)^3} F(s, E_+, E_-) v_{ij} f_i f_j}{\int \frac{d^3 \mathbf{p}_i}{(2\pi)^3} \frac{d^3 \mathbf{p}_j}{(2\pi)^3} f_i f_j},
\end{equation}
where $T_\alpha$ is the temperature of particles $\alpha$, $F(s, E_+, E_-)$ is a generic function, $s=(p_i+p_j)^2$, $v_{ij}$ is the M{\o}ller velocity given by
\begin{equation}\label{eq:MV}
  v_{ij} = \frac{\sqrt{(p_i\cdot p_j)^2 - m_i^2 m_j^2}}{E_i E_j} = \frac{2}{E_+^2 - E_-^2}\sqrt{(s - (m_i + m_j)^2)(s - (m_i - m_j)^2)},
\end{equation}
and $f_\alpha$ the Maxwell-Boltzmann distribution for particle $\alpha$
\begin{equation}\label{eq:MBdistribution}
  f_\alpha = e^{-\frac{E_\alpha}{T_\alpha}}.
\end{equation}
Since $v_{ij}$ can be expressed as a function of $E_+$, $E_-$ and $s$, the derived result will still be generic. The inclusion of the $v_{ij}$ factor is simply more convenient. The denominator is trivially given by
\begin{equation}\label{eq:Denominator}
  \int \frac{d^3 \mathbf{p}_i}{(2\pi)^3} \frac{d^3 \mathbf{p}_j}{(2\pi)^3} f_i f_j = \frac{n^{\text{eq}}_i}{g_i} \frac{n^{\text{eq}}_j}{g_j},
\end{equation}
where $n^{\text{eq}}_\alpha$ is the equilibrium number density of particle $\alpha$ at temperature $T_\alpha$
\begin{equation}\label{eq:Neq}
  n^{\text{eq}}_\alpha = \frac{T_\alpha}{2\pi^2}g_\alpha m_\alpha^2 K_2\left(\frac{m_\alpha}{T_\alpha}\right),
\end{equation}
where $K_n$ is the modified Bessel function of the second kind of order $n$. To simplify the treatment of the numerator, introduce the notation
\begin{equation}\label{eq:TSTA}
  T_S = \frac{2T_i T_j}{T_j + T_i}, \;\;\; T_A = \frac{2T_i T_j}{T_j - T_i}.
\end{equation}
This can be used to rewrite $f_i f_j$ in the more convenient form
\begin{equation}\label{eq:fifj}
  f_i f_j = e^{-\frac{E_+}{T_S}}e^{-\frac{E_-}{T_A}}.
\end{equation}
Considering that the only non-trivial angular dependence of the differential element is on the angle between the momenta of particles $i$ and $j$, it can be rewritten as
\begin{equation}\label{eq:DiffConv}
  \frac{d^3 \mathbf{p}_i}{(2\pi)^3}\frac{d^3 \mathbf{p}_j}{(2\pi)^3} = \frac{E_+^2 - E_-^2}{(2\pi)^4}\frac{dE_+ dE_- ds}{8},
\end{equation}
where the equality is as far as integration is concerned. In terms of these variables, the region of integration is given by
\begin{equation}\label{eq:Boundaries}
  s > (m_i + m_j)^2, \;\;\;
  E_+ > \sqrt{s},    \;\;\;
  E_-^{\text{min}} < E_- < E_-^{\text{max}},
\end{equation}
where
\begin{equation}\label{eq:Emaxmin}
  E_-^{\text{min/max}} = \frac{E_+(m_i^2 - m_j^2)}{s} \mp 2p_{ij}\sqrt{\frac{E_+^2 - s}{s}},
\end{equation}
where $p_{ij}$ is the norm of the center-of-mass (CM) three-dimensional momentum of particle $i$ or $j$ and is given by
\begin{equation}\label{eq:pij}
  p_{ij} = \frac{E_+^2 - E_-^2}{4\sqrt{s}}v_{ij} = \frac{\sqrt{(s - (m_i + m_j)^2)(s - (m_i - m_j)^2)}}{2\sqrt{s}}.
\end{equation}
With this change of variables, the numerator becomes
\begin{equation}\label{eq:Numerator}
  \int \frac{d^3 \mathbf{p}_i}{(2\pi)^3} \frac{d^3 \mathbf{p}_j}{(2\pi)^3} F(s, E_+, E_-) v_{ij} f_i f_j = \frac{1}{32\pi^4}\int ds dE_+ dE_- \sqrt{s} p_{ij}F(s, E_+, E_-) e^{-\frac{E_+}{T_S}}e^{-\frac{E_-}{T_A}}.
\end{equation}
Finally, the thermal average is given by
\begin{equation}\label{eq:TAGenerricFinal}
\langle F(s, E_+, E_-) v_{ij}\rangle_{i j \to m n}^{T_i, T_j} = \frac{\int ds dE_+ dE_- \sqrt{s} p_{ij} F(s, E_+, E_-) e^{-\frac{E_+}{T_S}}e^{-\frac{E_-}{T_A}}}{8T_i T_j m_i^2 m_j^2 K_2(\frac{m_i}{T_i})K_2(\frac{m_j}{T_j})}.
\end{equation}
To obtain the cross sections or exchange rates that appear in the cosmological evolution equations, it suffices to evaluate Eq.~\eqref{eq:TAGenerricFinal} with the proper $F(s, E_+, E_-)$. For the most part, this is trivial. The only exception is for $E_-'$, which we elaborate on in the next subsection. The final results are collected in Sec.~\ref{Sec:TAResults}.

\subsection{$E_-'$ computation}\label{sSec:Eminprime}
Consider a given collision $ij \to mn$. In addition to the `laboratory' frame, one can define a center-of-mass frame. Its three-velocity with respect to the `laboratory' is labelled as $\mathbf{v_{\text{CM}}}$ and has norm $v_{\text{CM}}$. Conversely, the three-velocity of the `laboratory' in the CM frame is labelled $\mathbf{v_{\text{lab}}}$ and has norm $v_{\text{lab}} = v_{\text{CM}}$. Quantities in the CM frame are labelled with a CM subscript. As long as the coordinate systems are properly aligned, the following holds
\begin{equation}\label{eq:vcm3vect}
  \frac{\mathbf{v}_{\text{lab}}}{v_{\text{lab}}} = -\frac{\mathbf{v_{\text{CM}}}}{v_{\text{CM}}} = -\frac{\mathbf{p_i} + \mathbf{p_j}}{|\mathbf{p_i} + \mathbf{p_j}|} = -\frac{\mathbf{p_+}}{|\mathbf{p_+}|},
\end{equation}
where $\mathbf{p}_\pm = \mathbf{p}_i \pm \mathbf{p}_j$ and
\begin{equation}\label{eq:pplusnorm}
  |\mathbf{p_+}| = \sqrt{E_+^2 - s}.
\end{equation}
The quantity $E_-'$ is then related to its CM value by a simple Lorentz transformation
\begin{equation}\label{eq:EminusPrime1}
  E_-' = \gamma_{\text{CM}}\left[(E_-')_{\text{CM}} - \mathbf{v}_{\text{lab}}\cdot(\mathbf{p}'_-)_{\text{CM}}\right],
\end{equation}
where $\mathbf{p}'_\pm = \mathbf{p}_m \pm \mathbf{p}_n$. The first term of Eq.~\eqref{eq:EminusPrime1} is easily evaluated in terms of standard $2 \to 2$ kinematics and gives
\begin{equation}\label{eq:Eminusterm1}
  (E_-')_{\text{CM}} = \frac{m_m^2 - m_n^2}{\sqrt{s}}.
\end{equation}
The second term can be evaluated as follows. First, decompose $(\mathbf{p}_m)_{\text{CM}}$ as
\begin{equation}\label{eq:pmdecomp}
  (\mathbf{p}_m)_{\text{CM}} = p_{mn} \cos\theta_m \frac{(\mathbf{p}_i)_{\text{CM}}}{p_{ij}} + (\mathbf{p}_m^\perp)_{\text{CM}},
\end{equation}
where $p_{mn} = |(\mathbf{p}_m)_{\text{CM}}|$ and $\theta_m$ is the angle between $(\mathbf{p}_m)_{\text{CM}}$ and $(\mathbf{p}_i)_{\text{CM}}$. The three-vector $(\mathbf{p}_m^\perp)_{\text{CM}}$ is the component of $(\mathbf{p}_m)_{\text{CM}}$ perpedicular to $(\mathbf{p}_i)_{\text{CM}}$. In $E_-'$, it leads to a term proportional to $\cos$ of an azimuthal angle. In the thermal averages, this term vanishes once integrated over that angle as long as axial symmetry is respected. We will ignore $(\mathbf{p}_m^\perp)_{\text{CM}}$ from now on. Then, we have
\begin{equation}\label{eq:vlabdotpminusCM1}
  \mathbf{v}_{\text{lab}}\cdot(\mathbf{p}'_-)_{\text{CM}} = 2 v_{\text{CM}}p_{mn}\cos\theta_{\text{lab}}\cos\theta_m,
\end{equation}
where $\theta_{\text{lab}}$ is the angle between $\mathbf{v}_{\text{lab}}$ and $(\mathbf{p}_i)_{\text{CM}}$ and we used the fact that $(\mathbf{p}_m)_{\text{CM}} = - (\mathbf{p}_n)_{\text{CM}}$. The quantity $\cos\theta_{\text{lab}}$ is given by
\begin{equation}\label{eq:costhetalab}
  \cos\theta_{\text{lab}} = \frac{(\mathbf{p}_i)_{\text{CM}}\cdot \mathbf{v}_{\text{lab}}}{|(\mathbf{p}_i)_{\text{CM}}||\mathbf{v}_{\text{lab}}|}
                          = \frac{(\mathbf{p}_-)_{\text{CM}}\cdot \mathbf{v}_{\text{lab}}}{2p_{ij}|\mathbf{v}_{\text{lab}}|}
                          = -\frac{(\mathbf{p}_-)_{\text{CM}}\cdot \mathbf{p}_+}{2p_{ij}|\mathbf{p}_+|}.
\end{equation}
The three-vector $(\mathbf{p}_-)_{\text{CM}}$ is related to its `laboratory' value by a Lorentz transformation
\begin{equation}\label{eq:pminuscm1}
  \begin{aligned}
    (\mathbf{p}_-)_{\text{CM}} &= \mathbf{p}_- + \frac{(\gamma_{\text{CM}} - 1)}{v_{\text{CM}}^2}(\mathbf{p}_-\cdot \mathbf{v}_{\text{CM}})\mathbf{v}_{\text{CM}} - \gamma_{\text{CM}}E_- \mathbf{v}_{\text{CM}}\\
                               &= \mathbf{p}_- + \frac{(\gamma_{\text{CM}} - 1)}{|\mathbf{p}_+|^2}(\mathbf{p}_-\cdot \mathbf{p}_+)\mathbf{p}_+ - \gamma_{\text{CM}}E_- v_{\text{CM}}\frac{\mathbf{p}_+}{|\mathbf{p}_+|}.
  \end{aligned}
\end{equation}
With the results
\begin{equation}\label{eq:vcmgammacm}
  v_{\text{CM}} = \frac{\sqrt{E_+^2 - s}}{E_+}, \;\;\; \gamma_{\text{CM}} = \frac{E_+}{\sqrt{s}}, \;\;\; \mathbf{p}_-\cdot \mathbf{p}_+ = E_+ E_- - m_i^2 + m_j^2,
\end{equation}
taking the dot product of Eq.~\eqref{eq:pminuscm1} and $\mathbf{p}_+$ leads to
\begin{equation}\label{eq:pminuscm2}
  (\mathbf{p}_-)_{\text{CM}}\cdot \mathbf{p}_+ = \frac{1}{\sqrt{s}}\left(E_- s - E_+ (m_i^2 - m_j^2)\right).
\end{equation}
Assembling everything finally leads to the main result of this section\footnote{We reiterate that a term proportial to $\cos$ of an azimuthal angle was dropped from Eq.~\eqref{eq:EminusPrime2} as it vanishes in all relevant thermal averages.}
\begin{equation}\label{eq:EminusPrime2}
  E_-' = \frac{E_+}{s}\left(m_m^2 - m_n^2\right) + \frac{\left(E_-s - E_+(m_i^2 - m_j^2)\right)}{s}\frac{p_{mn}}{p_{ij}}\cos\theta_m.
\end{equation}

\subsection{Results for thermal averages and energy exchange rates}\label{Sec:TAResults}
With the results of the previous two sections, it is a trivial matter to obtain the thermally averaged cross sections and energy exchange rates. It suffices to use Eq.~\eqref{eq:TAGenerricFinal} and then perform the integral over $E_-$, which can easily be done analytically. The results are
\begin{equation}\label{eq:ThermalAv2T}
  \begin{aligned}
    \langle\sigma v\rangle_{i j \to m n}^{T_i, T_j} &= \frac{T_A \int_{s_{\text{min}}}^\infty ds \int_{\sqrt{s}}^\infty dE_+\left[e^{-A_+} - e^{-A_-}\right]p_{ij}\sqrt{s}\sigma(s) }{8 T_i T_j m_i^2 m_j^2 K_2\left(\frac{m_i}{T_i}\right) K_2\left(\frac{m_j}{T_j}\right)},\\
    \langle\sigma v E_+\rangle_{i j \to m n}^{T_i, T_j} &= \frac{T_A \int_{s_{\text{min}}}^\infty ds \int_{\sqrt{s}}^\infty dE_+ E_+\left[e^{-A_+} - e^{-A_-}\right]p_{ij}\sqrt{s}\sigma(s) }{8 T_i T_j m_i^2 m_j^2 K_2\left(\frac{m_i}{T_i}\right) K_2\left(\frac{m_j}{T_j}\right)},\\
    \langle\sigma v E_-\rangle_{i j \to m n}^{T_i, T_j} &= \frac{T_A \int_{s_{\text{min}}}^\infty ds \int_{\sqrt{s}}^\infty dE_+\left[B_+ e^{-A_+} - B_- e^{-A_-}\right]p_{ij}\sqrt{s}\sigma(s) }{8 T_i T_j m_i^2 m_j^2 K_2\left(\frac{m_i}{T_i}\right) K_2\left(\frac{m_j}{T_j}\right)},\\
    \langle\sigma v E_+'\rangle_{i j \to m n}^{T_i, T_j} &= \frac{T_A \int_{s_{\text{min}}}^\infty ds \int_{\sqrt{s}}^\infty dE_+ E_+\left[e^{-A_+} - e^{-A_-}\right]p_{ij}\sqrt{s}\sigma(s) }{8 T_i T_j m_i^2 m_j^2 K_2\left(\frac{m_i}{T_i}\right) K_2\left(\frac{m_j}{T_j}\right)},\\
    \langle\sigma v E_-'\rangle_{i j \to m n}^{T_i, T_j} &= \frac{T_A \int_{s_{\text{min}}}^\infty ds \int_{\sqrt{s}}^\infty dE_+ \frac{(m_m^2 - m_n^2)}{s}E_+\left[e^{-A_+} - e^{-A_-}\right]p_{ij}\sqrt{s}\sigma(s) }{8 T_i T_j m_i^2 m_j^2 K_2\left(\frac{m_i}{T_i}\right) K_2\left(\frac{m_j}{T_j}\right)}\\
                                 & +\frac{T_A \int_{s_{\text{min}}}^\infty ds \int_{\sqrt{s}}^\infty dE_+\left[C_+ e^{-A_+} - C_- e^{-A_-}\right]p_{mn}\sqrt{s}\sigma^t(s) }{8 T_i T_j m_i^2 m_j^2 K_2\left(\frac{m_i}{T_i}\right) K_2\left(\frac{m_j}{T_j}\right)},
  \end{aligned}
\end{equation}
where the indices on $v_{ij}$ are now implicit, $s_{\text{min}}=\text{max}\{(m_i + m_j)^2, (m_m + m_n)^2\}$,
\begin{equation}\label{eq:ABC}
    A_\pm = \frac{E_+}{T_S} + \frac{E_-^{\text{min/max}}}{T_A},\qquad
    B_\pm = T_A + E_-^{\text{min/max}},\qquad
    C_\pm = T_A \mp 2p_{ij}\sqrt{\frac{E_+^2 - s}{s}},
\end{equation}
and
\begin{equation}\label{eq:FAM}
  \sigma^t = \int_{t_1}^{t_0} \frac{d\sigma}{dt}\cos\theta_m dt = \int_{t_1}^{t_0}\frac{d\sigma}{dt}\left[1 + \frac{t - t_0}{2p_{ij}p_{mn}} \right] dt,
\end{equation}
where $t$ is the standard Mandelstam variable and
\begin{equation}\label{eq:t0t1}
  t_0(t_1) = \left[\frac{m_i^2 - m_j^2 - m_m^2 + m_n^2}{2\sqrt{s}}\right]^2 - \left(p_{ij} \mp p_{mn}\right)^2.
\end{equation}
With these results, the computation of any exchange rate is trivial.\footnote{It is of course understood that $\langle\sigma v E_-'\rangle_{i j \to m n}^{T_i, T_j}$ is a shorthand notation for
\begin{equation*}
 \left\langle\sigma v \frac{E_+}{s}\left(m_m^2 - m_n^2\right)\right\rangle_{i j \to m n}^{T_i, T_j} + \left\langle\sigma^t v \frac{\left(E_-s - E_+(m_i^2 - m_j^2)\right)}{s}\frac{p_{mn}}{p_{ij}}\right\rangle_{i j \to m n}^{T_i, T_j}.
\end{equation*}}
In a given $2 \to 2$ process, it suffices to use Eqs.~\eqref{eq:ThermalAv2T} and Eq.~\eqref{eq:Energyvariables} to know exactly the rate at which a specific incoming or outgoing particle gains or loses energy. Knowing the rate at which particles of a given type either gain or lose energy is then trivial. In the limit of $T_i = T_j = T$, the results of Eq.~\eqref{eq:ThermalAv2T} reduce to
\begin{equation}\label{eq:ThermalAv1T}
  \begin{aligned}
    \langle\sigma v\rangle_{i j \to m n}^{T, T} &= \frac{\int_{s_{\text{min}}}^\infty \frac{1}{\sqrt{s}} (s - (m_i + m_j)^2) (s - (m_i - m_j)^2) \sigma(s) K_1\left(\frac{\sqrt{s}}{T}\right) ds}{8 T m_i^2 m_j^2 K_2\left(\frac{m_i}{T}\right) K_2\left(\frac{m_j}{T}\right)},\\
    \langle\sigma v E_+\rangle_{i j \to m n}^{T, T} &= \frac{\int_{s_{\text{min}}}^\infty (s - (m_i + m_j)^2) (s - (m_i - m_j)^2) \sigma(s) K_2\left(\frac{\sqrt{s}}{T}\right) ds}{8 T m_i^2 m_j^2 K_2\left(\frac{m_i}{T}\right) K_2\left(\frac{m_j}{T}\right)}, \\
    \langle\sigma v E_-\rangle_{i j \to m n}^{T, T} &= \frac{\int_{s_{\text{min}}}^\infty \frac{(m_i^2 - m_j^2)}{s}(s - (m_i + m_j)^2) (s - (m_i - m_j)^2) \sigma(s) K_2\left(\frac{\sqrt{s}}{T}\right) ds}{8 T m_i^2 m_j^2 K_2\left(\frac{m_i}{T}\right) K_2\left(\frac{m_j}{T}\right)},\\
    \langle\sigma v E_+'\rangle_{i j \to m n}^{T, T} &= \frac{\int_{s_{\text{min}}}^\infty (s - (m_i + m_j)^2) (s - (m_i - m_j)^2) \sigma(s) K_2\left(\frac{\sqrt{s}}{T}\right) ds}{8 T m_i^2 m_j^2 K_2\left(\frac{m_i}{T}\right) K_2\left(\frac{m_j}{T}\right)},\\
    \langle\sigma v E_-'\rangle_{i j \to m n}^{T, T} &= \frac{\int_{s_{\text{min}}}^\infty \frac{(m_m^2 - m_n^2)}{s}(s - (m_i + m_j)^2) (s - (m_i - m_j)^2) \sigma(s) K_2\left(\frac{\sqrt{s}}{T}\right) ds}{8 T m_i^2 m_j^2 K_2\left(\frac{m_i}{T}\right) K_2\left(\frac{m_j}{T}\right)}.
  \end{aligned}
\end{equation}

\section{Decay asymmetry}\label{Sec:DecayAsymmetry}
In this section, we compute the asymmetry between the decay of $N_2$ to baryons and antibaryons. Similar albeit partial results can be found in Refs.~\cite{Cui:2013bta, Arcadi:2015ffa, Beauchesne:2017jou}. In an effort to make the result applicable to more generic models, we do the computation for the toy Lagrangian
\begin{equation}\label{eq:ToyLagrangian}
  \begin{aligned}
    \mathcal{L} = & -\frac{1}{2}m_{N_1}\bar{N}_1 N_1 -\frac{1}{2}m_{N_2}\bar{N}_2 N_2 - m_{\phi}^2|\phi|^2\\
                  & + \lambda_3 \phi^\dagger \bar{d}' P_L u^c + \text{h.c.}\\
                  & + \phi^\dagger \bar{N}_a \left(L_a P_L + R_a P_R \right)d + \text{h.c.}
  \end{aligned}
\end{equation}
The down-type quarks $d$ and $d'$ are assumed distinct. Two assumptions are made: $d$ and $d'$ are massless and $\phi$ is heavy. These assumptions are made to simplify the calculations, but are not crucial to the mechanism. The mass of $u$ is labelled $m_u$ and is not neglected. The fermion $N_2$ is assumed heavier than both $N_1$ and $u$.

\begin{figure}[t!]
  \centering
   \captionsetup{justification=centering}
    \begin{subfigure}{0.33\textwidth}
    \centering
    \includegraphics[width=\textwidth]{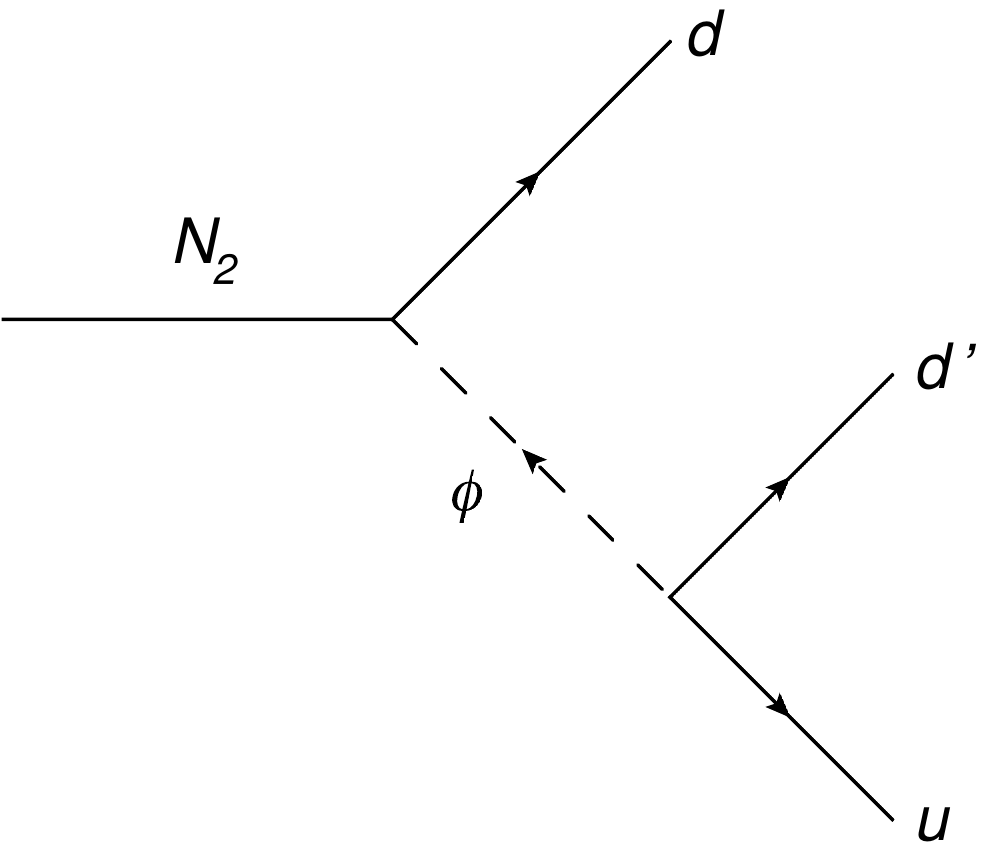}
    \caption{}
    \label{fig:D1}
  \end{subfigure}
   \hspace{0.5cm}
  \begin{subfigure}{0.45\textwidth}
    \centering
    \includegraphics[width=\textwidth]{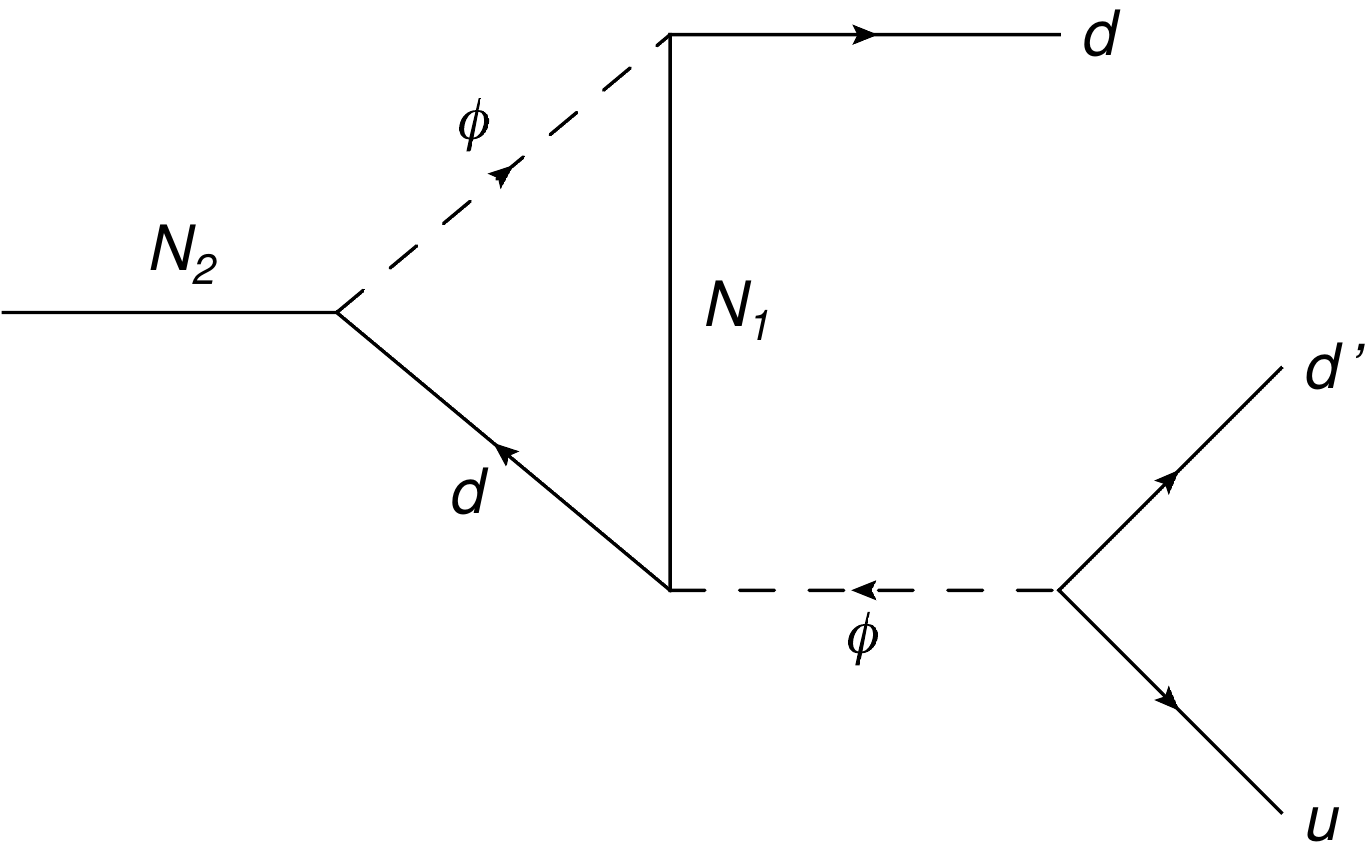}
    \caption{}
    \label{fig:D2}
  \end{subfigure}
\caption{(a) Tree-level decay of $N_2$ to three quarks. (b) NLO correction to that decay.}\label{fig:FMasymmetry}
\end{figure}

The leading order diagram for the decay of $N_2$ to three quarks is shown in Fig.~\ref{fig:D1} and the next-to-leading order diagram in Fig.~\ref{fig:D2}.
The interference term of these diagrams leads to an asymmetry in the decay to baryons and antibaryons $\Delta \Gamma^{N_2}_{udd'{}} \equiv \Gamma^{N_2}_{udd'{}} - \Gamma^{N_2}_{\bar{u}\bar{d}\bar{d}'{}}$. First, define the variables
\begin{equation}\label{eq:AandB}
  A = \frac{m_u}{m_{N_2}}, \;\;\;\; B = \frac{m_{N_1}}{m_{N_2}}, \;\;\;\; u = 1 + A^2 - x - y,
\end{equation}
where $x$ and $y$ are integration variables. Then, also define
\begin{equation}\label{eq:Functionstilde}
  \tilde{f}(x) = \min\left\{1 + A^2 - B^2 - x, \frac{(x - A^2)(1 - x)}{x}\right\}.
\end{equation}
Finally, define
\begin{equation}\label{eq:Gi}
  \begin{aligned}
    G_1(A, B) &= \int_{A^2}^{\min\left\{1, 1 +A^2 - B^2\right\}} dx \int_0^{\tilde{f}(x)}dy (1 - u)(u - A^2)(u - B^2)^2\frac{1}{u}  ,\\
    G_2(A, B) &= \int_{A^2}^{\min\left\{1, 1 +A^2 - B^2\right\}} dx \int_0^{\tilde{f}(x)}dy (1 - u)(u - A^2)(u - B^2)^2\frac{B}{u^2}.
  \end{aligned}
\end{equation}
When $B > A$, these functions are given by
\begin{equation}\label{eq:G1andG2B>A}
  \begin{aligned}
    G_1(A, B) &=  -\frac{A^4 B^6}{3}-3 A^4 B^4+4 A^4 B^4 \ln B+3 A^4 B^2+4 A^4 B^2 \ln B+\frac{A^4}{3}+\frac{A^2 B^8}{6}\\
              & -\frac{4 A^2 B^6}{3}+4 A^2 B^4 \ln B+\frac{4 A^2 B^2}{3}-\frac{A^2}{6}-\frac{B^{10}}{30}+\frac{B^8}{6}-\frac{B^6}{3}+\frac{B^4}{3}-\frac{B^2}{6}+\frac{1}{30},\\
    G_2(A, B) &= 3 A^4 B^5-2 A^4 B^5 \log B-8 A^4 B^3 \ln B-3 A^4 B-2 A^4 B \ln B+\frac{2 A^2 B^7}{3}\\
              & +6 A^2 B^5-8 A^2 B^5 \ln B-6 A^2 B^3-8 A^2 B^3 \ln B-\frac{2 A^2 B}{3}-\frac{B^9}{12}+\frac{2B^7}{3}\\
              & -2 B^5 \ln B-\frac{2 B^3}{3}+\frac{B}{12}.
  \end{aligned}
\end{equation}
When $A > B$, they are instead given by
\begin{equation}\label{eq:G1andG2B<A}
  \begin{aligned}
    G_1(A, B) &= -\frac{A^{10}}{30}+\frac{A^8 B^2}{6}+\frac{A^8}{6}-\frac{A^6 B^4}{3}-\frac{4 A^6 B^2}{3}-\frac{A^6}{3}-3 A^4 B^4 +4 A^4 B^4 \ln A\\
              &  +4 A^4 B^2 \ln A+\frac{A^4}{3}+3 A^2 B^4+4 A^2 B^4 \ln A+\frac{4 A^2 B^2}{3}-\frac{A^2}{6}+\frac{B^4}{3}-\frac{B^2}{6}+\frac{1}{30},\\
    G_2(A, B) &= -\frac{A^8 B}{12}+\frac{2 A^6 B^3}{3}+\frac{2 A^6 B}{3}+3 A^4 B^5-2 A^4 B^5 \ln A+6 A^4 B^3-8 A^4 B^3 \ln A\\
              &  -2 A^4 B \ln A-8 A^2 B^5 \ln A-6 A^2 B^3-8 A^2 B^3 \ln A-\frac{2 A^2 B}{3}-2B^5 \ln A-3 B^5\\
              &  -\frac{2 B^3}{3}+\frac{B}{12}.
  \end{aligned}
\end{equation}
With all this, we get the asymmetry
\begin{equation}\label{eq:AsymmetryTM}
  \Delta \Gamma^{N_2}_{udd'{}} = \frac{3|\lambda_3|^2}{2048\pi^4}\frac{m_{N_2}^7}{m_\phi^6}\left[2\text{Im}\left\{L_2^* L_1 R_2^* R_1\right\} G_1(A, B) + \text{Im}\left\{(L_2^* L_1)^2 + (R_2^* R_1)^2 \right\} G_2(A, B) \right].
\end{equation}

\section{Scattering asymmetries}\label{Sec:ScatteringAsymmetry}
In a similar fashion to decays, scattering processes of the form $N_i \bar{q} \to q q$ and $N_i q \to \bar{q}\bar{q}$ can present an asymmetry in their cross sections. This is due to variations of the diagrams of Fig.~\ref{fig:FMasymmetry}. We maintain the notation and assumptions of Sec.~\ref{Sec:DecayAsymmetry}. There are then two possibilities. First, there is the asymmetry
\begin{equation}\label{eq:ScatteringAsymmetry1}
  \begin{aligned}
     \Delta \sigma_{N_i \bar{d} \to d' u} &\equiv \frac{\sigma_{N_i \bar{d} \to d' u} - \sigma_{N_i d \to \bar{d}' \bar{u}}}{2}\\
                                          &= \sum_j\frac{|\lambda_3|^2}{512\pi^2 m_\phi^6}\frac{(s - m_u^2)^2(s - m_{N_j}^2)^2}{s^3}\theta(s - m_{N_j}^2)\\
                                          &\hspace{1cm}\times\left[2\text{Im}\left\{L_i^* L_j R_i^* R_j\right\} s + \text{Im}\left\{(L_i^* L_j)^2 + (R_i^* R_j)^2 \right\} m_{N_i} m_{N_j}\right].
  \end{aligned}
\end{equation}
Obviously, only terms where $i \neq j$ contribute. In practice, Eq.~\eqref{eq:ScatteringAsymmetry1} means that $N_2$ can always present an asymmetry in this scattering, but $N_1$ can only for a sufficiently large center-of-mass energy. Second, there is also the asymmetry
\begin{equation}\label{eq:ScatteringAsymmetry2}
  \begin{aligned}
     \Delta \sigma_{N_i \bar{d}' \to d u} &\equiv \frac{\sigma_{N_i \bar{d}' \to d u} - \sigma_{N_i d' \to \bar{d} \bar{u}}}{2}\\
                                          &= \sum_j\frac{|\lambda_3|^2}{512\pi^2 m_\phi^6}\frac{m_{N_i}^8}{(s - m_{N_i}^2)^2}\theta\left(\frac{m_{N_i}^2 m_u^2}{m_{N_j}^2} - s\right)\\
                                          &  \times\left[2\text{Im}\left\{L_i^* L_j R_i^* R_j\right\}G_3(A, B, C) + \text{Im}\left\{(L_i^* L_j)^2 + (R_i^* R_j)^2 \right\}G_4(A, B, C)\right],
  \end{aligned}
\end{equation}
where
\begin{equation}\label{eq:AandBanC}
  A = \frac{m_u}{m_{N_i}}, \;\;\;\; B = \frac{m_{N_j}}{m_{N_i}}, \;\;\;\; C = \frac{\sqrt{s}}{m_{N_i}},
\end{equation}
with
\begin{equation}\label{eq:Gi2}
  \begin{aligned}
    G_3(A, B, C) &= \int_{B^2}^{\frac{A^2}{C^2}}dx (x - 1)(x - A^2)(x - B^2)^2\frac{1}{x} ,\\
    G_4(A, B, C) &= \int_{B^2}^{\frac{A^2}{C^2}}dx (x - 1)(x - A^2)(x - B^2)^2\frac{B}{x^2}.
  \end{aligned}
\end{equation}
More concretely, these functions are given by
\begin{equation}\label{eq:G3andG4}
  \begin{aligned}
    G_3(A, B, C)& = \frac{A^8}{4 C^8}-\frac{A^8}{3 C^6}-\frac{2 A^6 B^2}{3 C^6}+\frac{A^6 B^2}{C^4}-\frac{A^6}{3 C^6}+\frac{A^6}{2 C^4}+\frac{A^4 B^4}{2 C^4}-\frac{A^4 B^4}{C^2}+\frac{A^4 B^2}{C^4} \\
                &   -\frac{2 A^4B^2}{C^2}+\frac{A^2 B^6}{3}-\frac{A^2 B^4}{C^2}-2 A^2 B^4 \ln C+\frac{3 A^2 B^4}{2}+2 A^2 B^4 \ln A\\
                &    -2 A^2 B^4 \ln B-\frac{B^8}{12}+\frac{B^6}{3},\\
    G_4(A, B, C)& = \frac{A^6 B}{3 C^6}-\frac{A^6 B}{2 C^4}-\frac{A^4 B^3}{C^4}+\frac{2 A^4 B^3}{C^2}-\frac{A^4 B}{2 C^4}+\frac{A^4 B}{C^2}+\frac{A^2 B^5}{C^2}+2 A^2 B^5 \ln C\\
                &    -\frac{3 A^2 B^5}{2}-2 A^2 B^5 \ln A+2 A^2 B^5 \ln B+\frac{2 A^2 B^3}{C^2}+4 A^2 B^3 \ln C-4 A^2 B^3 \ln A\\
                &    +4 A^2 B^3 \ln B-2 B^5 \ln A-\frac{B^7}{3}-B^5 C^2+2 B^5 \ln C-\frac{3 B^5}{2}+2 B^5 \ln B.
  \end{aligned}
\end{equation}
Because of the kinematics, only $N_2$ can present an asymmetry and only for $N_1$ lighter than $u$ and sufficiently low center-of-mass energy. The channel $N_i \bar{u} \to d d'$ does not present an asymmetry at this order of perturbation.

\section{Evolution equations}\label{Sec:Evolution}
In this section, we present the evolution equations that are used to compute the relic densities in Sec.~\ref{Sec:BNeffR}. The relevant processes are first introduced and some important properties are then discussed. To simplify the treatment, we will work with the Lagrangian
\begin{equation}\label{eq:LagrangianModelC4FermionsEvolution}
  \begin{aligned}
    \mathcal{L} = & - \frac{1}{2} m_{N_1} \bar{N}_1 N_{1} - \frac{1}{2} m_{N_2} \bar{N}_2 N_{2}  - m_{\phi^A}^2|\phi^A|^2 - m_{\phi^B}^2|\phi^B|^2\\
                  & + \lambda_3\left[(\phi^A)^\dagger \bar{d}'^A P_L (u^A)^c + (\phi^B)^\dagger \bar{d}'^B P_L (u^B)^c \right] + \text{h.c.}\\
                  & + \lambda^*_{4i}\left[(\phi^A)^\dagger \bar{N}_j P_R d^A + (\phi^B)^\dagger \bar{N} P_R d^B\right] + \text{h.c.},
  \end{aligned}
\end{equation}
where $d$ and $d'$ are distinct. This is equivalent to Eqs.~\eqref{eq:LagrangianModelC4Fermions} and \eqref{eq:LagrangianModelC2} with $\lambda_{3ij}$ and $\lambda_{4ij}$ each having only one combination of flavours for which they are non-zero. In this section, decay widths and cross sections are averaged over all incoming degrees of freedom and summed over all outgoing degrees of freedom, including particles and antiparticles when distinct. The only exception are the asymmetries which maintain their definitions. The scalars $\phi^M$ are again assumed heavy and the masses of $d^M$ and ${d'}^M$ are neglected unless stated otherwise.

\subsection{Decay: $N_2 \to N_1 d^M \bar{d}^M$}\label{sSec:N2->N1ddbar}
The decay width is
\begin{equation}\label{eq:GammaN2N1->ddbar}
  \Gamma^{N_2}_{N_1 d^M \bar{d}^M} = \frac{|\lambda_{41}\lambda_{42}|^2}{1024\pi^3}\frac{m_{N_2}^5}{m_{\phi^M}^4}\left[f_1\left(\frac{m_{N_1}}{m_{N_2}}\right) + 2\frac{m_{N_1}}{m_{N_2}}f_2\left(\frac{m_{N_1}}{m_{N_2}}\right)\cos 2\phi_{12} \right],
\end{equation}
with
\begin{equation}\label{eq:f1f2}
  \begin{aligned}
    f_1(x) &= 1 - 8 x^2 +8 x^6 -x^8 - 24 x^4 \ln x,\\
    f_2(x) &= 1 + 9 x^2 - 9 x^4 - x^6 + 12 x^2 (1 + x^2) \ln x,
  \end{aligned}
\end{equation}
and
\begin{equation}\label{eq:cosphi}
  \cos 2\phi_{ij} = \frac{\text{Re}\left\{(\lambda_{4i}^*\lambda_{4j})^2\right\}}{|\lambda_{4i}\lambda_{4j}|^2}.
\end{equation}
The average energy fraction of $N_1$ in the centre-of-mass frame is obtained by computing the expectation value of $m_{23}^2 = (p_d + p_{\bar{d}})^2$, which gives
\begin{equation}\label{eq:m232}
  \langle \frac{m_{23}^2}{m_{N_2}^2} \rangle = \frac{\frac{3}{10}f_3\left(\frac{m_{N_1}}{m_{N_2}}\right) + \frac{m_{N_1}}{m_{N_2}}f_4\left(\frac{m_{N_1}}{m_{N_2}}\right)\cos 2\phi_{12} }{f_1\left(\frac{m_{N_1}}{m_{N_2}}\right) + 2\frac{m_{N_1}}{m_{N_2}}f_2\left(\frac{m_{N_1}}{m_{N_2}}\right)\cos 2\phi_{12} },
\end{equation}
where
\begin{equation}\label{eq:f3f4}
  \begin{aligned}
    f_3(x) &= 1 - 15x^2 - 80x^4 + 80x^6 + 15x^8 - x^{10} -120x^4(1 + x^2)\ln x,\\
    f_4(x) &= 1 + 28 x^2 - 28 x^6 - x^8 + 24 x^2 (1 + 3 x^2 + x^4) \ln x.
  \end{aligned}
\end{equation}
It is then a basic exercise in kinematics to compute the average energy fraction of $N_1$ in the centre-of-mass frame, which gives
\begin{equation}\label{eq:E1onM}
  \langle\frac{E_{N_1}}{m_{N_2}}\rangle = \frac{1}{2}\left(1 + \left(\frac{m_{N_1}}{m_{N_2}}\right)^2 - \langle \frac{m_{23}^2}{m_{N_2}^2}\rangle\right).
\end{equation}

\subsection{Decay: $N_i \to u^M d^M {d'}^M$}\label{sSec:N2->udd}
The decay width is
\begin{equation}\label{eq:DecayNiudd}
  \Gamma^{N_i}_{u^M d^M {d'}^M} = \frac{|\lambda_3 \lambda_{4i}|^2}{512\pi^3}\frac{m_{N_i}^5}{m_{\phi^M}^4}f_1\left(\frac{m_{u^M}}{m_{N_i}}\right).
\end{equation}
In some regions of parameter space, the decay of $N_1$ to three quarks is forbidden. The particle $N_1$ is then forced to go through a four-body decay where $u^M$ is off-shell. This decay width is computed numerically, including the width of $u^M$ and the mass of $d^M$. The numerical result is also used for $m_{N_1}$ not too far removed from $m_{u^M}$, as the narrow width approximation is not necessarily a good approximation when these two masses are close.

\subsection{Scattering: $N_i N_j \to \bar{d}^M d^M$}\label{sSec:NiNj->dbard}
The cross section is
\begin{equation}\label{eq:CSNinj->dbard}
    \sigma_{N_i N_j \to \bar{d}^M d^M} =  \frac{|\lambda_{4i}\lambda_{4j}|^2}{64\pi m_{\phi^M}^4}\frac{\left(2s^2 - (m_{N_i}^2 + m_{N_j}^2)s - (m_{N_i}^2 - m_{N_j}^2)^2 - 6 m_{N_i} m_{N_j}s\cos 2\phi_{ij} \right)}{((s - (m_{N_i} + m_{N_j})^2)(s - (m_{N_i} - m_{N_j})^2))^{1/2}},
\end{equation}
and
\begin{equation}\label{eq:CStNinj->dbard}
  \sigma_{N_i N_j \to \bar{d}^M d^M}^t = 0.
\end{equation}

\subsection{Scattering: $N_i d^M \to N_j d^M$}\label{sSec:Nid->Njd}
The cross section is
\begin{equation}\label{eq:CSNid->Njd}
  \begin{aligned}
    \sigma_{N_i d^M \to N_j d^M} &= \frac{|\lambda_{4i}\lambda_{4j}|^2}{384\pi m_{\phi^M}^4}\frac{(s - m_{N_j}^2)^2}{s^3}\left(8s^2 + (m_{N_i}^2 + m_{N_j}^2)s + 2 m_{N_i}^2m_{N_j}^2\right.\\
    &\left.\hspace{4.5cm} + 6m_{N_i} m_{N_j}s\cos 2\phi_{ij} \right),
  \end{aligned}
\end{equation}
and
\begin{equation}\label{eq:CStNid->Njd}
  \sigma_{N_i d^M \to N_j d^M}^t = \frac{|\lambda_{4i}\lambda_{4j}|^2}{384\pi m_{\phi^M}^4}\frac{(s - m_{N_j}^2)^2}{s^3}\left(s^2 - m_{N_i}^2m_{N_j}^2 - 2m_{N_i} m_{N_j}s\cos 2\phi_{ij} \right).
\end{equation}

\subsection{Scattering: $N_i d^M \to \bar{d}'{}^M \bar{u}^M$}\label{sSec:NidM->dpMbaruMbar}
The cross section is
\begin{equation}\label{eq:CSNidM->dpMbaruMbar}
  \sigma_{N_i d^M \to \bar{d}'{}^M \bar{u}^M} = \frac{|\lambda_{3}\lambda_{4i}|^2}{32\pi m_{\phi^M}^4}\frac{(s - m_{u^M}^2)^2}{s},
\end{equation}
and
\begin{equation}\label{eq:CStNidM->dpMbaruMbar}
  \sigma_{N_i d^M \to \bar{d}'{}^M \bar{u}^M}^t = 0.
\end{equation}

\subsection{Scattering: $N_i d'{}^M \to \bar{d}^M \bar{u}^M$}\label{sSec:NidpM->dMbaruMbar}
The cross section is
\begin{equation}\label{eq:CSNidpM->dMbaruMbar}
  \sigma_{N_i d'{}^M \to \bar{d}^M \bar{u}^M} = \frac{|\lambda_{3}\lambda_{4i}|^2}{192\pi m_{\phi^M}^4}\frac{(s - m_{u^M}^2)^2(2s^2 + (m_{N_i}^2 + m_{u^M}^2)s + 2m_{N_i}^2 m_{u^M}^2)}{s^3},
\end{equation}
and
\begin{equation}\label{eq:CStNidpM->dMbaruMbar}
  \sigma_{N_i d'{}^M \to \bar{d}^M \bar{u}^M}^t = \frac{|\lambda_{3}\lambda_{4i}|^2}{192\pi m_{\phi^M}^4}\frac{(s - m_{u^M}^2)^2(s^2 - m_{N_i}^2 m_{u^M}^2)}{s^3}.
\end{equation}

\subsection{Scattering: $N_i u^M \to \bar{d}^M \bar{d}'{}^M$}\label{sSec:NiuM->dMbardpMbar}
The cross section is
\begin{equation}\label{eq:CSNiuM->dMbardpMbar}
  \sigma_{N_i u^M \to \bar{d}^M \bar{d}'{}^M} = \frac{|\lambda_{3}\lambda_{4i}|^2}{192\pi m_{\phi^M}^4}\frac{(2s^2 - (m_{N_i}^2 + m_{u^M}^2)s - (m_{N_i}^2 - m_{u^M}^2)^2)}{((s - (m_{N_i} + m_{u^M})^2)(s - (m_{N_i} - m_{u^M})^2))^{1/2}},
\end{equation}
and
\begin{equation}\label{eq:CStNiuM->dMbardpMbar}
  \sigma_{N_i u^M \to \bar{d}^M \bar{d}'{}^M}^t = -\frac{|\lambda_{3}\lambda_{4i}|^2}{192\pi m_{\phi^M}^4}s.
\end{equation}

\subsection{Scattering: $f^A {f'}^B \to f^A {f'}^B$}\label{sSec:fAfBp->fAfBp}
The cross section between an $A$ sector fermion and a $B$ sector fermion that is not necessarily its partner is
\begin{equation}\label{eq:CSfAfBp->fAfBp}
  \begin{aligned}
  \sigma_{f^A {f'}^B \to f^A {f'}^B} &= \frac{m_{f^A}^2 m_{{f'}^B}^2}{48\pi((v^A)^2 + (v^B)^2)^2 m_h^4}\frac{1}{(s - (m_{f^A} + m_{{f'}^B})^2)(s - (m_{f^A} -m_{{f'}^B})^2)s^3}\\ 
  &\hspace{-0.3cm}\times\left(s^6 - 9(m_{f^A}^2 - m_{{f'}^B}^2)^2 s^4 + 16(m_{f^A}^6 - 3m_{f^A}^4 m_{{f'}^B}^2 - 3m_{f^A}^2 m_{{f'}^B}^4 + m_{{f'}^B}^6)s^3\right.\\
  &\hspace{0.3cm}\left.- 9(m_{f^A}^2 - m_{{f'}^B}^2)^4 s^2 + (m_{f^A}^2 - m_{{f'}^B}^2)^6\right),
  \end{aligned}
\end{equation}
and
\begin{equation}\label{eq:CStfAfBp->fAfBp}
  \begin{aligned}
  \sigma_{f^A {f'}^B \to f^A {f'}^B}^t = &-\frac{m_{f^A}^2 m_{{f'}^B}^2}{96\pi((v^A)^2 + (v^B)^2)^2 m_h^4}\frac{(s^2 - (m_{f^A} + m_{{f'}^B})^4)(s^2 - (m_{f^A} - m_{{f'}^B})^4)}{s^3}.
  \end{aligned}
\end{equation}

\subsection{Scattering: $f^A \bar{f}^A \to {f'}^B \bar{f}'{}^B$}\label{sSec:fAfAbar->fBpfBpbar}
The cross section is
\begin{equation}\label{eq:CSfAfAbar->fBpfBpbar}
  \sigma_{f^A \bar{f}^A \to {f'}^B \bar{f}'{}^B} = \frac{m_{f^A}^2 m_{{f'}^B}^2}{32\pi((v^A)^2 + (v^B)^2)^2 m_h^4}\frac{N^c_{{f'}^B}}{N^c_{{f}^A}}\frac{(s - 4m_{{f'}^B}^2)^{3/2}}{(s - 4m_{f^A}^2)^{1/2}},
\end{equation}
where $N^c_p$ is the number of colours of particle $p$ and
\begin{equation}\label{eq:CStfAfAbar->fBpfBpbar}
  \begin{aligned}
  \sigma_{f^A \bar{f}^A \to {f'}^B \bar{f}'{}^B}^t = 0.
  \end{aligned}
\end{equation}

\subsection{Combined evolution equations}\label{sSec:Combined}
We now combine the different results together. The number density of particle $p$ is labelled as $n_p$, its energy density as $\rho_p$ and its pressure as $P_p$.\footnote{When considering their densities, quarks and antiquarks are treated as a single particle with twice as many degrees of freedom, i.e. they have $g_p = 12$.} The differences between the baryon number densities are labelled as $\Delta B_M$. The equations are
\begin{align*}
  \frac{d n_{N_1}}{dt} = &
      - 3Hn_{N_1}\\
    & +  \sum_M^{A, B}\langle\Gamma^{N_2}_{N_1 d^M \bar{d}^M}\rangle^{T_{N_2}} n_{N_2}\\
    & -  \sum_M^{A, B}\left[\langle\Gamma^{N_1}_{u^M d^M d'{}^M}\rangle^{T_{N_1}} n_{N_1} - \langle\Gamma^{N_1}_{u^M d^M d'{}^M}\rangle^{T_M} n^{\text{eq}}_{N_1}(T_M)\right]\\
    & -  \sum_M^{A, B}\left[\langle\sigma v\rangle_{N_1 N_1 \to \bar{d}^M d^M}^{T_{N_1}, T_{N_1}} n_{N_1}^2 -  \langle\sigma v\rangle_{N_1 N_1 \to \bar{d}^M d^M}^{T_M, T_M} (n^{\text{eq}}_{N_1}(T_M))^2 \right]\\
    & -  \sum_M^{A, B}\left[\langle\sigma v\rangle_{N_1 N_2 \to \bar{d}^M d^M}^{T_{N_1}, T_{N_2}} n_{N_1} n_{N_2} -  \langle\sigma v\rangle_{N_1 N_2 \to \bar{d}^M d^M}^{T_M, T_M} n^{\text{eq}}_{N_1}(T_M) n^{\text{eq}}_{N_2}(T_M)\right]\\
    & -  \sum_M^{A, B}\left[\langle\sigma v\rangle_{N_1 d^M \to N_2 d^M}^{T_{N_1}, T_M}n_{N_1}- \langle\sigma v\rangle_{N_2 d^M \to N_1 d^M}^{T_{N_2}, T_M}n_{N_2}\right]n^{\text{eq}}_{d^M}(T_M) \\
    & -  \sum_M^{A, B}\left[\langle\sigma v\rangle_{N_1 d^M \to \bar{d}'{}^M \bar{u}^M}^{T_{N_1}, T_M}n_{N_1} - \langle\sigma v\rangle_{N_1 d^M \to \bar{d}'{}^M \bar{u}^M}^{T_M, T_M}n^{\text{eq}}_{N_1}(T_M) \right]n^{\text{eq}}_{d^M}(T_M)\\
    & -  \sum_M^{A, B}\left[\langle\sigma v\rangle_{N_1 d'{}^M \to \bar{d}^M \bar{u}^M}^{T_{N_1}, T_M}n_{N_1} - \langle\sigma v\rangle_{N_1 d'{}^M \to \bar{d}^M \bar{u}^M}^{T_M, T_M}n^{\text{eq}}_{N_1}(T_M) \right]n^{\text{eq}}_{d'{}^M}(T_M)\\
    & -  \sum_M^{A, B}\left[\langle\sigma v\rangle_{N_1 u^M \to \bar{d}^M \bar{d}'{}^M}^{T_{N_1}, T_M}n_{N_1} - \langle\sigma v\rangle_{N_1 u^M \to \bar{d}^M \bar{d}'{}^M}^{T_M, T_M}n^{\text{eq}}_{N_1}(T_M) \right]n^{\text{eq}}_{u^M}(T_M),
\end{align*}

\begin{align*}
  \frac{d n_{N_2}}{dt} = &
      - 3Hn_{N_2}\\
    & -  \sum_M^{A, B}\langle\Gamma^{N_2}_{N_1 d^M \bar{d}^M}\rangle^{T_{N_2}} n_{N_2}\\
    & -  \sum_M^{A, B}\left[\langle\Gamma^{N_2}_{u^M d^M d'{}^M}\rangle^{T_{N_2}} n_{N_2} - \langle\Gamma^{N_2}_{u^M d^M d'{}^M}\rangle^{T_M} n^{\text{eq}}_{N_2}(T_M)\right]\\
    & -  \sum_M^{A, B}\left[\langle\sigma v\rangle_{N_2 N_2 \to \bar{d}^M d^M}^{T_{N_2}, T_{N_2}} n_{N_2}^2 -  \langle\sigma v\rangle_{N_2 N_2 \to \bar{d}^M d^M}^{T_M, T_M} (n^{\text{eq}}_{N_2}(T_M))^2 \right]\\
    & -  \sum_M^{A, B}\left[\langle\sigma v\rangle_{N_1 N_2 \to \bar{d}^M d^M}^{T_{N_1}, T_{N_2}} n_{N_1} n_{N_2} -  \langle\sigma v\rangle_{N_1 N_2 \to \bar{d}^M d^M}^{T_M, T_M} n^{\text{eq}}_{N_1}(T_M) n^{\text{eq}}_{N_2}(T_M)\right]\\
    & -  \sum_M^{A, B}\left[\langle\sigma v\rangle_{N_2 d^M \to N_1 d^M}^{T_{N_2}, T_M}n_{N_2} - \langle\sigma v\rangle_{N_1 d^M \to N_2 d^M}^{T_{N_1}, T_M}n_{N_1}\right]n^{\text{eq}}_{d^M}(T_M)\\
    & -  \sum_M^{A, B}\left[\langle\sigma v\rangle_{N_2 d^M \to \bar{d}'{}^M \bar{u}^M}^{T_{N_2}, T_M}n_{N_2} - \langle\sigma v\rangle_{N_2 d^M \to \bar{d}'{}^M \bar{u}^M}^{T_M, T_M}n^{\text{eq}}_{N_2}(T_M) \right]n^{\text{eq}}_{d^M}(T_M)\\
    & -  \sum_M^{A, B}\left[\langle\sigma v\rangle_{N_2 d'{}^M \to \bar{d}^M \bar{u}^M}^{T_{N_2}, T_M}n_{N_2} - \langle\sigma v\rangle_{N_2 d'{}^M \to \bar{d}^M \bar{u}^M}^{T_M, T_M}n^{\text{eq}}_{N_2}(T_M) \right]n^{\text{eq}}_{d'{}^M}(T_M)\\
    & -  \sum_M^{A, B}\left[\langle\sigma v\rangle_{N_2 u^M \to \bar{d}^M \bar{d}'{}^M}^{T_{N_2}, T_M}n_{N_2} - \langle\sigma v\rangle_{N_2 u^M \to \bar{d}^M \bar{d}'{}^M}^{T_M, T_M}n^{\text{eq}}_{N_2}(T_M) \right]n^{\text{eq}}_{u^M}(T_M),
\end{align*}

\begin{align*}
  \frac{d \rho_{N_1}}{dt} & =
      - 3H(\rho_{N_1} + P_{N_1})\\
    & +  \sum_M^{A, B}\Gamma^{N_2}_{N_1 d^M \bar{d}^M} m_{N_2} n_{N_2}\langle\frac{E_{N_1}}{m_{N_2}}\rangle\\
    & -  \sum_M^{A, B}\left[\Gamma^{N_1}_{u^M d^M d'{}^M} m_{N_1} n_{N_1} - \Gamma^{N_1}_{u^M d^M d'{}^M} m_{N_1} n^{\text{eq}}_{N_1}(T_M)\right]\\
    & -  \frac{1}{2}\sum_M^{A, B}\left[\langle\sigma vE_+\rangle_{N_1 N_1 \to \bar{d}^M d^M }^{T_{N_1}, T_{N_1}} n_{N_1}^2 -  \langle\sigma vE_+\rangle_{N_1 N_1 \to \bar{d}^M d^M}^{T_M, T_M} (n^{\text{eq}}_{N_1}(T_M))^2 \right]\\
    & -  \sum_M^{A, B}\left[\langle\sigma v E_{N_1}\rangle_{N_1 N_2 \to \bar{d}^M d^M}^{T_{N_1}, T_{N_2}} n_{N_1} n_{N_2} -  \langle\sigma v E_{N_1}\rangle_{N_1 N_2 \to \bar{d}^M d^M}^{T_M, T_M} n^{\text{eq}}_{N_1}(T_M) n^{\text{eq}}_{N_2}(T_M)\right]\\
    & + \sum_M^{A, B}\langle\sigma v (E_{N_1}^{\text{out}} - E_{N_1}^{\text{in}})\rangle_{N_1 d^M \to N_1 d^M}^{T_{N_1}, T_M}n_{N_1}n^{\text{eq}}_{d^M}(T_M)\\
    & - \sum_M^{A, B}\left[\langle\sigma v E_{N_1}\rangle_{N_1 d^M \to N_2 d^M}^{T_{N_1}, T_M}n_{N_1} - \langle\sigma v E_{N_1}\rangle_{N_2 d^M \to N_1 d^M}^{T_{N_2}, T_M}n_{N_2}\right]n^{\text{eq}}_{d^M}(T_M)\\
    & -  \sum_M^{A, B}\left[\langle\sigma vE_{N_1}\rangle_{N_1 d^M \to \bar{d}'{}^M \bar{u}^M}^{T_{N_1}, T_M}n_{N_1} - \langle\sigma v E_{N_1}\rangle_{N_1 d^M \to \bar{d}'{}^M \bar{u}^M}^{T_M, T_M}n^{\text{eq}}_{N_1}(T_M) \right]n^{\text{eq}}_{d^M}(T_M)\\
    & -  \sum_M^{A, B}\left[\langle\sigma vE_{N_1}\rangle_{N_1 d'{}^M \to \bar{d}^M \bar{u}^M}^{T_{N_1}, T_M}n_{N_1} - \langle\sigma v E_{N_1}\rangle_{N_1 d'{}^M \to \bar{d}^M \bar{u}^M}^{T_M, T_M}n^{\text{eq}}_{N_1}(T_M) \right]n^{\text{eq}}_{d'{}^M}(T_M)\\
    & -  \sum_M^{A, B}\left[\langle\sigma vE_{N_1}\rangle_{N_1 u^M \to \bar{d}^M \bar{d}'{}^M}^{T_{N_1}, T_M}n_{N_1} - \langle\sigma v E_{N_1}\rangle_{N_1 u^M \to \bar{d}^M \bar{d}'{}^M}^{T_M, T_M}n^{\text{eq}}_{N_1}(T_M) \right]n^{\text{eq}}_{u^M}(T_M),
\end{align*}

\begin{align*}
  \frac{d \rho_{N_2}}{dt} &=
      - 3H(\rho_{N_2} + P_{N_2})\\
    & -  \sum_M^{A, B}\Gamma^{N_2}_{N_1 d^M \bar{d}^M} m_{N_2} n_{N_2}\\
    & -  \sum_M^{A, B}\left[\Gamma^{N_2}_{u^M d^M d'{}^M} m_{N_2} n_{N_2} - \Gamma^{N_2}_{u^M d^M d'{}^M} m_{N_2} n^{\text{eq}}_{N_2}(T_M)\right]\\
    & -  \frac{1}{2}\sum_M^{A, B}\left[\langle\sigma v E_+\rangle_{N_2 N_2 \to \bar{d}^M d^M}^{T_{N_2}, T_{N_2}} n_{N_2}^2 -  \langle\sigma v E_+\rangle_{N_2 N_2 \to \bar{d}^M d^M}^{T_M, T_M} (n^{\text{eq}}_{N_2}(T_M))^2 \right]\\
    & -  \sum_M^{A, B}\left[\langle\sigma v E_{N_2}\rangle_{N_1 N_2 \to \bar{d}^M d^M}^{T_{N_1}, T_{N_2}} n_{N_1} n_{N_2} -  \langle\sigma v E_{N_2}\rangle_{N_1 N_2 \to \bar{d}^M d^M}^{T_M, T_M} n^{\text{eq}}_{N_1}(T_M) n^{\text{eq}}_{N_2}(T_M)\right]\\
    & + \sum_M^{A, B}\langle\sigma v (E_{N_2}^{\text{out}} - E_{N_2}^{\text{in}})\rangle_{N_2 d^M \to N_2 d^M}^{T_{N_2}, T_M}n_{N_2}n^{\text{eq}}_{d^M}(T_M)\\
    & + \sum_M^{A, B}\left[\langle\sigma v E_{N_2}\rangle_{N_1 d^M \to N_2 d^M}^{T_{N_1}, T_M}n_{N_1} - \langle\sigma v E_{N_2}\rangle_{N_2 d^M \to N_1 d^M}^{T_{N_2}, T_M}n_{N_2}\right]n^{\text{eq}}_{d^M}(T_M)\\
    & -  \sum_M^{A, B}\left[\langle\sigma vE_{N_2}\rangle_{N_2 d^M \to \bar{d}'{}^M \bar{u}^M}^{T_{N_2}, T_M}n_{N_2} - \langle\sigma vE_{N_2}\rangle_{N_2 d^M \to \bar{d}'{}^M \bar{u}^M}^{T_M, T_M}n^{\text{eq}}_{N_2}(T_M) \right]n^{\text{eq}}_{d^M}(T_M)\\
    & -  \sum_M^{A, B}\left[\langle\sigma vE_{N_2}\rangle_{N_2 d'{}^M \to \bar{d}^M \bar{u}^M}^{T_{N_2}, T_M}n_{N_2} - \langle\sigma vE_{N_2}\rangle_{N_2 d'{}^M \to \bar{d}^M \bar{u}^M}^{T_M, T_M}n^{\text{eq}}_{N_2}(T_M) \right]n^{\text{eq}}_{d'{}^M}(T_M)\\
    & -  \sum_M^{A, B}\left[\langle\sigma vE_{N_2}\rangle_{N_2 u^M \to \bar{d}^M \bar{d}'{}^M}^{T_{N_2}, T_M}n_{N_2} - \langle\sigma vE_{N_2}\rangle_{N_2 u^M \to \bar{d}^M \bar{d}'{}^M}^{T_M, T_M}n^{\text{eq}}_{N_2}(T_M) \right]n^{\text{eq}}_{u^M}(T_M),
\end{align*}

\begin{align*}
    \frac{d \rho_A}{dt} &=
        - 4H\left(\frac{1 + \frac{1}{4}T_A \frac{g_{*A}'}{g_{*A}}}{1 + \frac{1}{3}T_A \frac{g_{*A}'}{g_{*A}}}\right)\rho_A\\ 
      &\quad + \Gamma^{N_2}_{N_1 d^A \bar{d}^A} m_{N_2} n_{N_2}\left(1 - \langle\frac{E_{N_1}}{m_{N_2}}\rangle\right)\\
      &\quad + \sum_{i=1}^2\left[\Gamma^{N_i}_{u^A d^A d'{}^A} m_{N_i} n_{N_i} - \Gamma^{N_i}_{u^A d^A d'{}^A} m_{N_i} n^{\text{eq}}_{N_i}(T_A)\right]\\
      &\quad + \frac{1}{2}\sum_{i, j=1}^2\left[\langle\sigma vE_+\rangle_{N_i N_j \to \bar{d}^A d^A}^{T_{N_i}, T_{N_j}} n_{N_i} n_{N_j} -  \langle\sigma vE_+\rangle_{N_i N_j \to \bar{d}^A d^A}^{T_A, T_A} n^{\text{eq}}_{N_i}(T_A) n^{\text{eq}}_{N_j}(T_A)\right]\\
      &\quad + \sum_{i, j =1}^2\langle\sigma v (E_{d^A}^{\text{out}} - E_{d^A}^{\text{in}})\rangle_{N_i d^A \to  N_j d^A}^{T_{N_i}, T_A}n_{N_i}n^{\text{eq}}_{d^A}(T_A)\\
      &\quad + \sum_{i=1}^2\left[\langle\sigma vE_{N_i}\rangle_{N_i d^A \to \bar{d}'{}^A \bar{u}^A}^{T_{N_i}, T_A}n_{N_i} - \langle\sigma vE_{N_i}\rangle_{N_i d^A \to \bar{d}'{}^A \bar{u}^A}^{T_A, T_A}n^{\text{eq}}_{N_i}(T_A) \right]n^{\text{eq}}_{d^A}(T_A)\\
      &\quad + \sum_{i=1}^2\left[\langle\sigma vE_{N_i}\rangle_{N_i d'{}^A \to \bar{d}^A \bar{u}^A}^{T_{N_i}, T_A}n_{N_i} - \langle\sigma vE_{N_i}\rangle_{N_i d'{}^A \to \bar{d}^A \bar{u}^A}^{T_A, T_A}n^{\text{eq}}_{N_i}(T_A) \right]n^{\text{eq}}_{d'{}^A}(T_A)\\
      &\quad + \sum_{i=1}^2\left[\langle\sigma vE_{N_i}\rangle_{N_i u^A \to \bar{d}^A \bar{d}'{}^A}^{T_{N_i}, T_A}n_{N_i} - \langle\sigma vE_{N_i}\rangle_{N_i u^A \to \bar{d}^A \bar{d}'{}^A}^{T_A, T_A}n^{\text{eq}}_{N_i}(T_A) \right]n^{\text{eq}}_{u^A}(T_A)\\
      &\quad + \sum_{f^A, {f'}^B}\langle \sigma v(E_{f^A}^{\text{out}} - E_{f^A}^{\text{in}})\rangle_{f^A {f'}^B \to f^A {f'}^B}^{T_A, T_B} n^{\text{eq}}_{f^A}(T_A)n^{\text{eq}}_{{f'}^B}(T_B)\\
      &\quad - \frac{1}{2}\sum_{f^A, {f'}^B}\left[\langle\sigma v E_+\rangle_{f^A \bar{f}^A \to {f'}^B \bar{f}'{}^B}^{T_A, T_A}(n^{\text{eq}}_{f^A}(T_A))^2  - \langle\sigma v E_+\rangle_{f^A \bar{f}^A \to {f'}^B \bar{f}'{}^B}^{T_B, T_B}(n^{\text{eq}}_{f^A}(T_B))^2 \right],
\end{align*}

\begin{align*}
    \frac{d \rho_B}{dt} &=
             - 4H\left(\frac{1 + \frac{1}{4}T_B \frac{g_{*B}'}{g_{*B}}}{1 + \frac{1}{3}T_B \frac{g_{*B}'}{g_{*B}}}\right)\rho_B\\ 
      &\quad + \Gamma^{N_2}_{N_1 d^B \bar{d}^B} m_{N_2} n_{N_2}\left(1 - \langle\frac{E_{N_1}}{m_{N_2}}\rangle\right)\\
      &\quad + \sum_{i=1}^2\left[\Gamma^{N_i}_{u^B d^B d'{}^B} m_{N_i} n_{N_i} - \Gamma^{N_i}_{u^B d^B d'{}^B} m_{N_i} n^{\text{eq}}_{N_i}(T_B)\right]\\
      &\quad + \frac{1}{2}\sum_{i, j=1}^{2}\left[\langle\sigma vE_+\rangle_{N_i N_j \to \bar{d}^B d^B}^{T_{N_i}, T_{N_j}} n_{N_i} n_{N_j} -  \langle\sigma v E_+\rangle_{N_i N_j \to \bar{d}^B d^B}^{T_B, T_B} n^{\text{eq}}_{N_i}(T_B) n^{\text{eq}}_{N_j}(T_B)\right]\\
      &\quad + \sum_{i, j=1}^2\langle\sigma v (E_{d^B}^{\text{out}} - E_{d^B}^{\text{in}})\rangle_{N_i d^B \to N_j d^B}^{T_{N_i}, T_B}n_{N_i}n^{\text{eq}}_{d^B}(T_B)\\
      &\quad + \sum_{i=1}^2\left[\langle\sigma vE_{N_i}\rangle_{N_i d^B \to \bar{d}^B \bar{u}^B}^{T_{N_i}, T_B}n_{N_i} - \langle\sigma vE_{N_i}\rangle_{N_i d^B \to \bar{d}^B \bar{u}^B}^{T_B, T_B}n^{\text{eq}}_{u^B}(T_B) \right]n^{\text{eq}}_{d^B}(T_B)\\
      &\quad + \sum_{i=1}^2\left[\langle\sigma vE_{N_i}\rangle_{N_i d'{}^B \to \bar{d}^B \bar{u}^B}^{T_{N_i}, T_B}n_{N_i} - \langle\sigma vE_{N_i}\rangle_{N_i d'{}^B \to \bar{d}^B \bar{u}^B}^{T_B, T_B}n^{\text{eq}}_{N_i}(T_B) \right]n^{\text{eq}}_{d'{}^B}(T_B)\\
      &\quad + \sum_{i=1}^2\left[\langle\sigma vE_{N_i}\rangle_{N_i u^B \to \bar{d}^B \bar{d}'{}^B}^{T_{N_i}, T_B}n_{N_i} - \langle\sigma vE_{N_i}\rangle_{N_i u^B \to \bar{d}^B \bar{d}'{}^B}^{T_B, T_B}n^{\text{eq}}_{N_i}(T_B) \right]n^{\text{eq}}_{u^B}(T_B)\\
      &\quad - \sum_{f^A, {f'}^B}\langle \sigma v(E_{f^A}^{\text{out}} - E_{f^A}^{\text{in}})\rangle_{f^A {f'}^B \to f^A {f'}^B}^{T_A, T_B} n^{\text{eq}}_{f^A}(T_A)n^{\text{eq}}_{{f'}^B}(T_B)\\
      &\quad + \frac{1}{2}\sum_{f^A, {f'}^B}\left[\langle\sigma v E_+\rangle_{f^A \bar{f}^A \to {f'}^B \bar{f}'{}^B}^{T_A, T_A}(n^{\text{eq}}_{f^A}(T_A))^2  - \langle\sigma v E_+\rangle_{f^A \bar{f}^A \to {f'}^B \bar{f}'{}^B}^{T_B, T_B}(n^{\text{eq}}_{f^A}(T_B))^2 \right],
\end{align*}

\begin{align*}
   \frac{d \Delta B_A}{dt} &= 
        -3H\Delta B_A \\
      & + \langle\Delta\Gamma^{N_2}_{u^A d^A d'{}^A}\rangle^{T_{N_2}} n_{N_2} - \langle\Delta\Gamma^{N_2}_{u^A d^A d'{}^A}\rangle^{T_A} n_{N_2}^{\text{eq}}(T_A)\\
      & - 9\sum_{i=1}^2\langle\Gamma^{N_i}_{u^A d^A d'{}^A}\rangle^{T_A} \frac{n_{N_i}^{\text{eq}}(T_A)}{n_{q^A}^{\text{eq}}(T_A)}\Delta B_A\\
      & + \sum_{i=1}^2\left[\langle\Delta\sigma v\rangle_{N_i \bar{d}^A \to d'{}^A u^A}^{T_{N_i}, T_A}n_{N_i}
             -                   \langle\Delta\sigma v\rangle_{N_i \bar{d}^A \to d'{}^A u^A}^{T_A,     T_A}n_{N_i}^{\text{eq}}(T_A)\right]n_{d^A}^{\text{eq}}(T_A)\\
      & + \left[\langle\Delta\sigma v\rangle_{N_2 \bar{d}'{}^A \to d{}^A u^A}^{T_{N_2}, T_A}n_{N_2}
             -       \langle\Delta\sigma v\rangle_{N_2 \bar{d}'{}^A \to d{}^A u^A}^{T_A,     T_A}n_{N_2}^{\text{eq}}(T_A)\right]n_{d'{}^A}^{\text{eq}}(T_A)\\
      & - 3\sum_{i=1}^2\left[\langle\sigma v\rangle_{N_i d^A \to \bar{d}'{}^A \bar{u}^A}^{T_{N_i}, T_A}n_{N_i}
             + 2                  \langle\sigma v\rangle_{N_i d^A \to \bar{d}'{}^A \bar{u}^A}^{T_A,     T_A}n_{N_i}^{\text{eq}}(T_A)\right]\frac{n_{d^A   }^{\text{eq}}(T_A)}{n_{q^A}^{\text{eq}}(T_A)}\Delta B_A\\
      & - 3\sum_{i=1}^2\left[\langle\sigma v\rangle_{N_i d'{}^A \to \bar{d}^A \bar{u}^A}^{T_{N_i}, T_A}n_{N_i}
             + 2                  \langle\sigma v\rangle_{N_i d'{}^A \to \bar{d}^A \bar{u}^A}^{T_A,     T_A}n_{N_i}^{\text{eq}}(T_A)\right]\frac{n_{d'{}^A}^{\text{eq}}(T_A)}{n_{q^A}^{\text{eq}}(T_A)}\Delta B_A\\
      & - 3\sum_{i=1}^2\left[\langle\sigma v\rangle_{N_i u^A \to \bar{d}^A \bar{d}'{}^A}^{T_{N_i}, T_A}n_{N_i}
             + 2                  \langle\sigma v\rangle_{N_i u^A \to \bar{d}^A \bar{d}'{}^A}^{T_A,     T_A}n_{N_i}^{\text{eq}}(T_A)\right]\frac{n_{u^A   }^{\text{eq}}(T_A)}{n_{q^A}^{\text{eq}}(T_A)}\Delta B_A,
\end{align*}

\begin{align*}
  \frac{d \Delta B_B}{dt} &= 
       -3H\Delta B_B \\
      & + \langle\Delta\Gamma^{N_2}_{u^B d^B d'{}^B}\rangle^{T_{N_2}} n_{N_2} - \langle\Delta\Gamma^{N_2}_{u^B d^B d'{}^B}\rangle^{T_B} n_{N_2}^{\text{eq}}(T_B)\\
      & - 9\sum_{i=1}^2\langle\Gamma^{N_i}_{u^B d^B d'{}^B}\rangle^{T_B} \frac{n_{N_i}^{\text{eq}}(T_B)}{n_{q^B}^{\text{eq}}(T_B)}\Delta B_B\\
      & + \sum_{i=1}^2\left[\langle\Delta\sigma v\rangle_{N_i \bar{d}^B \to d'{}^B u^B}^{T_{N_i}, T_B}n_{N_i}
             -                   \langle\Delta\sigma v\rangle_{N_i \bar{d}^B \to d'{}^B u^B}^{T_B,     T_B}n_{N_i}^{\text{eq}}(T_B)\right]n_{d^B}^{\text{eq}}(T_B)\\
      & + \left[\langle\Delta\sigma v\rangle_{N_2 \bar{d}'{}^B \to d{}^B u^B}^{T_{N_2}, T_B}n_{N_2}
             -       \langle\Delta\sigma v\rangle_{N_2 \bar{d}'{}^B \to d{}^B u^B}^{T_B,     T_B}n_{N_2}^{\text{eq}}(T_A)\right]n_{d'{}^B}^{\text{eq}}(T_B)\\
      & - 3\sum_{i=1}^2\left[\langle\sigma v\rangle_{N_i d^B \to \bar{d}'{}^B \bar{u}^B}^{T_{N_i}, T_B}n_{N_i}
             + 2                  \langle\sigma v\rangle_{N_i d^B \to \bar{d}'{}^B \bar{u}^B}^{T_B,     T_B}n_{N_i}^{\text{eq}}(T_B)\right]\frac{n_{d^B   }^{\text{eq}}(T_B)}{n_{q^B}^{\text{eq}}(T_B)}\Delta B_B\\
      & - 3\sum_{i=1}^2\left[\langle\sigma v\rangle_{N_i d'{}^B \to \bar{d}^B \bar{u}^B}^{T_{N_i}, T_B}n_{N_i}
             + 2                  \langle\sigma v\rangle_{N_i d'{}^B \to \bar{d}^B \bar{u}^B}^{T_B,     T_B}n_{N_i}^{\text{eq}}(T_B)\right]\frac{n_{d'{}^B}^{\text{eq}}(T_B)}{n_{q^B}^{\text{eq}}(T_B)}\Delta B_B\\
      & - 3\sum_{i=1}^2\left[\langle\sigma v\rangle_{N_i u^B \to \bar{d}^B \bar{d}'{}^B}^{T_{N_i}, T_B}n_{N_i}
             + 2                  \langle\sigma v\rangle_{N_i u^B \to \bar{d}^B \bar{d}'{}^B}^{T_B,     T_B}n_{N_i}^{\text{eq}}(T_B)\right]\frac{n_{u^B   }^{\text{eq}}(T_B)}{n_{q^B}^{\text{eq}}(T_B)}\Delta B_B.
\end{align*}
A few comments are in order: 
\begin{itemize}
\item We defined $n^{\text{eq}}_{q^M}(T_M)$ via
\begin{equation}\label{eq:nqM}
  n_{q^M}^{\text{eq}}(T_M) = \sum_{i \in \{\text{quarks $M$}\}} n^{\text{eq}}_i(T_M).
\end{equation}
These appear in the equations because the baryon asymmetries redistribute themselves amongst the different quarks of a given sector.
\item We defined $\langle\Gamma\rangle^T$ via
\begin{equation}\label{eq:EonMexp}
  \frac{\langle\Gamma\rangle^T}{\Gamma} = \langle\frac{m}{E}\rangle^T = \langle\frac{1}{\gamma}\rangle^T = \langle \sqrt{1 - v^2} \rangle^T = \frac{K_1(m/T)}{K_2(m/T)}.
\end{equation}
In practice, this is simply the decay rate corrected by the fact that boosted particles decay more slowly.
\item In equilibrium at temperature $T$, massive particles respect
\begin{equation}\label{eq:nrhop}
  \begin{aligned}
    n^\text{eq}(T)    &= \frac{gm^2T  }{2\pi^2}K_2(m/T)\\
    \rho^\text{eq}(T) &= \frac{gm^2T  }{2\pi^2}\left(mK_1(m/T) + 3TK_2(m/T)\right)\\
    P^\text{eq}(T)    &= \frac{gm^2T^2}{2\pi^2}K_2(m/T)
  \end{aligned}
\end{equation}
These equations do not hold out-of-equilibrium, but ratios such as $\rho/n$ remain unchanged.  For massive particles, this means that we can obtain their temperature from the ratio of their energy and number densities. At leading order, we have
\begin{equation}\label{eq:TfromrhoandnT}
  T = \frac{2m}{3}\left(\frac{\rho}{nm} - 1\right).
\end{equation}
This is the result expected from the equipartition theorem.
\item Assume a process $ij \to mn$ and its inverse $mn \to ij$. At a given temperature $T$, the definition of thermal equilibrium implies
\begin{equation}\label{eq:CSconverstion}
  \langle \sigma v\rangle_{ij \to mn}^{T,T} n_i^{\text{eq}}(T) n_j^{\text{eq}}(T) = \langle \sigma v\rangle_{mn \to ij}^{T,T} n_m^{\text{eq}}(T) n_n^{\text{eq}}(T).
\end{equation}
This equation was used to simplify the evolution equations. Generalization to energy transfers and decays is trivial.
\item The Hubble constant is given by
\begin{equation}\label{eq:Hubble}
  H^2 = \frac{8\pi G\rho_{\text{tot}}}{3},
\end{equation}
where $\rho_{\text{tot}} = \rho_{N_1} + \rho_{N_2} + \rho_A + \rho_B$ is the total energy density.
\item The quantity $g_{*M}$ corresponds to the effective number of relativistic degrees of freedom in sector $M$ and is computed following standard procedure. A prime represents a derivative with respect to $T_M$.
\item The QCD and mirror QCD coupling constants are assumed to unify at high enough scale and are run at one loop order.
\item In the $d\rho_M/dt$ equations, the first term is simply $3H(\rho_M + P_M)$, where we took into account that $P_M$ is not exactly $\rho_M/3$ when particles are not fully relativistic. The correction factor takes values in the range $[3/4, 1]$. This is interesting as, when $g_*'$ goes to infinity, the energy density scales as the number density. From Eq.~\eqref{eq:TfromrhoandnT}, this means that the temperature remains constant during that time. This is why a loss of degrees of freedom in one sector results in the temperature of that sector rising with respect to the other sector.
\item The inverse decay of $N_2$ to $N_1$ and quarks or mirror quarks is neglected. Its treatment is complicated, but Boltzmann suppression ensures that it is negligible. 
\item In some regions of parameter space, the decay $u^M \to N_i \bar{d}^M \bar{d}'{}^M$ is allowed. In the $B$ sector, the large mass of the mirror top and Boltzmann suppression render this effect negligible. In the $A$ sector, the decay width to this channel is typically sufficiently small that it would only come into play once the top density is negligible. As such, this effect is neglected.
\item At sufficiently high temperatures, certain processes like $Z^A Z^B \to Z^A Z^B$ can contribute significantly to energy exchange between the $A$ and $B$ sectors. However, such high temperatures result in the $A$ and $B$ sectors having almost identical temperatures and their inclusion would have a negligible effect. This is the same reason why the Higgs can be assumed heavy in Eqs.~\eqref{eq:CSfAfBp->fAfBp}, \eqref{eq:CStfAfBp->fAfBp}, \eqref{eq:CSfAfAbar->fBpfBpbar} and \eqref{eq:CStfAfAbar->fBpfBpbar}.
\item The evolution equations are not very stiff in the regions of parameter space studied in this paper. The only exception is for annihilation/scattering between fermions of both sectors via Higgs exchange. At very high temperatures, this process can take place at a rate too high to easily manage numerically. Thankfully, this also means that the two sectors have extremely close temperatures. As such, the problem can be circumvented by treating the $A$ and $B$ sectors as a single population. This is done when the rate at which a sector can exchange energy with the other sector via Higgs exchange is much larger than the rate it receives energy from other sources.
\end{itemize}

\section{Higgs signal strengths constraints}\label{Sec:HiggsSS}
In this section, we discuss how the bounds on the Higgs couplings are applied. We follow the procedure of Ref.~\cite{Beauchesne:2020mih} which is based on the $\kappa$ formalism \cite{Heinemeyer:2013tqa}, albeit the present situation is considerably simpler. Assume a production mechanism $i$ with cross section $\sigma_i$ or decay process $i$ with width $\Gamma_i$. The parameter $\kappa_i$ is defined such that 
\begin{equation}\label{eq:Defkappa}
  \kappa_i^2 = \frac{\sigma_i}{\sigma_i^{\text{SM}}} \quad \text{or} \quad \kappa_i^2 = \frac{\Gamma_i}{\Gamma_i^{\text{SM}}},
\end{equation}
where $\sigma_i^{\text{SM}}$ and $\Gamma_i^{\text{SM}}$ are the corresponding SM quantities. For the Mirror Twin Higgs, all $\kappa_i$ are equal at leading order and given by
\begin{equation}\label{eq:kappaGeneral}
  \kappa = \frac{v^B}{\sqrt{(v^A)^2 + (v^B)^2}}.
\end{equation}
In addition, the Higgs can also decay to mirror particles that escape the detector unseen. The decay width to a pair of mirror fermions $f^B$ is
\begin{equation}\label{eq:hffmirror}
  \Gamma^h_{\bar{f}^B f^B} = \frac{N_c}{8\pi}\frac{m_{f^A}^2}{(v^A)^2 + (v^B)^2}\frac{(m_h^2 - 4 m_{f^B}^2)^{3/2}}{m_h^2},
\end{equation}
where $N_c$ is the number of mirror colours of $f^B$. The decay width to a pair of mirror gluons is given by
\begin{equation}\label{eq:hbbmirror}
  \Gamma^h_{g^B g^B} = \frac{(\alpha_S^B)^2 m_h^3}{128\pi^3}\left(\frac{v^A}{v^B}\right)^2\frac{1}{(v^A)^2 + (v^B)^2}\left|\sum_{f^B} F\left(\frac{4m_{f^B}^2}{m_h^2}\right)\right|^2,
\end{equation}
where
\begin{equation}\label{eq:FFunction}
  F(\tau) = -2\tau(1 + (1 - \tau)f(\tau)),
\end{equation}
with
\begin{equation}\label{eq:dFunction}
  f(\tau) = \Bigg\{ \begin{tabular}{cc} $\arcsin^2 \sqrt{\frac{1}{\tau}}$ & if $\tau \geq 1$,\\ $-\frac{1}{4}\left[\ln\left(\frac{1 + \sqrt{1 - \tau}}{1 - \sqrt{1 - \tau}}\right) - i\pi\right]^2$ & if $\tau < 1$, \end{tabular}
\end{equation}
and $\alpha_S^B$ is the mirror strong coupling constant. Because of the constraints on $v^B/v^A$, decays to mirror massive gauge bosons require both gauge bosons to be off-shell and can therefore be neglected. All other decays to mirror particles are also negligible.

\begin{figure}[t]
  \begin{center}
    \includegraphics[width=0.9\textwidth]{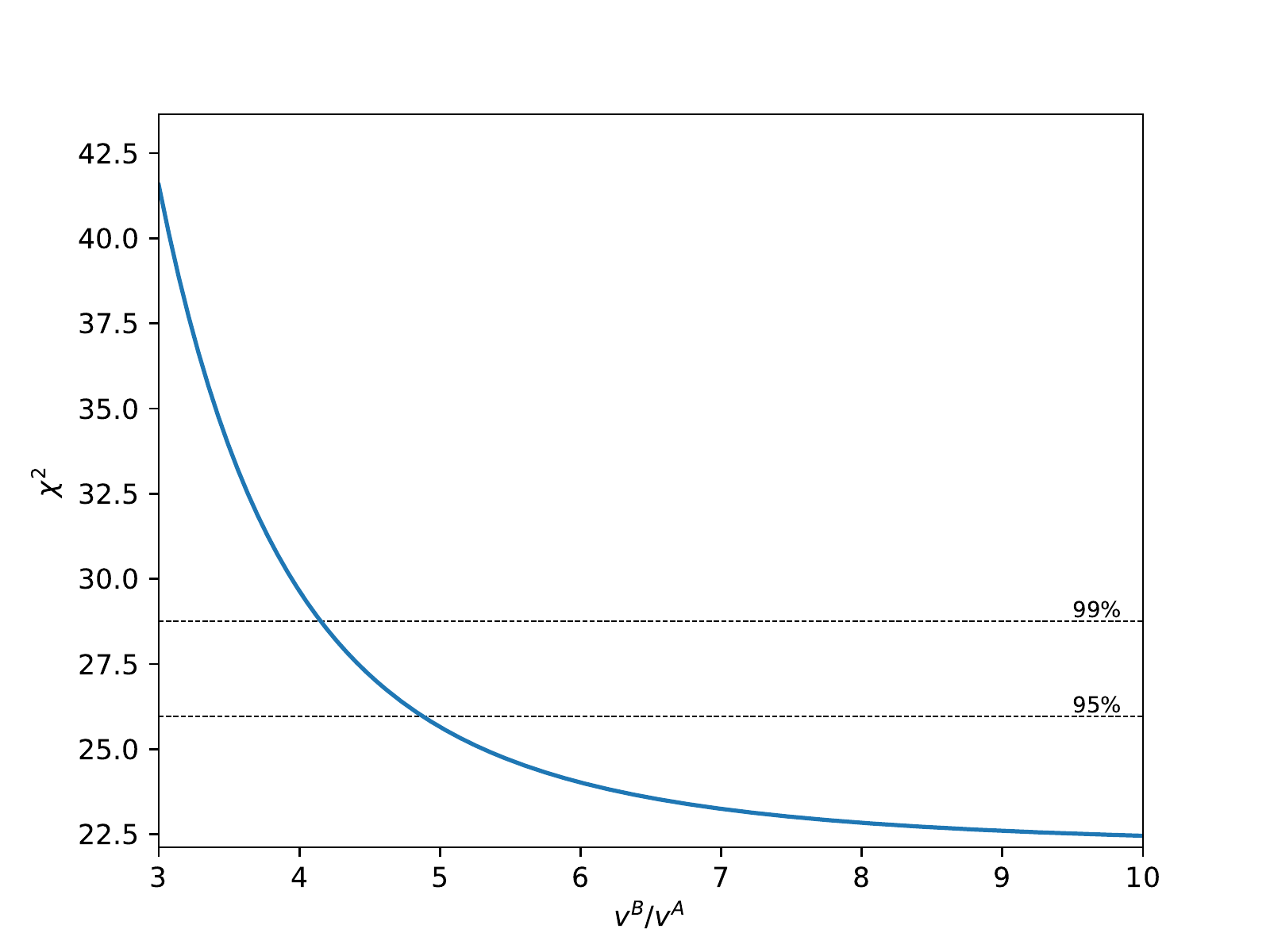}
  \end{center} 
  \caption{$\chi^2$ fit of the Higgs signal strengths of the Twin Higgs based on the measurements of Refs.~\cite{Aad:2019mbh} and~\cite{CMS-PAS-HIG-19-005}. The horizontal lines correspond to the 95\% and 99\% confidence level limits on $\chi^2$.}\label{fig:HiggsSS}
\end{figure}

With the above results, limits on the ratio $v^B/v^A$ can be obtained using the searches of Ref.~\cite{Aad:2019mbh} by ATLAS and Ref.~\cite{CMS-PAS-HIG-19-005} by CMS. These are the most up-to-date available global fits of the Higgs signal strengths and provide all the information necessary to perform a $\chi^2$ fit within the $\kappa$ formalism. The branching ratio to invisible is not directly constrained by these searches, but indirectly via the reduction of the signal strengths of visible channels. In the allowed range of $v^B/v^A$, this branching ratio is far below current constraints (see for example Refs.~\cite{Aaboud:2019rtt, Sirunyan:2018owy}), a fact that was already noted for the Twin MSSM in Ref.~\cite{Craig:2013fga}. A simple $\chi^2$ fit is performed combining the results of the two experiments and assuming no correlations between them. The results are shown in Fig.~\ref{fig:HiggsSS} and give the following limits
\begin{equation}\label{eq:vBonvAlimits}
  95\%: \frac{v^B}{v^A} > 4.86, \;\;\;\;\; 99\%: \frac{v^B}{v^A} > 4.13.
\end{equation}

\section{Computation of the mirror atom abundance}\label{App:XDAComp}
In this section, we explain how the mirror atom abundance is computed under the assumption that the mirror proton is heavier than the mirror neutron. We follow the procedure of Ref.~\cite{Beauchesne:2020mih}, from which we summarize the most important elements and to which we refer for more details. Three quantities first need to be computed.
\begin{itemize}
  \item The mirror QCD scale is computed by requesting that the strong coupling constants of both sectors unify at high enough scale using one loop beta functions.
  \item The binding energy of mirror deuteron $B_{D^B}$ is computed via
  \begin{equation}\label{eq:BD}
    \frac{B_{D^B}}{\Lambda_\text{QCD}^B} = B_1 \frac{m_{\pi^B}}{\Lambda_\text{QCD}^B} + B_2,
  \end{equation}
  with $B_1\approx 0.033$ and $B_2\approx -0.011$. This equation is obtained by a fit of the lattice QCD results of Refs.~\cite{Orginos:2015aya, Savage:2015eya, Beane:2011iw, Beane:2012vq, Yamazaki:2012hi, Yamazaki:2015asa} which compute the binding energy of deuteron for different pion masses and the appropriate rescaling. Note that these results contain large uncertainties and as such we will limit all computations to simple approximations.
  \item The difference between the masses of the mirror proton and neutron is given by
  \begin{equation}\label{eq:Deltpn}
    m_{pn}^B = m_{p^B} - m_{n^B} = C_0\left(C_1(m_{u^B} - m_{d^B}) + C_2 \alpha_{\text{EM}} \Lambda_\text{QCD}^B\right),
  \end{equation}
  where $\alpha_{\text{EM}}$ is the fine structure constant, $C_1\approx 0.86$, $C_2\approx 0.54$ and $C_0$ is fixed to reproduce the equivalent SM value of $m_{pn}^A$. This result also comes from lattice QCD and is obtained from Fig.~3 of Ref.~\cite{Borsanyi:2014jba} or alternatively Table~2.
\end{itemize}

With these three quantities, the ratio of mirror proton and mirror neutron abundances can be computed following Refs.~\cite{Kolb:1990vq, Mukhanov:2003xs, Chacko:2018vss}. At high temperatures, collisions with electrons and neutrinos maintain the protons and neutrons in equilibrium. This proceeds at a rate
\begin{equation}\label{eq:Conversion}
  \Gamma_{p^B e^B \to n^B \nu_e^B}= \frac{1 + 3g_A^2}{2\pi^3}(G_F^B)^2 (m_{pn}^B)^5 J\left(-\infty, -\frac{m_{e^B}}{m_{pn}^B}\right),
\end{equation}
where
\begin{equation}
  J(a, b) = \int_a^b\sqrt{1 - \frac{(m_{e^B}/m_{pn}^B)^2}{q^2}}\frac{q^2(q - 1)^2 dq}{\left(1 + e^{\frac{m_{pn}^B}{T_\nu^B}(q - 1)}\right)\left(1 + e^{-\frac{m_{pn}^B}{T_B}q}\right)},
\end{equation}
where $g_A = 1.27$, $G_F^B$ the mirror Fermi constant and $T_\nu^B$ the temperature of the mirror neutrinos, i.e. $T_B$ before and $(4/11)^{1/3}T_B$ after electron recombination. Conversion freezes-out at a temperature of the mirror sector $T^{\text{FO}^B}_B$ at which this rate is equal to the Hubble expansion rate. At this time, the ratio of abundances is $n_{p^B}/n_{n^B}\approx f_1 \approx \exp(-m_{pn}^B/T^{\text{FO}^B}_B)$. As long as they are unstable, free protons continue to decay until deuterium formation at a rate of
\begin{equation}\label{eq:GammapB}
  \Gamma^{p^B}_{n^B e^B \nu_e^B} = \frac{1 + 3g_A^2}{2\pi^3}(G_F^B)^2 m_{e^B}^5\lambda_0(m_{pn}^B/m_{e^B}),
\end{equation}
where
\begin{equation}\label{eq:gamma0}
  \lambda_0(Q) = \int_1^Q dq q(q - Q)^2(q^2 - 1)^{1/2}.
\end{equation}
The mirror deuterium bottleneck is crossed when the mirror sector reaches the temperature $T_B^{\text{DB}^B} \approx (B_{D^B}/B_{D^A}) T_A^{\text{DB}^A}$, where $T_A^{\text{DB}^A}$ is 0.08~MeV \cite{Mukhanov:2003xs}. This occurs at $t^{\text{DB}^B} \approx 0.301 g_\star^{-1/2} m_{\text{Pl}}r_T^2/(T_B^{\text{DB}^B})^2$, meaning that $n_{p^B}/n_{n^B}$ further decreased by a factor of $f_2 \approx \exp(-\Gamma^{p^B}_{n^B e^B \nu_e^B}t^{\text{DB}^B})$. At the onset of deuterium formation, the proton to neutron abundance ratio is then $(n_{p^B}/n_{n^B})^{\text{DB}^B} \approx f_1 \times f_2$. Almost all mirror protons are quickly absorbed into Helium-4 nuclei. If the splitting between the mirror up and mirror down is not too extreme, Helium-4 can safely be assumed to be stable and almost all neutrons end up in this isotope. If the splitting is very large, the abundance of mirror protons is bound to be far below experimental constraints and its exact abundance is irrelevant to us. As such, we perform the computation assuming mirror Helium-4 to be stable and mention that the results might not be accurate for extremely low abundances. The final fraction of dark atoms $X_{\text{DA}}$ is then
\begin{equation}\label{eq:FracAtom}
  X_{\text{DA}} \approx \frac{2(n_{p^B}/n_{n^B})^{\text{DB}^B}}{1 + (n_{p^B}/n_{n^B})^{\text{DB}^B}}.
\end{equation}

\bibliography{biblio}
\bibliographystyle{utphys}

\end{document}